\documentclass[11pt, a4paper]{article}
\usepackage{jheppub}
\usepackage{latexsym}
\usepackage{multirow}
\usepackage{color}
\usepackage{graphicx}
\usepackage{grffile}
\usepackage{epsfig}
\usepackage{epsf}
\usepackage{dcolumn}
\usepackage{bm}
\usepackage{textcomp}
\usepackage{float}
\usepackage{subfig}
\usepackage{hypcap}
\usepackage[]{hyperref}
\usepackage{makecell}
\usepackage{color}
\usepackage{pifont}
\usepackage{appendix}
\usepackage{amsmath}
\usepackage{multirow}
\usepackage{bigdelim}  
\usepackage{lineno}
\usepackage[normalem]{ulem}
\usepackage{adjustbox}
\usepackage{tablefootnote}
\usepackage{tabularx}
\usepackage{mathrsfs}
\usepackage{slashed}
\usepackage[11pt]{moresize}
\usepackage[usenames, dvipsnames, table]{xcolor}

\hypersetup{
 bookmarks=true,         
 unicode=false,          
 pdftoolbar=true,        
 pdfmenubar=true,        
 pdffitwindow=true,      
 pdfstartview={FitH},    
 pdfsubject={Neutrino Oscillations Phenomenology},   
 pdfnewwindow=true,      
 pdfcreator={RevTeX},
 colorlinks=true,        
 linkcolor=red,          
 citecolor=blue,         
 filecolor=black,        
 urlcolor=blue,          
}

\def\bra#1{\langle{#1}|}
\def\ket#1{|{#1}\rangle}
\newcommand{\be}{\begin{equation}}
\newcommand{\ee}{\end{equation}}
\newcommand{\ba}{\begin{eqnarray}}
\newcommand{\ea}{\end{eqnarray}}
\newcommand{\ldm}{\ensuremath{{\Delta m_{31}^2}}}         
\newcommand{\sdm}{\ensuremath{{\Delta m_{21}^2}}}

\newcommand{\capdef}{}
\newcommand{\mycaption}[2][\capdef]{\renewcommand{\capdef}{#2}
       \caption[#1]{{\footnotesize #2}}}
\makeatletter
\renewcommand{\fnum@table}{\textbf{\tablename~\thetable}}
\renewcommand{\fnum@figure}{\textbf{\figurename~\thefigure}}
\makeatother

\newenvironment{conditions*}
{\par\vspace{\abovedisplayskip}\noindent
  \tabularx{\columnwidth}{>{$}l<{$} @{${}={}$} >{\raggedright\arraybackslash}X}}
{\endtabularx\par\vspace{\belowdisplayskip}}

\preprint{IP/BBSR/2021-06}

\title{Probing Lorentz Invariance Violation with Atmospheric Neutrinos at INO-ICAL}
\author[a,b]{Sadashiv Sahoo}
\author[a,b,c]{Anil Kumar,}
\author[a,b,d]{Sanjib Kumar Agarwalla}

\affiliation[a]{Institute of Physics, Sachivalaya Marg, Sainik School Post,
  Bhubaneswar 751005, India}
\affiliation[b]{Homi Bhabha National Institute, Anushakti Nagar,
  Mumbai 400094, India}
\affiliation[c]{Applied Nuclear Physics Division, Saha Institute of
  Nuclear Physics, Block AF, Sector 1, Bidhannagar, Kolkata 700064, India}
\affiliation[d]{International Centre for Theoretical Physics,
  Strada Costiera 11, 34151 Trieste, Italy}

\emailAdd{sadashiv.sahoo@iopb.res.in (ORCID: 0000-0001-6719-7723)}
\emailAdd{anil.k@iopb.res.in (ORCID: 0000-0002-8367-8401)}
\emailAdd{sanjib@iopb.res.in (ORCID: 0000-0002-9714-8866)}

\abstract{
 The possibility of Lorentz Invariance Violation (LIV) may appear in unified theories, such as string theory, which allow the existence of a new space-time structure at the Planck scale ($M_p \sim 10^{19}$ GeV). This effect can be observed at low energies with a strength of $\sim 1/M_p$ using the perturbative approach. In the minimal Standard Model extension (SME) framework, the neutrino mass-induced flavor oscillation gets modified in the presence of LIV. The Iron Calorimeter (ICAL) detector at the proposed India-based Neutrino Observatory (INO) offers a unique window to probe these LIV parameters by observing atmospheric neutrinos and antineutrinos separately over a wide range of baselines in the multi-GeV energy range. In this paper, for the first time, we study in detail how the CPT-violating LIV parameters $(a_{\mu\tau}, a_{e\mu}, a_{e\tau})$ can alter muon survival probabilities and expected $\mu^-$ and $\mu^+$ event rates at ICAL. Using 500 kt$\cdot$yr exposure of ICAL, we place stringent bounds on these CPT-violating LIV parameters at 95\% C.L., which are slightly better than the present Super-Kamiokande limits. We demonstrate the advantage of incorporating hadron energy information and charge identification capability at ICAL while constraining these LIV parameters. Further, the impact of the marginalization over the oscillation parameters and choice of true values of $\sin^2\theta_{23}$ on LIV constraints is described. We also study the impact of these LIV parameters on mass ordering determination and precision measurement of atmospheric oscillation parameters.
}

\arxivnumber{2110.13207}

\begin{document}
\maketitle
\flushbottom

\section{Introduction and motivation}
\label{sec:introduction}

The Lorentz symmetry has been preserved in the fundamental theories of physics, such as the general theory of relativity and the quantum field theory. However, this symmetry could be challenged at the Planck scale physics ($M_p \sim 10^{19}$ GeV), where the unification of gravity with the gauge fields of the Standard Model (SM) of particle physics is expected. Quantum loop gravity~\cite{Gambini:1998it, Alfaro:2002xz, Sudarsky:2002ue, Amelino-Camelia:2002aqz, Ng:2003jk} and String theory~\cite{Polyakov:1987ez, Kostelecky:1988zi, Kostelecky:1991ak, Kostelecky:1995qk, Kostelecky:2000hz, Kostelecky:1999mu} attempt such unification by allowing small perturbation of Lorentz symmetry breaking, so-called Lorentz Invariance Violation (LIV). Even an introduction of non-commutativity in space-time structure at local fields~\cite{Carroll:2001ws, Mocioiu:2001fx} can give rise to a breaking of charge, parity, and time-reversal (CPT) symmetry, so-called CPT violation, which is indeed a special case of LIV~\cite{Greenberg:2002uu}. At the low energy observables, the coupling strength of LIV is expected to be suppressed by an order of ($1/M_p$)~\cite{Myers:2003fd, Mavromatos:2003qc}. In order to accommodate all possible LIV interactions with the presently known SM physics, the suitable framework is the Standard Model extension (SME)~\cite{Colladay:1998fq, Kostelecky:2003fs, Bluhm:2005uj, Colladay:1996iz}, where the LIV terms act as observer scalars to the SM fields. In principle, the presence of LIV can be probed via three basic mechanisms like coherence, interference, and extreme effects. In the interference category, the mass-induced neutrino flavor transition, so-called neutrino oscillation~\cite{Pontecorvo:1967fh}, is a potential candidate to probe LIV~\cite{Super-Kamiokande:2014exs,Diaz:2015dxa,Arguelles:2016rkg,IceCube:2017qyp,Katori:2019xpc,KumarAgarwalla:2019gdj}.

The neutrino oscillations allow neutrinos to change their flavor while traveling. The Super-Kamiokande (Super-K) experiment discovered the phenomenon of neutrino oscillation in their atmospheric neutrino data for the first time in 1998~\cite{Fukuda:1998mi}. The neutrino oscillation is parameterized in terms of three mixing angles ($\theta_{12}$, $\theta_{13}$, $\theta_{23}$), two mass-squared differences ($\Delta m^2_{21}$, $\Delta m^2_{32}$), and one CP-violating phase ($\delta_{\rm CP}$). There has been significant progress in the past two decades, and now, neutrino oscillation is a well-established model which has entered into an era of precision measurement. In the standard three-flavor neutrino oscillation framework, only a few parameters have been left to be measured precisely, such as CP-violating phase ($\delta_{\rm CP}$), atmospheric mixing angle ($\theta_{23}$), and neutrino mass ordering. This is also important to note that neutrino oscillations can only be explained if neutrinos have non-zero degenerate masses. The non-zero neutrino masses provide one of the strongest hints towards physics beyond the Standard Model (BSM). Thus, neutrino oscillation is a legitimate area to look for hints towards BSM physics.

The atmospheric neutrinos provide an avenue to study neutrino oscillations in the multi-GeV range of energies over a wide range of baselines starting from 10 km to 10\textsuperscript{4} km. The upward-going neutrinos travel deep inside the Earth and experience Earth's matter effect~\cite{Wolfenstein:1977ue,Mikheev:1986gs,Mikheev:1986wj,Petcov:1998su,Chizhov:1998ug,Petcov:1998sg,Chizhov:1999az,Chizhov:1999he,Akhmedov:1998ui,Akhmedov:1998xq} due to the interactions with ambient electrons. Apart from providing crucial information on standard three-flavor oscillation parameters, the atmospheric neutrinos can also play an important role to probe various BSM scenarios like non-standard interactions (NSI)~\cite{Wolfenstein:1977ue,Guzzo:1991hi,Super-Kamiokande:2011dam,Ohlsson:2012kf,Esmaili:2013fva,Miranda:2015dra,Salvado:2016uqu,IceCube:2017zcu,Farzan:2017xzy,Esteban:2018ppq,Bhupal:2019qno,Khatun:2019tad,Arguelles:2019xgp,KhanChowdhury:2020xev,Blot:2020ekg,Kumar:2021lrn,KumarAgarwalla:2021twp,IceCube:2021abg,HernandezRey:2021qac}, sterile neutrinos~\cite{Razzaque:2011ab,Abazajian:2012ys,Super-Kamiokande:2014ndf,IceCube:2016rnb,IceCube:2017ivd,Blennow:2018hto,Denton:2018dqq,Miranda:2018buo,ANTARES:2018rtf,Moulai:2019gpi,KhanChowdhury:2020qqu,IceCube:2020phf,IceCube:2020tka,Razzaque:2021cft,Schneider:2021wzs,KM3NeT:2021uez}, neutrino decay~\cite{Chikashige:1980ui,Gelmini:1980re,Gelmini:1983ea}, LIV~\cite{Super-Kamiokande:2014exs,Arguelles:2016rkg,IceCube:2017qyp,Katori:2019xpc,KumarAgarwalla:2019gdj}, and several other new physics models~\cite{Super-Kamiokande:2011dgc,Coelho:2017cwp}. The signature of LIV, which is the topic of this paper, can be explored at several $L/E$ values accessible in the case of atmospheric neutrinos. An atmospheric neutrino detector having good resolutions in energy and direction will be able to observe the possible modifications in standard three-flavor oscillations due to LIV.

The proposed 50 kt Iron Calorimeter (ICAL) detector at the India-based Neutrino Observatory (INO)~\cite{Kumar:2017sdq} aims to measure neutrino mass ordering by separately detecting atmospheric neutrinos and antineutrinos in the multi-GeV range of energies over a wide range of baselines. Harnessing the magnetic field of 1.5 T~\cite{Behera:2014zca}, the ICAL detector would be able to identify $\mu^-$ and $\mu^+$ events separately. The ICAL has an excellent muon energy resolution of about 10 to 15\% in the reconstructed muon energy range of 1 to 25 GeV~\cite{Chatterjee:2014vta}. As far as the muon direction is concerned, the ICAL detector is capable of providing muon angular resolution of less than $1^\circ$. Using this good detector response, the ICAL Collaboration has performed various analyses involving standard oscillations using atmospheric neutrino~\cite{GOSWAMI2009198,Ghosh:2012px,Thakore:2013xqa,Ghosh:2013mga,Devi:2014yaa,Kumar:2017sdq,Mohan:2016gxm,Rebin:2018fdl,Datta:2019uwv,Kumar:2020wgz,Kumar:2021faw}. ICAL provides a unique avenue to probe a plethora of BSM scenarios~\cite{Khatun:2019tad,Kumar:2021lrn,Thakore:2018lgn,Dash:2014fba,Behera:2016kwr,Khatun:2017adx,Choubey:2017eyg,Khatun:2018lzs,Choubey:2017vpr,Tiwari:2018gxz}. In the context of searching the signature of CPT violation at ICAL, a dedicated work has been performed in Ref.~\cite{Chatterjee:2014oda} by introducing the new CPT-violating parameters in the effective Hamiltonian as suggested by the authors in Ref.~\cite{Coleman:1998ti}. Since ICAL can measure the atmospheric mass-squared differences using atmospheric neutrinos and antineutrinos separately, any possible differences in their measurements would be a smoking gun signature of the CPT-violation, which may also indicate the possible violation of Lorentz Invariance. This possibility has been explored by the INO Collaboration in Refs.~\cite{Kaur:2017dpd,Dar:2019mnk}. In the present work, for the first time, we explore in detail the impact of CPT-violating LIV parameters in the SME framework using the atmospheric neutrinos at the ICAL detector. The sensitivity of ICAL towards the presence of CPT-violating LIV parameters is estimated using the ICAL detector with 500 kt$\cdot$yr exposure, and stringent constraints are placed.  We will also demonstrate the impact of the presence of LIV on the measurement of neutrino mass ordering and oscillation parameters.

In section~\ref{sec:liv_theory}, we describe the theoretical background of LIV in the context of the neutrino oscillations. The experimental attempts to probe LIV are summarized in section~\ref{sec:LIV_history}. The impacts of LIV on neutrino oscillograms are presented in section~\ref{sec:oscillograms}. In section~\ref{sec:event_generation}, we elaborate the method to simulate neutrino events at the ICAL detector and modifications in the event distributions due to the presence of LIV. The method to perform the statistical analysis is explained in section~\ref{sec:statistical_analysis} which is followed by the results in section~\ref{sec:results} where we constrain the CPT-violating LIV parameters and show the impact of LIV on the measurement of mass ordering and precision measurement. Finally, we summarize our findings and conclude in section~\ref{sec:conclusion}. In appendix~\ref{app:L_CPTV}, we describe the properties of gauge invariant LIV parameters. The effect of LIV parameters on the appearance channel is discussed in appendix~\ref{app:Peu_emu_etau}. The appendix~\ref{app:effective_regions_emu_etau} presents some additional plots identifying effective regions in the plane of energy and direction of reconstructed muons while constraining LIV parameters. 

\section{Lorentz and CPT violation in neutrino oscillations}
\label{sec:liv_theory}

The study of atmospheric neutrinos provides an avenue to probe neutrino oscillations as an interferometer with a length scale of the diameter of Earth, which may also get affected by the presence of LIV. In order to understand the effect of LIV on neutrino oscillation probabilities, we first look at the modified Hamiltonian in the presence of LIV keeping only the terms which are gauge-invariant and renormalizable under the minimal SME framework. In the ultra-relativistic limit, the effective Hamiltonian $\big(\mathcal{H}_{\rm eff}\big)_{ij}$ describing the propagation of left-handed neutrino in vacuum can be written in the following fashion:
  \begin{align}
  \big(\mathcal{H}_{\rm eff}\big)_{ij} & = E \delta_{ij}
  + \frac{m^2_{ij}}{2E}
  + \frac{1}{E}\big(a^\mu_{L} p_\mu
  - c^{\mu\nu}_{L} p_\mu p_\nu \big)_{ij}\,,
  \label{eq:liv2}
  \end{align}
where, $i$ and $j$ are the neutrino flavor indices, whereas $p_\mu$ and $E$ are the four momenta and energy of neutrino, respectively. In Eq.~\ref{eq:liv2}, the first two terms correspond to the standard kinematics\footnote{Here, $m^2_{ij} = U\cdot {\rm diag}(m^2_1,\,m^2_2,\,m^2_3)\cdot U^\dagger$, where $U$ is the unitary neutrino mixing matrix known as the Pontecorvo-Maki-Nakagawa-Sakata (PMNS) matrix ~\cite{Pontecorvo:1957qd, Maki:1962mu, Pontecorvo:1967fh}.}, whereas the last two terms denote the LIV interactions governed by the parameters $a^\mu_L$ (CPT-violating LIV parameters) and $c^{\mu\nu}_L$ (CPT-conserving LIV parameters). While going from neutrino to antineutrino, the CPT-violating LIV parameters change their sign, whereas the sign of the CPT-conserving parameters remain unchanged (see the discussion in detail in appendix~\ref{app:L_CPTV}). 
  
So far, we consider the neutrino propagation in vacuum, but in reality, the neutrinos propagate through Earth's matter. The electron neutrinos undergo $W$-mediated charged-current interactions with the ambient electrons. This coherent and forward-scattering of $\nu_e$ with matter electrons gives rise to the so-called ``Wolfenstein matter potential''~\cite{Wolfenstein:1977ue, Mikheyev:1985zog, Mikheev:1986wj}. Note that the Wolfenstein matter potential doesn't get affected by the presence of non-zero LIV parameters. In the present work, we focus on the isotropic\footnote{One of the choices for the isotropic frame is the Sun-centered celestial-equatorial frame~\cite{Kostelecky:2008ts}. In this frame, we ignore the anisotropy generated due to the Earth's boost, which is suppressed by a factor of $\sim 10^{-4}$.} component ($\mu = 0,\,\nu = 0)$ of the LIV parameters and denote  
\begin{equation}
a^0_{L} \equiv a,\,\, c^{00}_{L} \equiv c, \,\, p^0 \equiv E.
\end{equation}
Therefore, the effective Hamiltonian $\big(\mathcal{H}_{\rm eff}\big)_{ij}$ for left handed neutrino boils down to
\begin{align}
\mathcal{H}_{\rm eff}  = &\frac{1}{2E} 
U\left(\begin{array}{ccc}
0 & 0 & 0                 \\
0 & \Delta m^{2}_{21} & 0 \\
0 & 0 & \Delta m^{2}_{31} \\
\end{array}\right)U^{\dagger}
+ 
\left(\begin{array}{ccc}
a_{ee} & a_{e\mu} & a_{e\tau} \\
a^*_{e\mu} & a_{\mu\mu} & a_{\mu\tau} \\
a^*_{e\tau} & a^*_{\mu\tau} & a_{\tau\tau}
\end{array} \right)
-\frac{4}{3}
E \left(
\begin{array}{ccc}
c_{ee} & c_{e\mu} & c_{e\tau} \\
c^*_{e\mu} & c_{\mu\mu} & c_{\mu\tau} \\
c^*_{e\tau} & c^*_{\mu\tau} & c_{\tau\tau}
\end{array}
\right) \nonumber \\
&+ \sqrt{2}G_{F}N_{e}
\left(\begin{array}{ccc}
1 & 0 & 0 \\ 
0 & 0 & 0 \\ 
0 & 0 & 0
\end{array}\right),
\label{eq:Heff_liv}
\end{align}
here, $U$ is the standard 3$\times$3 neutrino mixing matrix called PMNS matrix, $\Delta m^2_{ab}\,(\equiv m^2_a - m^2_b)$ are the mass-squared splittings of the neutrino mass eigenstates, and $\sqrt{2}G_{F}N_{e}$ is matter potential where $G_F$ is Fermi weak coupling constant, and $N_e$ is the electron number density inside the Earth's matter. The contributions to $\mathcal{H}_{\rm eff}$ from LIV
are given by the terms containing CPT-violating parameters $a_{\alpha\beta}$ and CPT-conserving parameters\footnote{In Eq.~\ref{eq:Heff_liv}, there is a multiplicative factor 4/3 in the terms containing $c_{\alpha\beta}$. This factor arises due to the choice of a frame which is rotational invariant. See Eq.~\ref{eq:c00-1} and the related discussion in appendix~\ref{app:L_CPTV}.} $c_{\alpha\beta}$. Note that the effective Hamiltonian $\mathcal{H}_{\rm eff}$ in Eq.~\ref{eq:Heff_liv} is written for neutrino, if we go to antineutrino then, $U \rightarrow U^*$, $\sqrt{2}G_{F}N_{e} \rightarrow -\sqrt{2}G_{F}N_{e}$, $a_{\alpha\beta}\rightarrow-a_{\alpha\beta}^*$, and $c_{\alpha\beta} \rightarrow c_{\alpha\beta}^*$, where the negative sign for CPT-violating LIV parameter $a_{\alpha\beta}$ appears as a multiplicative factor due to the construction of LIV formalism as discussed in appendix~\ref{app:L_CPTV}. However, the sign of CPT-conserving LIV parameter $c_{\alpha\beta}$ does not change when we go to antineutrino (see appendix~\ref{app:L_CPTV}).

In this work, we focus our attention on CPT-violating LIV parameters ($a_{\alpha\beta}$), whereas the CPT-conserving LIV parameters ($c_{\alpha\beta}$) will be studied separately in another work. As far as the CPT-violating LIV parameters are concerned, in the present work, we only focus on the off-diagonal parameters: $a_{\mu\tau}\equiv|a_{\mu\tau}|e^{i\phi_{\mu\tau}}$, $a_{e\mu}\equiv|a_{e\mu}|e^{i\phi_{e\mu}}$, and $a_{e\tau}\equiv|a_{e\tau}|e^{i\phi_{e\tau}}$. Our analysis is mainly dominated by $\nu_\mu \rightarrow \nu_\mu$ and $\bar\nu_\mu \rightarrow \bar\nu_\mu$ disappearance oscillation channels  where  the off-diagonal LIV parameter $a_{\mu\tau}$ appears only in the form of $|a_{\mu\tau}| \cos\phi_{\mu\tau}$ at the leading order (one can see a similar discussion in the context of NSI in Ref.~\cite{Kopp:2007ne}). 
Therefore, a complex phase only modifies the effective value of $a_{\mu\tau}$ to a real number between $-|a_{\mu\tau}|$ to $|a_{\mu\tau}|$ at the leading order. We exploit this observation by considering only real values of $a_{\mu\tau}$ in the range of $-|a_{\mu\tau}|$ to $|a_{\mu\tau}|$ for the present work. From these arguments, we can say that our method effectively covers the entire range of complex values of $a_{\mu\tau}$ which is equivalent to the variation of $\phi_{\mu\tau}$ in the full range of $-\pi$ to $\pi$. As far as the off-diagonal LIV parameters $a_{e\mu}$ and $a_{e\tau}$ are concerned, they appear at the subleading order in $\nu_\mu \rightarrow \nu_\mu$ survival channel and are always $\theta_{13}$-suppressed. The imaginary parts associated with the phases are further suppressed in the survival channel compared to the real parts and have a negligible impact ($\ll 1\%$). Thus, the real values of $a_{e\mu}$ and $a_{e\tau}$ cover the entire complex parameter space. 

Note that the CPT-violating LIV parameters ($a_{\alpha\beta}$) appear in an analogous fashion to that of neutral-current (NC) NSI parameters ($\varepsilon_{\alpha\beta}$) in the effective Hamiltonian (see Eq.~\ref{eq:Heff_liv}). One can obtain the following relation between the NC-NSI and CPT-violating LIV parameters~\cite{Diaz:2015dxa,KumarAgarwalla:2019gdj}
\begin{align}
\varepsilon_{\alpha\beta} = \frac{a_{\alpha\beta}}{\sqrt{2}G_{F}N_{e}} \label{eq:NSI_LIV}.
\end{align}
Here, we would like to mention that though these two new physics scenarios show a correspondence as given in the above equation, their physics origins are completely different. Further, the NC-NSI which appears during neutrino propagation is driven by the non-standard matter effects and it does not exist in vacuum. On the other hand, the LIV is an intrinsic effect that can be present even in vacuum. Note that for long-baseline experiments such as DUNE~\cite{DUNE:2015lol}, where a line-averaged constant Earth matter density may be a valid approximation, this one-to-one correspondence between CPT-violating LIV and NC-NSI parameters as shown in Eq.~\ref{eq:NSI_LIV} may hold. But in the present paper, our focus is on atmospheric neutrino experiments where we deal with a wide range of baselines spanning from 15 km to 12757 km. For each of these baselines, the scaling between CPT-violating LIV and NC-NSI parameters is expected to be different. Moreover, for a large fraction of these baselines (5721 km to 12757 km), neutrinos pass through the inner mantle or even core, where the density changes significantly following the PREM~\cite{Dziewonski:1981xy} profile with a sharp jump in the density at the core-mantle boundary. Therefore, for these trajectories, the line-averaged constant Earth matter density approximation is no longer valid and a simple scaling between  CPT-violating LIV and NC-NSI parameters does not work.

While showing the approximate analytical expressions of neutrino oscillation probabilities in the presence of non-zero CPT-violating LIV parameters $a_{\alpha\beta}$, we use the existing expressions in the literature for NC-NSI~\cite{Kopp:2007ne,Kumar:2021lrn} by replacing the NC-NSI parameters $\varepsilon_{\alpha\beta}$ with the CPT-violating LIV parameters $a_{\alpha\beta}$ following Eq.~\ref{eq:NSI_LIV}. Note that  to study the impact of CPT-violating LIV parameters on the atmospheric neutrinos at INO-ICAL, we indeed perform a full-fledged numerical simulation from scratch considering three-flavor neutrino oscillation probabilities in the presence of the PREM profile of Earth and obtain all the results that we show in the present paper.

In this work, we focus on the survival channel ($\nu_\mu \rightarrow \nu_\mu$) and appearance channel ($\nu_e \rightarrow \nu_\mu$) because these two channels have significant contribution to the neutrino events at the ICAL detector. The survival channel probability $P(\nu_\mu \rightarrow \nu_\mu)$ is dominantly affected by LIV parameter $a_{\mu\tau}$ whereas the effects of $a_{e\mu}$ and $a_{e\tau}$ are subdominant. Using approximations of one mass scale dominance (OMSD) [$\Delta m^2_{21} L/(4E) \ll \Delta m^2_{32} L/(4E)$] in the limit of $\theta_{13} \to 0$ with constant matter density, the survival probability for $\nu_\mu$ can be expressed as~\cite{GonzalezGarcia:2004wg, Kumar:2021lrn},
\begin{equation}
P(\nu_\mu\rightarrow\nu_\mu) = 1 - \sin^2 2\theta_{\rm eff} \,
\sin^2\left[ \xi \,\frac{\Delta m^2_{32} L_\nu}{4 E_\nu} \right]\,,
\label{eq:pmumu-nsi-mutau-omsd}
\end{equation}
where
\begin{equation}
\sin^2 2\theta_{\rm eff} =
\frac{|\sin 2\theta_{23} +  2 \beta \, \eta_{\mu\tau}|^2}{\xi^2}\,,
\label{eq:pmumu-thetaeff-omsd}
\end{equation}
\begin{equation}
\xi = \sqrt{|\sin 2\theta_{23} + 2 \beta \, \eta_{\mu\tau}|^2
  + \cos^2 2\theta_{23}}\,,
\label{eq:pmumu-xi-omsd}
\end{equation}
and
\begin{equation}
\eta_{\mu\tau} = \frac{ 2 E_\nu \,\omega \,a_{\mu\tau}}{|\Delta m^2_{32}|}
\,,
\end{equation}
where, we have replaced $\varepsilon_{\mu\tau}$ with $a_{\mu\tau}$ using Eq.~\ref{eq:NSI_LIV}. For a given value of CPT-violating parameter $a_{\alpha\beta}$, the parameter $\omega = +1$ for neutrino and $\omega = -1$ for antineutrino due to the construction of LIV formalism as explained in appendix~\ref{app:L_CPTV}. Here, $\beta \equiv {\rm sgn}(\Delta m^2_{32})$ which is the sign of atmospheric mass-squared difference. We have $\beta = +1$ for normal ordering (NO, $m_1<m_2<m_3$) and $\beta = -1$ for inverted ordering (IO, $m_3<m_1<m_2$). In the limit of maximal mixing ($\theta_{23} = 45^\circ$), the Eq.~\ref{eq:pmumu-nsi-mutau-omsd} boils down to \cite{Mocioiu:2014gua}, 
\begin{equation}
P(\nu_\mu\rightarrow\nu_\mu) = \cos^2 \left[L_\nu \left(
\frac{\Delta m^2_{32}}{4E_\nu} + \omega \, a_{\mu\tau} \right)\right]\,.
\label{eq:pmumu-nu-final-omsd}
\end{equation}

As far as the effects of LIV parameters $a_{e\mu}$ and $a_{e\tau}$ on survival channel ($\nu_\mu \rightarrow \nu_\mu$) are concerned, it appears in subleading order terms and non-trivial to express analytically. The effects of $a_{e\mu}$ and $a_{e\tau}$ with similar strengths become dominant in appearance channel ($\nu_e \rightarrow \nu_\mu$) because it occurs at leading order terms. We provide the expression for $P(\nu_e \rightarrow \nu_\mu)$ in the presence of both LIV parameters $a_{e\mu}$ and $a_{e\tau}$ at-a-time in appendix~\ref{app:Peu_emu_etau}. The effect of LIV parameter $a_{\mu\tau}$ on $P(\nu_e \rightarrow \nu_\mu)$ is not significant, hence, it is not present in Eq.~\ref{eq:Pmue-mat-LIV}. Note that these approximate expressions for oscillation probabilities are just for understanding purposes. However, in this work, we use numerically calculated full three-flavor oscillation probabilities with LIV in the presence of Earth's matter with the PREM profile~\cite{Dziewonski:1981xy}.

\section{A brief history of the search for LIV in neutrino oscillations}
\label{sec:LIV_history}

In this section, we discuss how the search for LIV parameters in neutrino oscillations progressed. In 2004, a general formalism for LIV and CPT-violation was developed in the neutrino oscillation sector, where possible definitive signals are predicted in the light of prevailing and proposed neutrino experiments~\cite{Kostelecky:2003cr}. After that, a few sample and global models are proposed to illustrate the key physical effects of LIV by considering both with and without neutrino mass, along that, a few generalized models with operators of arbitrary dimension are discussed to accommodate the signals observed in various neutrino experiments ~\cite{Kostelecky:2004hg,Kostelecky:2003xn,Katori:2006mz,Barger:2007dc,Diaz:2009qk,Diaz:2011tx,Barger:2011qj,Diaz:2010ft,Kostelecky:2011gq,Diaz:2011ia}. Now, we summarize the experimental attempts to probe the LIV:

\begin{itemize}
  \item A proposal of possible LIV signal was made in the context of the results obtained from the short-baseline experiments, and in 2005, the first experimental attempt to probe Lorentz and CPT violation was made by the liquid scintillator neutrino detector (LSND) experiment~\cite{Auerbach:2005tq} to accommodate the oscillation excess. But null results of LIV in LSND signal put bounds on $(a_L)$ and $(E \times c_L)$ of the order of $10^{-19}$ GeV, which is in an expected  scale of suppression. 
  \item  In 2008, a search for the sidereal modulation in the MINOS near detector neutrino data was performed~\cite{Adamson:2008aa}. No significant evidence was found, resulting in the bounds on the sidereal components of LIV parameters to the order of $10^{-20}$ GeV. Consequently, in 2010, the same search was performed with the MINOS  far detector neutrino data (Run I, II, and III)~\cite{Adamson:2010rn}. No signature of sidereal effect was found, and the bounds are further improved to the order of $10^{-23}$ GeV.
  \item In 2010, the IceCube experiment~\cite{Abbasi:2010kx} had reported the results of the search for a Lorentz-violating sidereal signal with the atmospheric neutrinos. No direction-dependent variation was found, and the constraints improved by a factor of 3 for $(a_L)$ and an order of 3 for $(c_L)$.
  \item In 2011, the result from the MiniBooNE Collaboration~\cite{AguilarArevalo:2011yi} was consistent with MINOS near detector bounds. 
  \item In 2012, the MINOS Collaboration searched the sidereal variation with muon type antineutrino data at the near detector~\cite{Adamson:2012hp}, and the result was consistent with their earlier work with neutrino data~\cite{Adamson:2008aa}.
  \item In the later part of 2012, the Double Chooz Collaboration~\cite{Abe:2012gw} showed their results of the search for LIV with reactor antineutrinos. Data indicated no sidereal variations, so their work set the first limits on 14 LIV coefficients associated with $e$-$\tau$ sector and set two competitive limits associated with $e$-$\mu$ sectors.
  \item The presence of LIV allows the neutrino and antineutrino mixings, hence to see such a signal, in early 2013, MINOS has performed a search analysis~\cite{Rebel:2013vc}. There was no evidence of such a signal, resulting in the limit of the appropriate 66 LIV coefficients ($H$ and $g$)\footnote{These two parameters are the elements of the off-diagonal block in the effective Hamiltonian, see Refs.~\cite{Kostelecky:2003cr, Diaz:2009qk}.}, which govern the neutrino-antineutrino mixing.
  \item Further search for neutrino-antineutrino mixing was performed using the disappearance of reactor antineutrinos in the Double Chooz experiment~\cite{Diaz:2013iba} and set limits on another 15 LIV coefficients ($H$ and $g$). A similar study in the context of solar neutrinos was performed by the authors in Ref.~\cite{Diaz:2016fqd}.
  \item In 2015, the Super-Kamiokande (Super-K) Collaboration published their result to search Lorentz Invariance with atmospheric neutrinos with a large range of baselines and wide range of energies~\cite{Super-Kamiokande:2014exs}. No evidence of Lorentz violation was observed, so limits were set on the renormalizable isotropic LIV coefficients in the $e$-$\mu$, $\mu$-$\tau$, and $e$-$\tau$ sectors, with an improvement of seven orders of magnitude. 
  \item In 2017, the T2K Collaboration searched LIV using sidereal time dependence of neutrino flavor transitions with the T2K on-axis near detector~\cite{Abe:2017eot}. The results were consistent with the limits put by other short-baseline experiments. 
  \item In 2018, a simultaneous measurement of LIV coefficients was done using the Daya Bay reactor antineutrino experiment~\cite{Adey:2018qsd}. The bounds on the appropriate parameters were agreed to the suppression of the Planck mass scale ($M_p$).
  \item The IceCube experiment analyzed the LIV hypothesis using the high-energy part of atmospheric neutrinos in the year 2018~\cite{IceCube:2017qyp}. They perform the analysis using a perturbative approach in an effective two-flavor neutrino oscillation framework and obtained the limits on individual LIV coefficients ($a_L$ and $c_L$) with mass dimensions ranging from 3 to 8. These are the most stringent limits on Lorentz violation (in the $\mu$-$\tau$ sector) set by any physical experiment.
\end{itemize} 

The current constraints on all the relevant LIV/CPT-violating parameters are nicely summarized in Ref.~\cite{Kostelecky:2008ts}. In a recent review~\cite{Antonelli:2020nhn}, the authors have discussed the phenomenological effects of CPT and Lorentz symmetry violations in the context of particle and astroparticle physics. 

\section{Effects of Lorentz Invariance Violation on oscillograms}
\label{sec:oscillograms}

In this section, we describe how does the neutrino oscillation probabilities get affected due to the presence of LIV. The atmospheric neutrinos are produced as a result of the interaction of primary cosmic rays with the nuclei of the atmosphere and mostly consist of electron and muon flavors. The atmospheric neutrinos with the multi-GeV energy range have access to a wide range of baselines starting from about 15 km to 12757 km. The upward-going neutrinos with longer baselines experience the Earth's matter effect, which results in the modification of oscillation probabilities. The effect of LIV on neutrino oscillations can be explored at several $L/E$ values available for atmospheric neutrinos.

\begin{table}[t]
  \centering
  \begin{tabular}{|c|c|c|c|c|c|c|}
    \hline
    $\sin^2 2\theta_{12}$ & $\sin^2\theta_{23}$ & $\sin^2 2\theta_{13}$ & $\Delta m^2_
    \text{eff}$ (eV$^2$) & $\Delta m^2_{21}$ (eV$^2$) & $\delta_{\rm CP}$ & Mass Ordering\\
    \hline
    0.855 & 0.5 & 0.0875 & $2.49\times 10^{-3}$ & $7.4\times10^{-5}$ & 0 & Normal (NO)\\
    \hline 
  \end{tabular}
  \mycaption{The values of benchmark neutrino oscillation parameters used in this analysis. These values are consistent with the present global fits of neutrino oscillation parameters~\cite{Marrone:2021,NuFIT,Esteban:2020cvm,deSalas:2020pgw}.}
  \label{tab:osc-param-value}
\end{table}

In the present analysis, we numerically calculate the neutrino oscillation probability in the three-flavor paradigm with Earth's matter effect using the PREM profile~\cite{Dziewonski:1981xy}. We use the benchmark values of oscillation parameters given in Table~\ref{tab:osc-param-value}. The value of $\Delta m^2_{31}$ is obtained from the effective mass-squared difference\footnote{The effective mass-squared difference is defined in terms of $\Delta m^2_{31}$ as follows~\cite{deGouvea:2005hk,Nunokawa:2005nx}:
\begin{equation}\label{eq:eff_dmsq}
  \Delta m^2_\text{eff} = \Delta m^2_{31} - \Delta m^2_{21} (\cos^2\theta_{12} - \cos \delta_\text{CP} \sin\theta_{13}\sin2\theta_{12}\tan\theta_{23}).
\end{equation}} 
$\Delta m^2_\text{eff}$. To consider NO as mass ordering, we use positive value of $\Delta m^2_\text{eff}$ whereas for IO, $\Delta m^2_\text{eff}$ is taken as negative with the same magnitude.

\begin{figure}[t]
  \centering
  \includegraphics[width=0.32\textwidth]{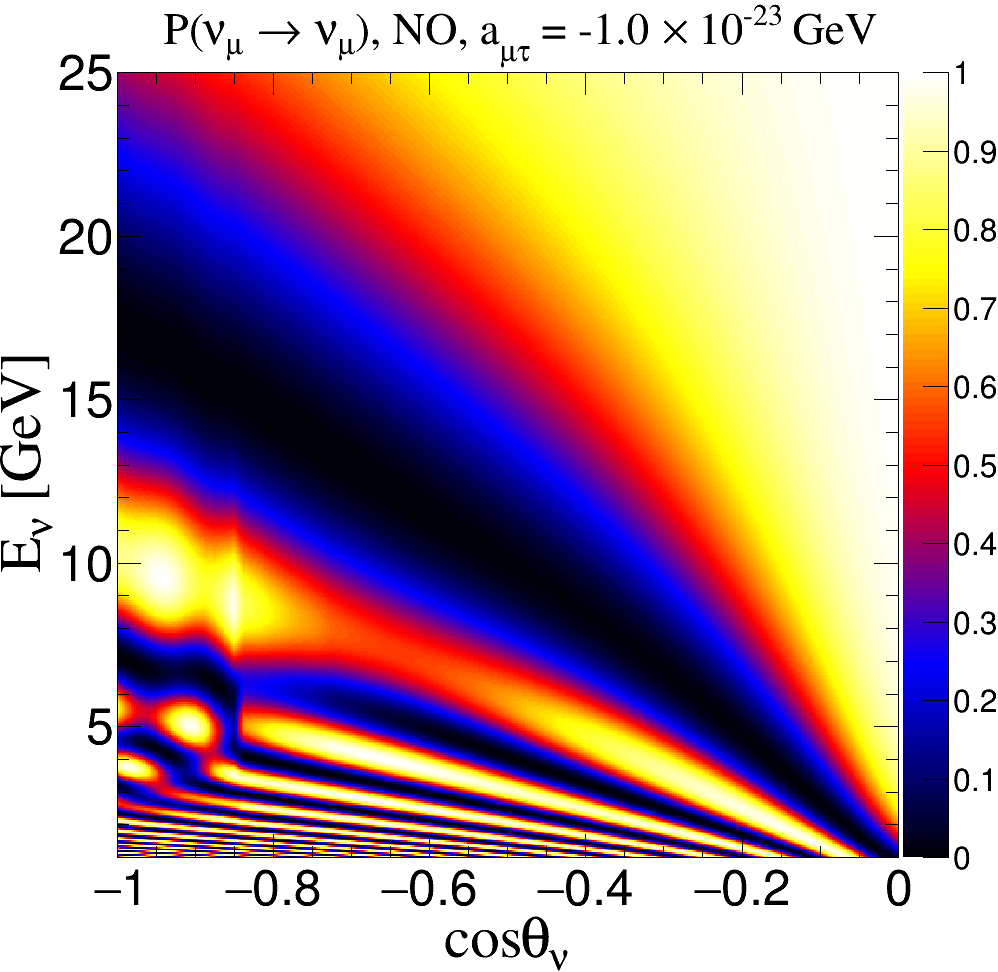}
  \includegraphics[width=0.32\textwidth]{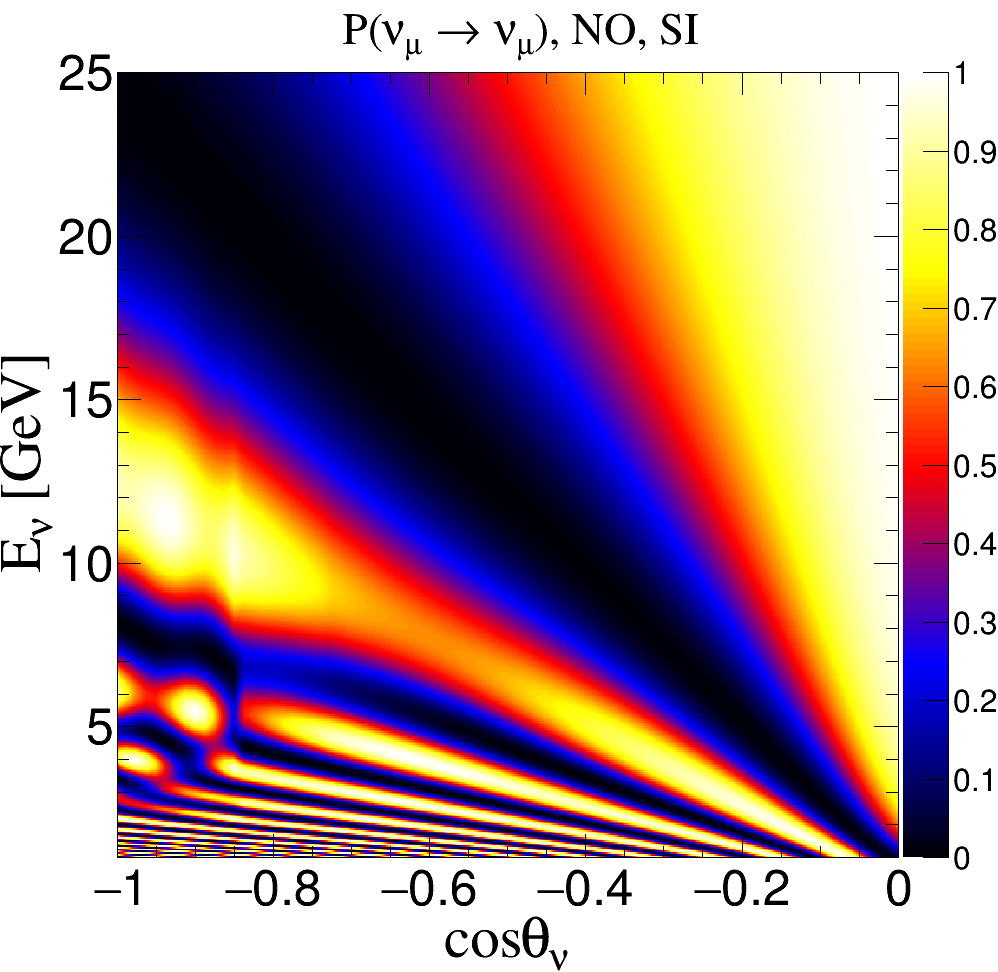}
  \includegraphics[width=0.32\textwidth]{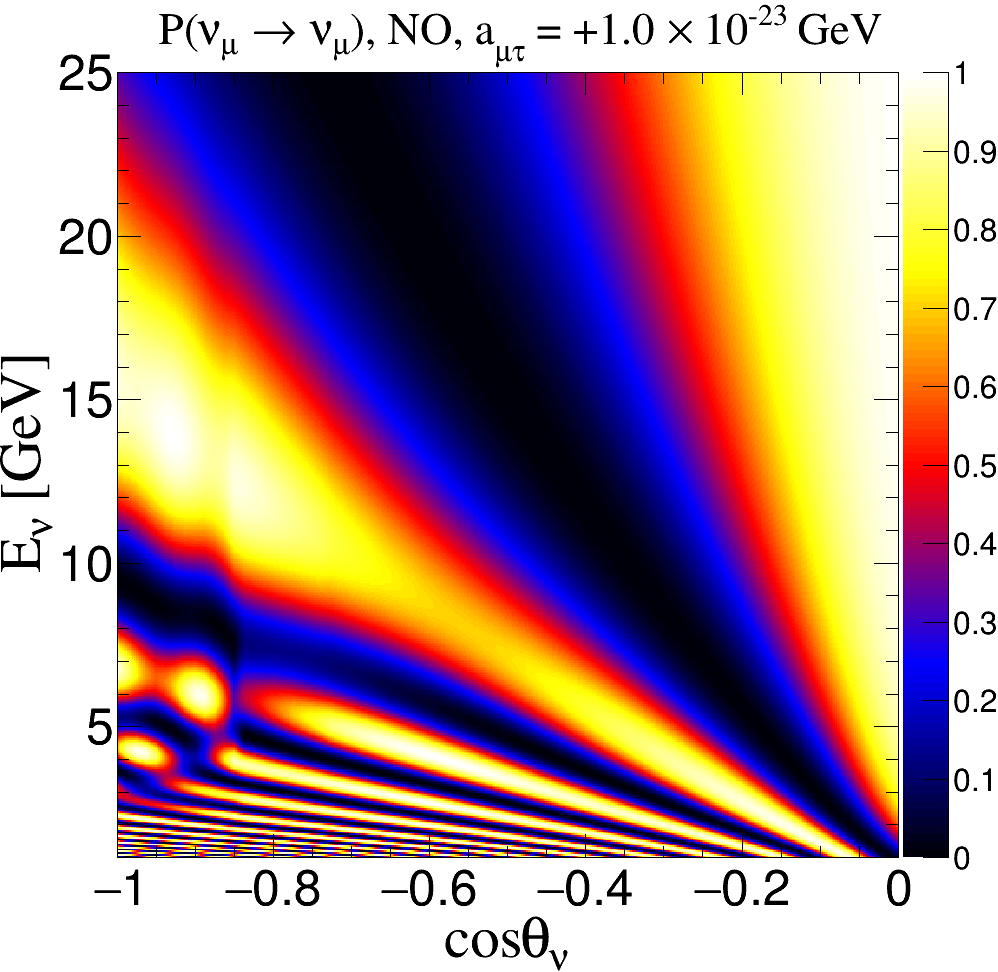}
  \includegraphics[width=0.32\textwidth]{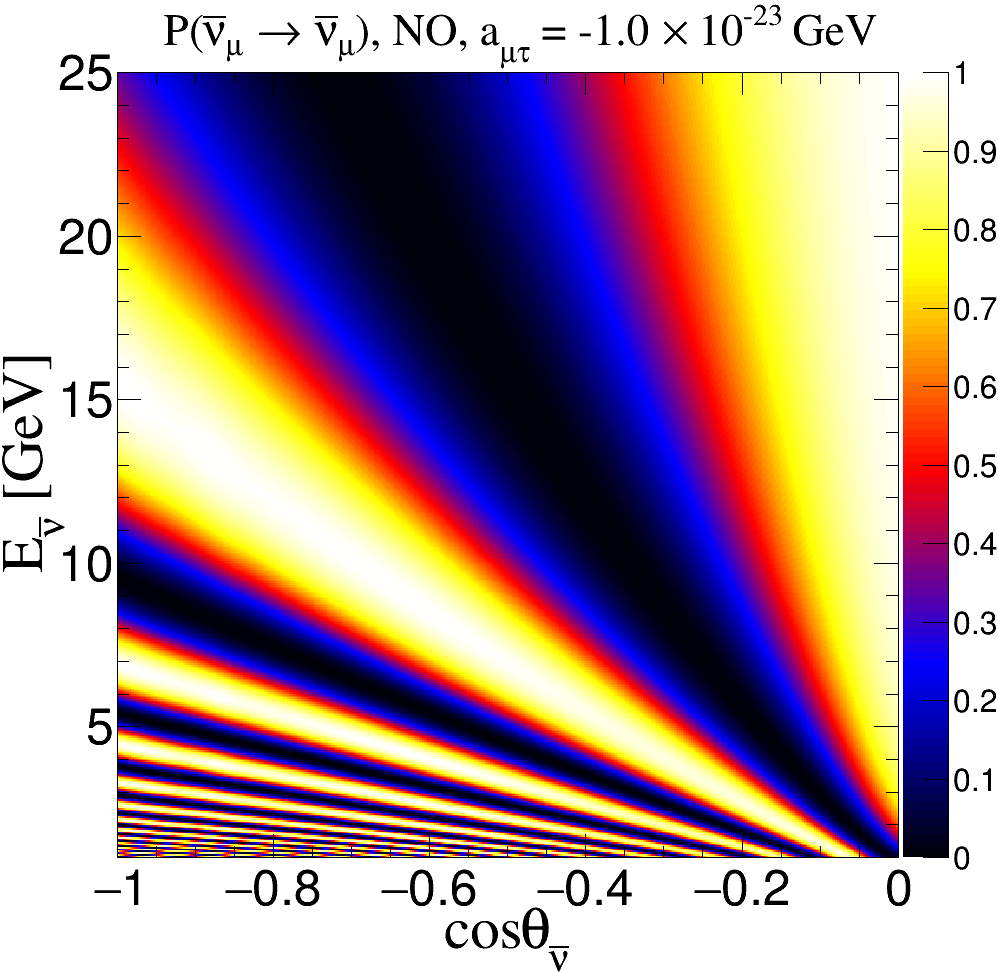}
  \includegraphics[width=0.32\textwidth]{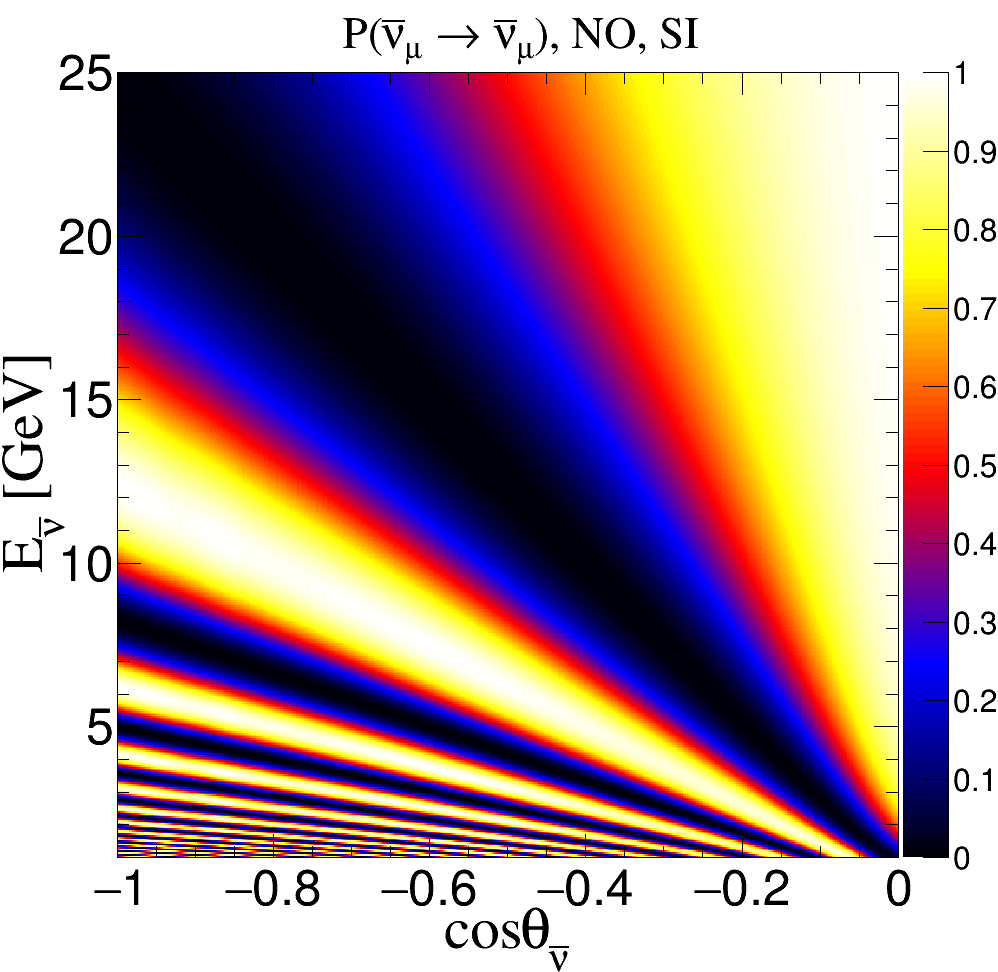}
  \includegraphics[width=0.32\textwidth]{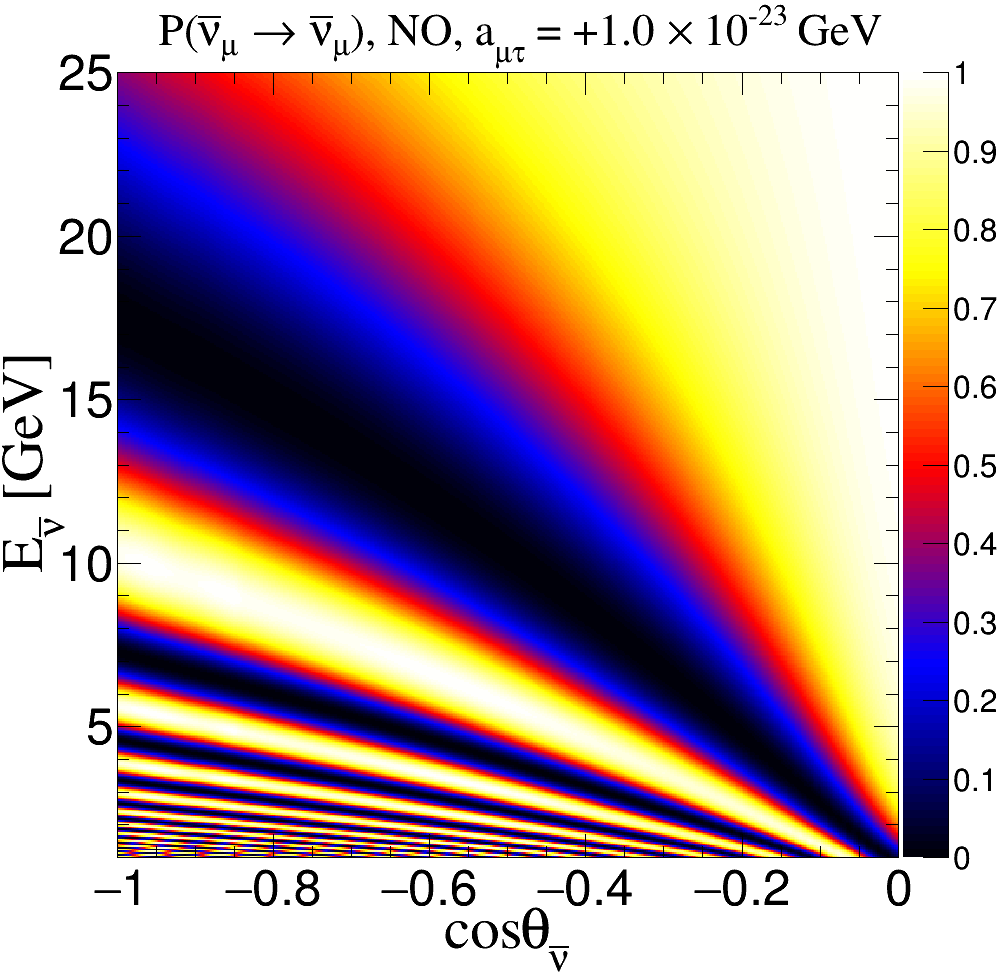}
  \mycaption{The $\nu_\mu$ survival probability oscillograms in ($E_\nu$, $\cos\theta_\nu$) plane for neutrino (top panels) and antineutrino (bottom panels) with Earth's matter effect considering the PREM profile. We use NO and benchmark oscillation parameters given in Table~\ref{tab:osc-param-value}. We consider the values of $a_{\mu\tau} = -1.0 \times 10^{-23}$ GeV, $0$, and $1.0 \times 10^{-23}$ GeV in the left, middle and right panels, respectively.}
  \label{fig:oscillograms_mutau}
\end{figure}

The neutrino events at ICAL are contributed by both disappearance ($\nu_\mu \rightarrow \nu_\mu$) channel as well as appearance ($\nu_{\rm e} \rightarrow \nu_\mu$) channel. Since, more than 98\% events at ICAL are contributed by disappearance ($\nu_\mu \rightarrow \nu_\mu$) channel, we discuss the impact of LIV on P$(\nu_\mu \rightarrow \nu_\mu)$ in this section. In Fig.~\ref{fig:oscillograms_mutau}, we show the impact of non-zero value of CPT-violating LIV parameter $a_{\mu\tau}$ on $\nu_\mu \rightarrow \nu_\mu$ survival probability oscillograms in the plane of energy ($E_\nu$) and direction\footnote{The relation between neutrino zenith angle $\theta_\nu$ and neutrino baseline $L_\nu$ is given as 
\begin{equation}
  L_\nu = \sqrt{(R+h)^2 - (R-d)^2\sin^2\theta_\nu} \,-\, (R-d)\cos\theta_\nu \, ,
\end{equation}
where, $R$, $h$, and $d$ represent the radius of Earth, the average production height for neutrinos, and the depth of the detector, respectively. In the present study, we consider $R = 6371$ km, $h = 15$ km, and $d = 0$ km. } 
($\cos\theta_{\nu}$) of neutrino. We consider three different choices of $a_{\mu\tau}$ which are $-1.0 \times 10^{-23}$ GeV, $0$, and $1.0 \times 10^{-23}$ GeV as shown in the left, middle and right columns, respectively, in Fig.~\ref{fig:oscillograms_mutau}. The panels in top and bottom rows correspond to neutrinos and antineutrinos, respectively.

In Fig.~\ref{fig:oscillograms_mutau}, the effect of LIV parameter $a_{\mu\tau}$ can be observed at energies above 10 GeV and baselines with $\cos\theta_\nu < -0.6$. This region is dominated by vacuum oscillations. The dark blue diagonal band, which is termed as oscillation valley in Ref.~\cite{Kumar:2020wgz,Kumar:2021lrn}, bends in the presence of LIV parameter $a_{\mu\tau}$. The direction of bending depends on the sign of $a_{\mu\tau}$. For example, the oscillation valley for neutrino bends in the downward direction for negative $a_{\mu\tau}$ in the top left panel, whereas for positive $a_{\mu\tau}$, the bending is in the upward direction in the top right panel. For a given value of $a_{\mu\tau}$, the oscillation valley curves in the opposite directions for neutrinos and antineutrinos. In summary, we can infer that the bending of oscillation valley for neutrino with positive $a_{\mu\tau}$ (top right panel) is similar to that for antineutrino with negative $a_{\mu\tau}$ (bottom left panel) and vice versa.

We now discuss how the modification of the oscillation valley due to non-zero $a_{\mu\tau}$ can be understood analytically. The oscillation valley corresponds to the first oscillation minimum in $\nu_\mu$ survival probability in Eq.~\ref{eq:pmumu-nu-final-omsd} which results in the following relation between $E_\nu$ and $\cos\theta_\nu$~\cite{Kumar:2021lrn},
\begin{equation}
E_\nu |_{\rm valley} \approx \frac{| \Delta m^2_{32}| }{
  (\pi / |R \cos\theta_\nu|) \;  
  - 4 \beta \, \omega\, a_{\mu\tau} }\, .
\label{eq:osc-min-Ecostheta-wounit}
\end{equation}
where, we assume  $L_\nu \approx 2R |\cos\theta_\nu|$. For the case of standard interactions (SI) with $a_{\mu\tau} = 0$, Eq.~\ref{eq:osc-min-Ecostheta-wounit} results in a linear relation between $E_\nu$ and $\cos\theta_\nu$ which is manifested as oscillation valley with straight line for SI case in the middle panels of Fig.~\ref{fig:oscillograms_mutau}.

In the case of NO ($\beta = +1$), if $a_{\mu\tau}$ is positive then for neutrino ($\omega = +1$), $E_\nu$ increases compared to that for SI case for a given $\cos\theta_\nu$ as observed in the top right panel of Fig.~\ref{fig:oscillograms_mutau}. If we consider negative value of $a_{\mu\tau}$ then for neutrino, $E_\nu$ decreases as seen in the top left panel of Fig.~\ref{fig:oscillograms_mutau}. As far as antineutrino is concerned, where $\omega = -1$, positive (negative) value of $a_{\mu\tau}$ results in decrease (increase) in $E_\nu$ as shown in the bottom right (left) panel of Fig.~\ref{fig:oscillograms_mutau}. Since, $\beta$ and $\omega$ appear as product, the effect of positive (negative) $a_{\mu\tau}$ with neutrino is same as that of negative (positive) $a_{\mu\tau}$ with antineutrino. If we consider the case of IO ($\beta = -1$), then all these effects are reversed with the opposite curvatures of oscillation valley.

\begin{figure}[t]
  \centering
  \includegraphics[width=0.32\textwidth]{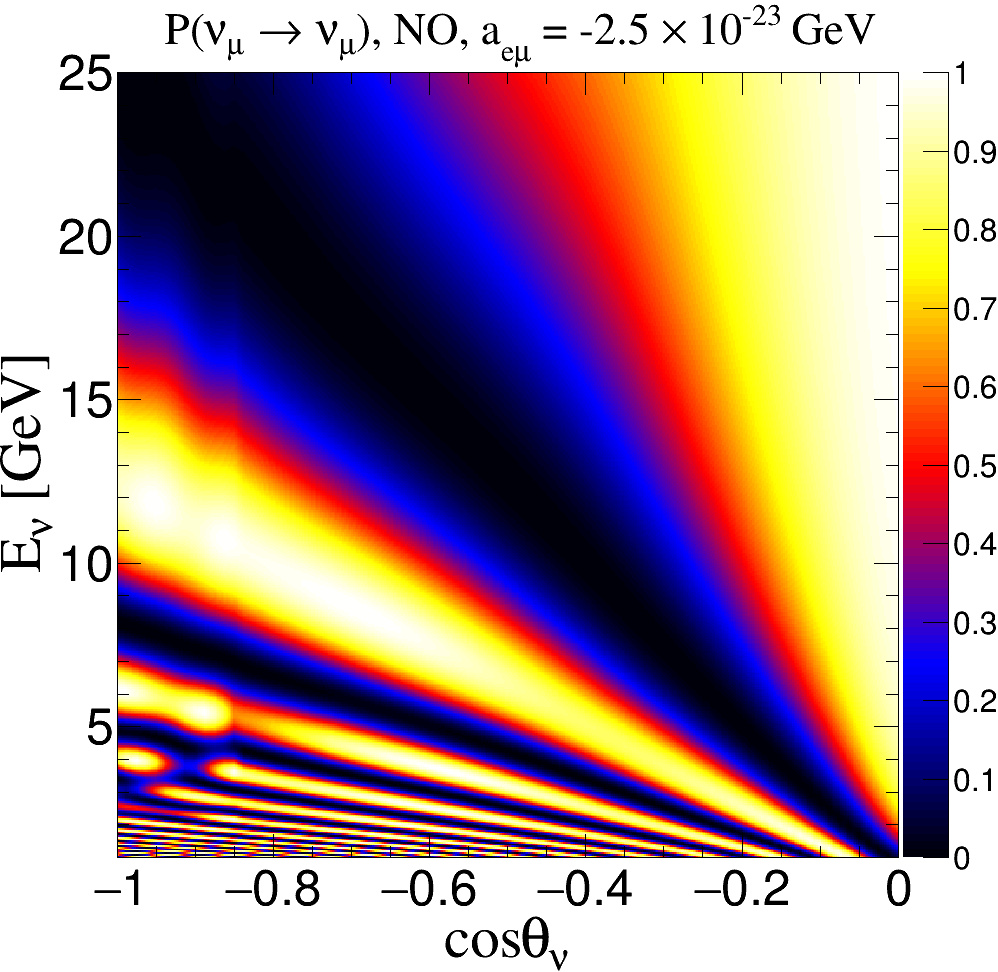}
  \includegraphics[width=0.32\textwidth]{images/Oscillogram/SI/MuMu_SI.png}
  \includegraphics[width=0.32\textwidth]{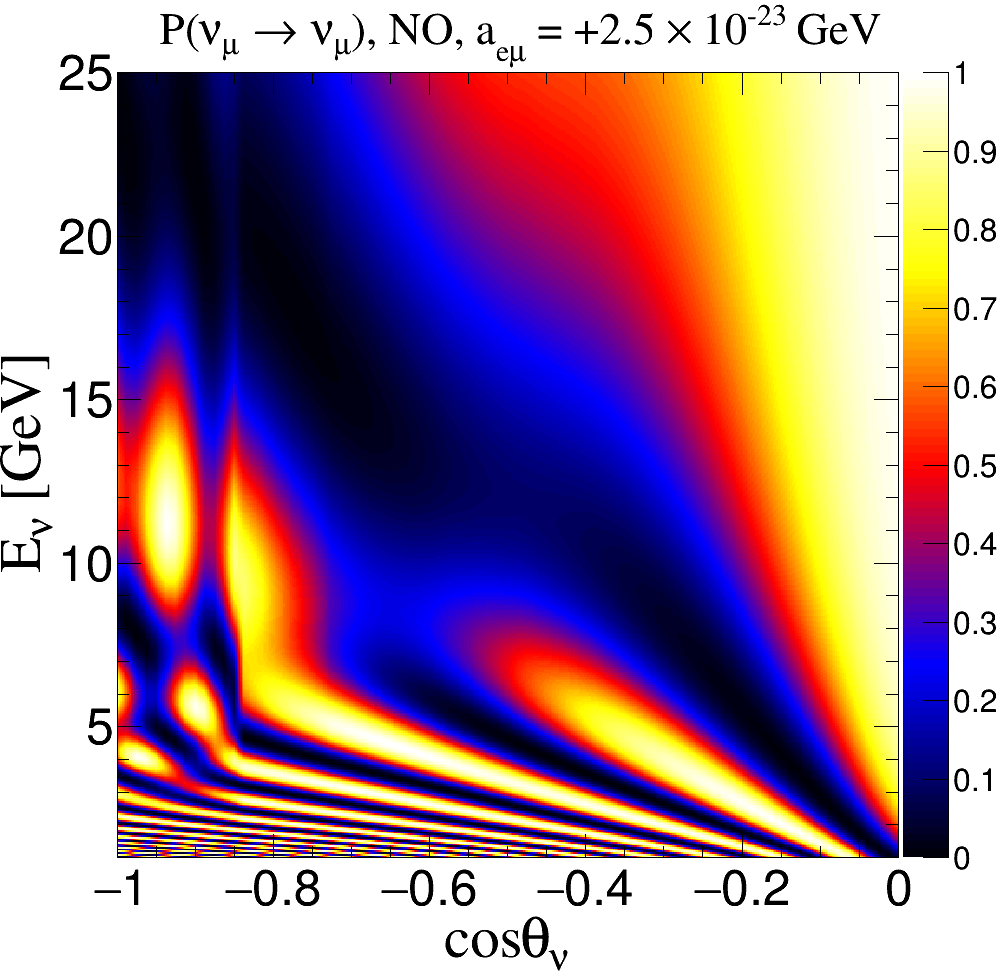}
  \includegraphics[width=0.32\textwidth]{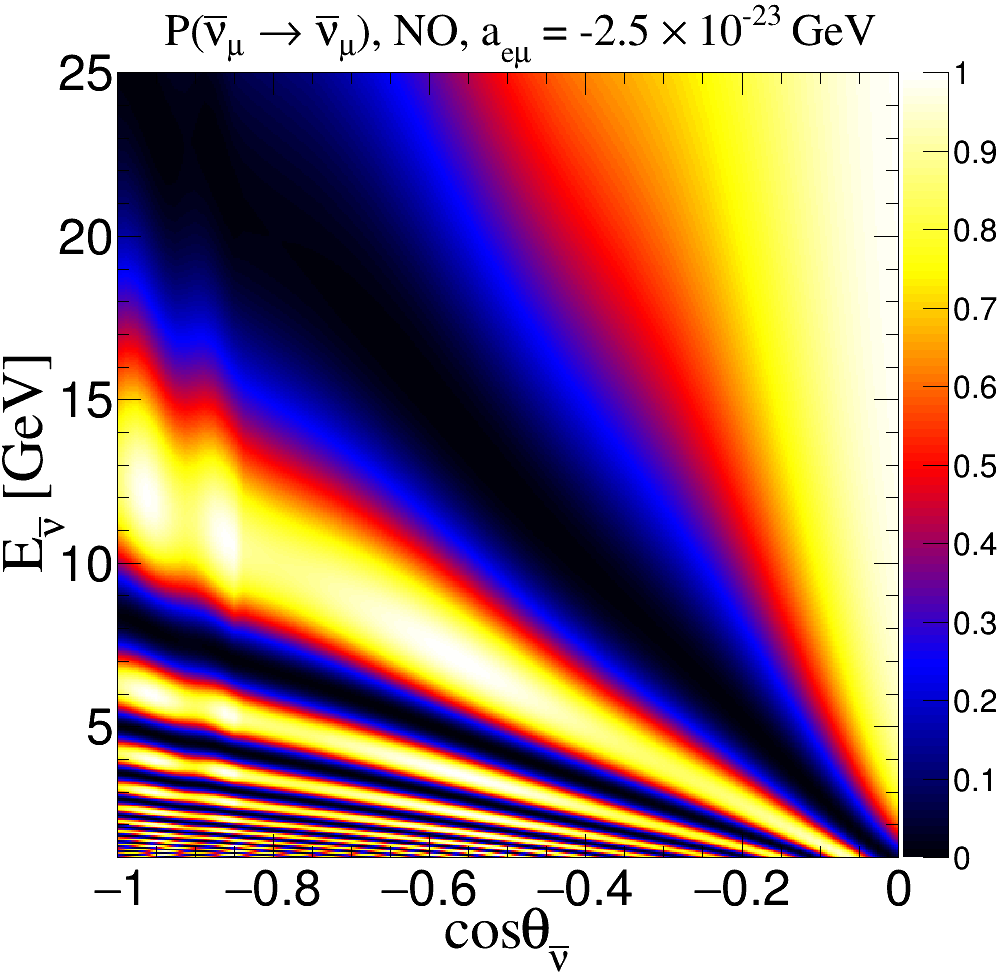}
  \includegraphics[width=0.32\textwidth]{images/Oscillogram/SI/MubarMubar_SI.png}
  \includegraphics[width=0.32\textwidth]{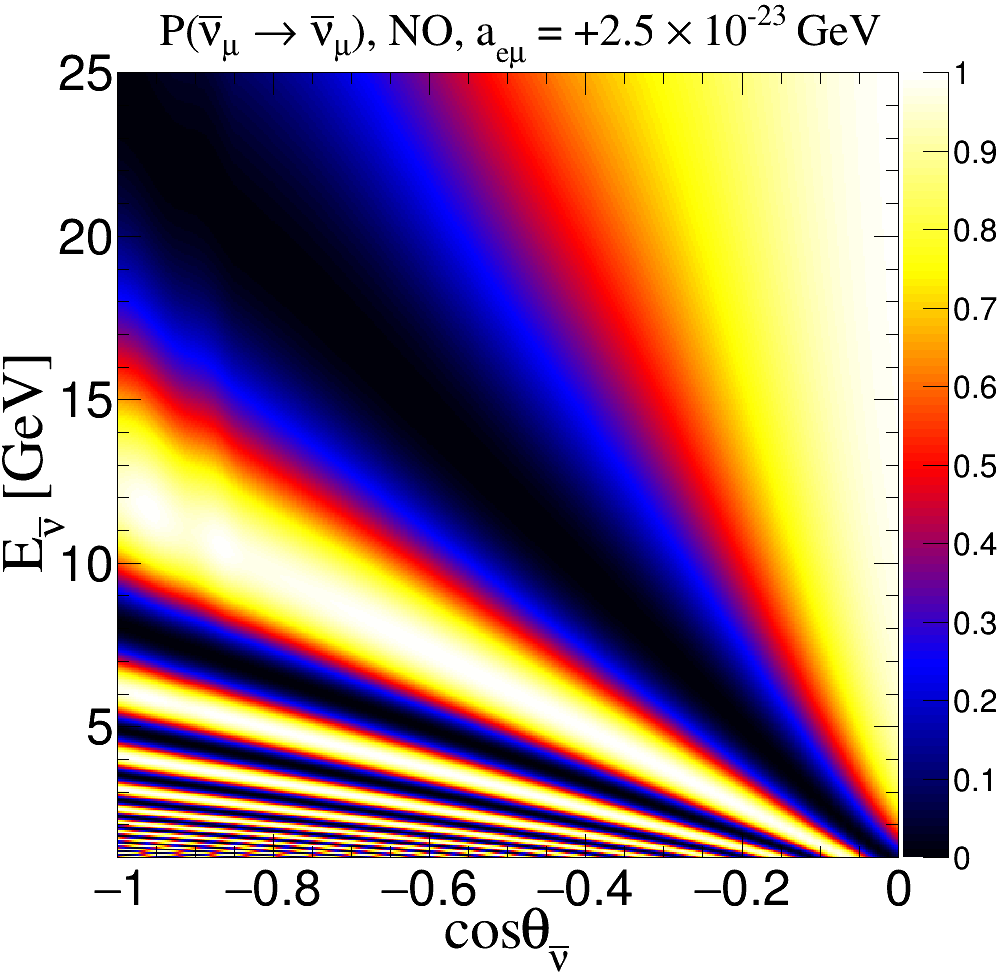}
  \mycaption{The $\nu_\mu$ survival probability oscillograms in ($E_\nu$, $\cos\theta_\nu$) plane for neutrino (top panels) and antineutrino (bottom panels) with Earth's matter effect considering the PREM profile. We use NO and benchmark oscillation parameters given in Table~\ref{tab:osc-param-value}. We consider the values of $a_{e\mu} = -2.5 \times 10^{-23}$ GeV, $0$, and $2.5 \times 10^{-23}$ GeV in the left, middle and right panels, respectively.}
  \label{fig:oscillograms_emu}
\end{figure}

In Fig.~\ref{fig:oscillograms_emu}, we present the impact of CPT-violating LIV parameter $a_{e\mu}$ on $\nu_\mu \rightarrow \nu_\mu$ survival probability oscillograms in the plane of neutrino energy ($E_\nu$) and direction ($\cos\theta_\nu$). We assume three different values of $a_{e\mu}$ which are $-2.5 \times 10^{-23}$ GeV, $0$, and $2.5 \times 10^{-23}$ GeV as shown in the left, middle and right columns, respectively, in Fig.~\ref{fig:oscillograms_emu}. In the top and bottom panels, we portray plots for neutrinos and antineutrinos, respectively. Unlike the case of $a_{\mu\tau}$, here, we do not observe any bending in the oscillation valley due to the presence of non-zero $a_{e\mu}$. However, a distortion in the oscillation valley can be observed due to $a_{e\mu}$. The regions with matter effect also get modified. For neutrino, the effect due to positive $a_{e\mu}$ is significantly more than that due to negative $a_{e\mu}$. We also observe that the effect of $a_{e\mu}$ for antineutrino is not significant. 
  
We would like to highlight that the effect of $a_{e\mu}$ appears at the subleading order in the expression of $P(\nu_\mu \rightarrow \nu_\mu)$, which are non-trivial to express analytically. We can analyze the same by studying the analytical expression for the appearance channel $P(\nu_e \rightarrow \nu_\mu)$ as given in Eq.~\ref{eq:Pmue-mat-LIV} in appendix~\ref{app:Peu_emu_etau}, where the effect of $a_{e\mu}$ appears in the leading order terms. The effect of $a_{e\mu}$ is dominantly contributed by the fifth term in Eq.~\ref{eq:Pmue-mat-LIV}, which has the following form 
\begin{equation}
  + 16\omega a_{e\mu} E_\nu \tilde{s}_{13} s_{23}^{3} \frac{1}{\ldm - a_{\rm CC}}
  \sin^{2} \frac{(\ldm - a_{\rm CC})L_\nu}{4E_\nu},
\end{equation}  
where $(\ldm - a_{\rm CC})$ factor in the denominator causes the matter-driven resonance effect for the case of neutrino with NO. Since this term has positive sign for the case of neutrino ($\omega = +1$), the positive value of $a_{e\mu}$ increases $P(\nu_e \rightarrow \nu_\mu)$. This can be translated as the decrease in  $P(\nu_\mu \rightarrow \nu_\mu)$ around the region of matter effect as shown in the top right panel of Fig.~\ref{fig:oscillograms_emu} because we have,
\begin{equation}
P(\nu_\mu \rightarrow \nu_\mu) = 1 - P(\nu_\mu \rightarrow \nu_e) - P(\nu_\mu \rightarrow \nu_\tau),
\end{equation}
where $P(\nu_\mu \rightarrow \nu_e) = P(\nu_e \rightarrow \nu_\mu)$ for $\delta_{\rm CP} = 0$. Note that $\nu_\mu \rightarrow \nu_\tau$ oscillation channel also gets affected due to matter effects in certain ranges of energies and baselines (see Ref.~\cite{Gandhi:2004md}). But as far as the impact of LIV parameter $a_{e\mu}$ is concerned, it appears only at the subleading order in $\nu_\mu \rightarrow \nu_\tau$ oscillation channel. For negative value of  $a_{e\mu}$, $P(\nu_e \rightarrow \nu_\mu)$ decreases and $P(\nu_\mu \rightarrow \nu_\mu)$ increases, hence we observe dilution of matter effect in the top left panel of Fig.~\ref{fig:oscillograms_emu}. We can see that the effect of $a_{e\mu}$ is larger in the case of neutrino than antineutrino. This happens because $a_{\rm CC}$ becomes negative for antineutrino and matter-driven resonance condition is not fulfilled for NO due to which the above-mentioned term does not contribute significantly.

\begin{figure}[t]
  \centering
  \includegraphics[width=0.32\textwidth]{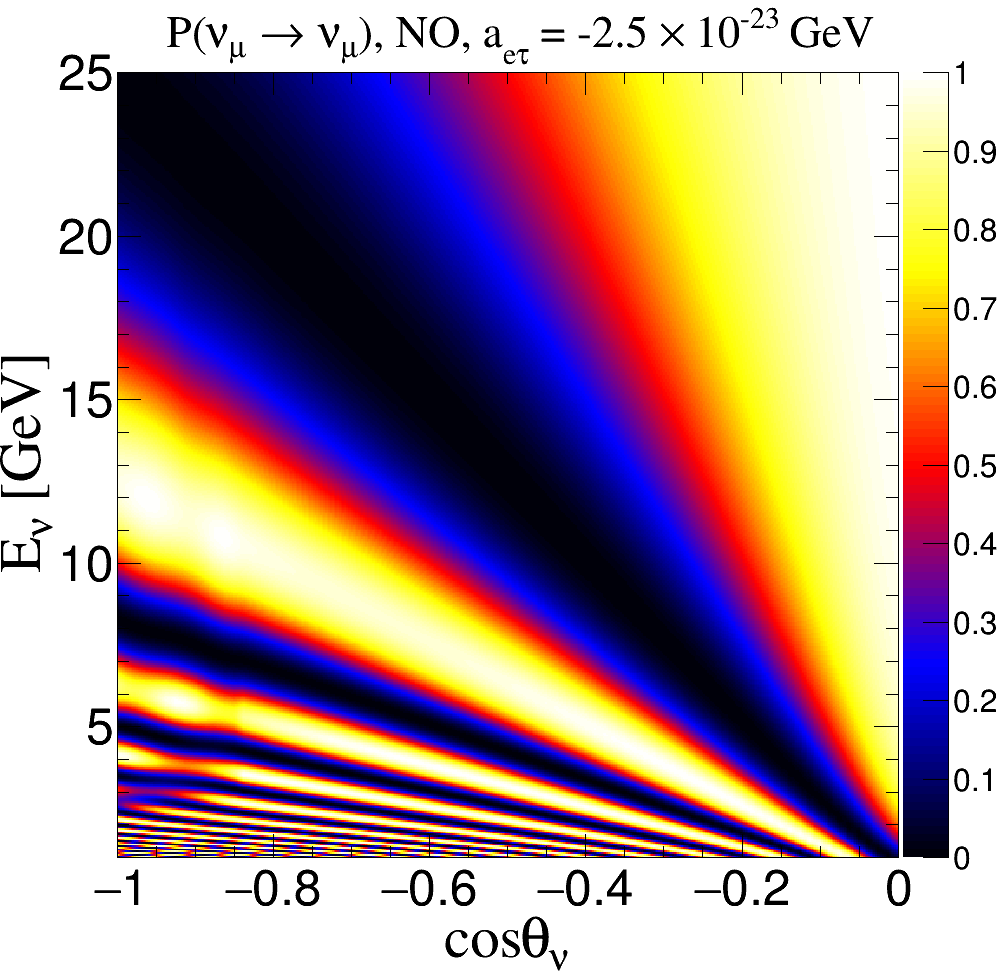}
  \includegraphics[width=0.32\textwidth]{images/Oscillogram/SI/MuMu_SI.png}
  \includegraphics[width=0.32\textwidth]{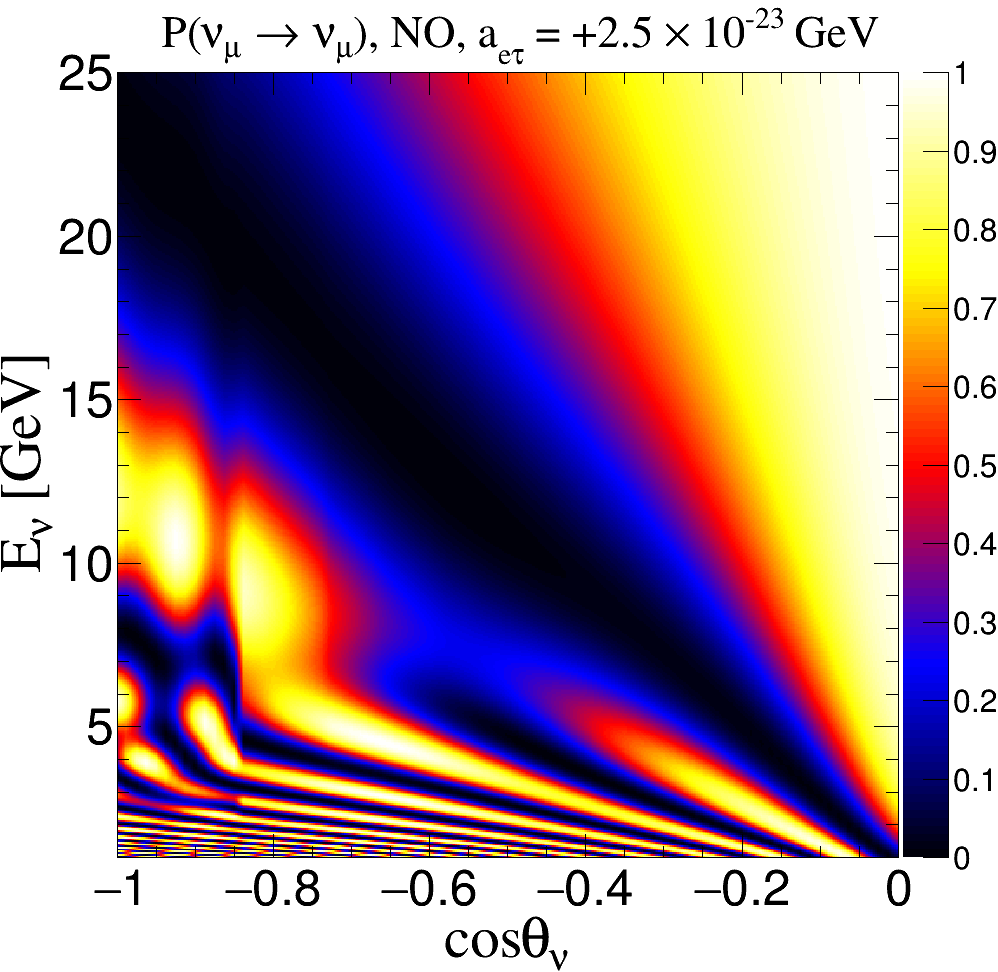}
  \includegraphics[width=0.32\textwidth]{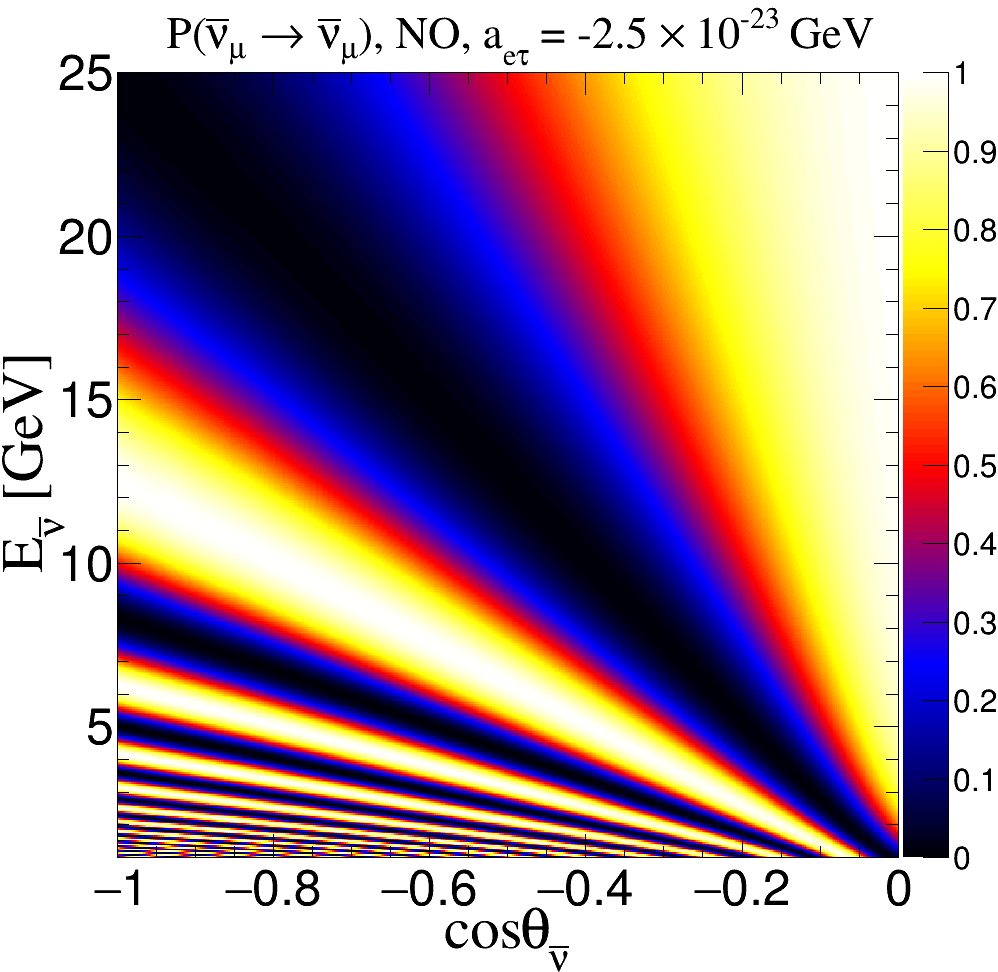}
  \includegraphics[width=0.32\textwidth]{images/Oscillogram/SI/MubarMubar_SI.png}
  \includegraphics[width=0.32\textwidth]{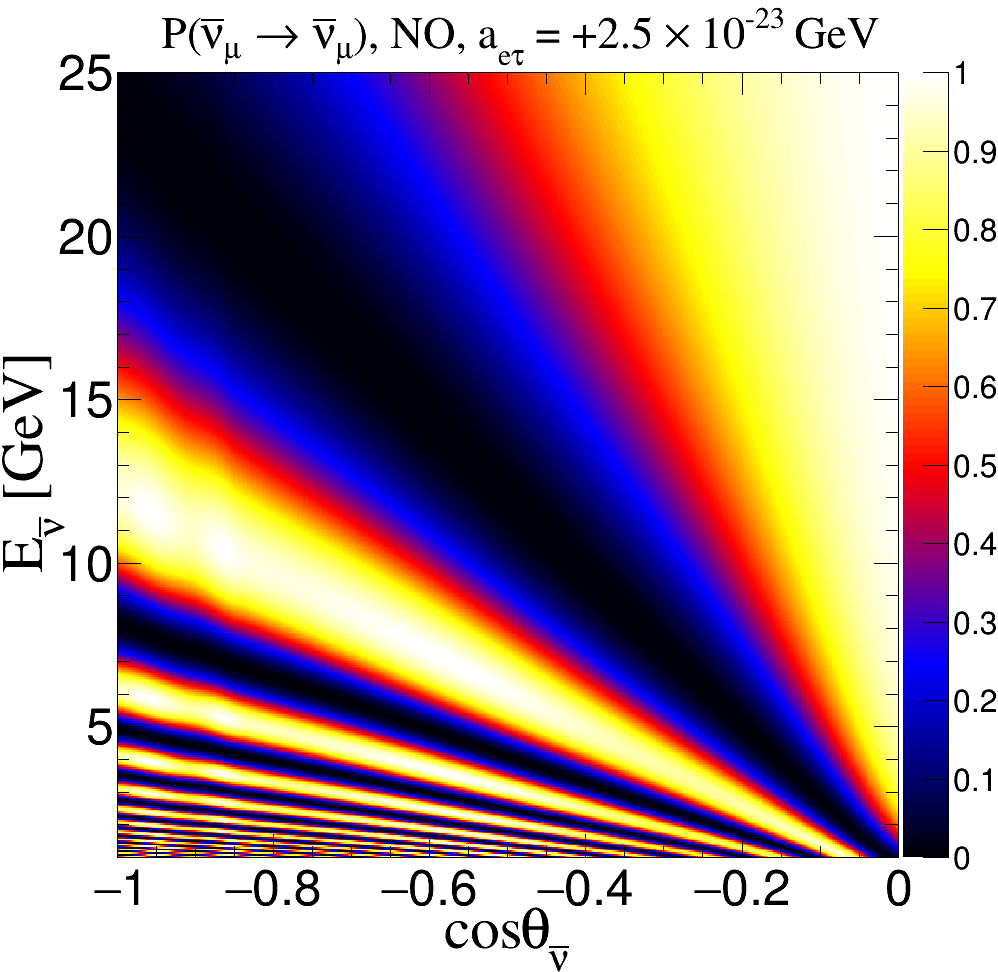}
  \mycaption{The $\nu_\mu$ survival probability oscillograms in ($E_\nu$, $\cos\theta_\nu$) plane for neutrino (top panels) and antineutrino (bottom panels) with Earth's matter effect considering the PREM profile. We use NO and benchmark oscillation parameters given in Table~\ref{tab:osc-param-value}. We consider the values of $a_{e\tau} = -2.5 \times 10^{-23}$ GeV, $0$, and $2.5 \times 10^{-23}$ GeV in the left, middle and right panels, respectively.}
  \label{fig:oscillograms_etau}
\end{figure}

Similar to LIV parameter $a_{e\mu}$, now we depict the impact of  $a_{e\tau}$  on $\nu_\mu \rightarrow \nu_\mu$ survival probability oscillograms in the plane of neutrino energy ($E_\nu$) and direction ($\cos\theta_\nu$) in Fig.~\ref{fig:oscillograms_etau}.  We take three different values of $a_{e\tau}$ which are $-2.5 \times 10^{-23}$ GeV, $0$, and $2.5 \times 10^{-23}$ GeV as shown in the left, middle and right columns, respectively, in Fig.~\ref{fig:oscillograms_etau}. The features present in the oscillograms due to $a_{e\tau}$ in Fig.~\ref{fig:oscillograms_etau} are analogous to that due to $a_{e\mu}$ in Fig.~\ref{fig:oscillograms_emu}. Here also, we can observe the distortions in the oscillation valley as well as the matter effect regions. The effect of $a_{e\tau}$ is more for neutrinos than that for antineutrino. If we focus on neutrinos, we see that the effect of positive $a_{e\tau}$ (top right panel) is larger than that of negative $a_{e\tau}$ (top left panel). Following the similar arguments as for $a_{e\mu}$ in the previous paragraph, these features can be explained using the seventh term in Eq.~\ref{eq:Pmue-mat-LIV}. These asymmetric effects for CPT-violating LIV parameter $a_{e\mu}$ and $a_{e\tau}$  can also be observed in our result section where we constrain them using 500 kt$\cdot$yr exposure at the ICAL detector.

Till now, we have described the impact of CPT-violating LIV parameters at the probability level, where we observe some interesting features. In the next section, we explain the procedure to simulate neutrino events at the ICAL detector and explore how the event distribution modifies in the presence of LIV.

\section{Event generation at ICAL}
\label{sec:event_generation}

The proposed ICAL detector at INO~\cite{Kumar:2017sdq} consists of a 50 kt magnetized iron stack of size 48 m $\times$ 16 m $\times$ 14.5 m  which would be able to detect atmospheric neutrinos and antineutrinos separately in the multi-GeV range of energies over a wide range of baselines. The vertically stacked 151 layers of iron of thickness 5.6 cm act as the target material for neutrino interactions. The charged-current interactions of neutrinos and antineutrinos with iron nuclei produce charged muons, $\mu^-$ and $\mu^+$ respectively. The 2 m $\times $ 2 m Resistive Plate Chambers (RPCs)~\cite{SANTONICO1981377,Bheesette:2009yrp,Bhuyan:2012zzc} sandwiched between iron layers act as active detector elements. The charged particles deposit energies while passing through RPCs, which result in the production of electronic signals. The perpendicularly arranged pickup strips of RPC give the X and Y coordinates of the hit, whereas the Z coordinate is obtained from the layer number of RPC.

The muons are minimum ionizing particles in the multi-GeV range of energies, and hence, they pass through many layers producing hits in each of those layers in the form of tracks. The ICAL can efficiently reconstruct muon energies and directions in the reconstructed muon energy range of 1 to 25 GeV~\cite{Chatterjee:2014vta}. Thanks to the magnetic field of 1.5 Tesla~\cite{Behera:2014zca}, ICAL will be able to distinguish $\mu^-$ and $\mu^+$ from the opposite curvatures of their tracks. This charge identification (CID) capability of ICAL helps us to separately identify neutrinos ($\nu_\mu$) and antineutrinos ($\bar{\nu}_\mu$). Due to ns time resolution of RPCs~\cite{Dash:2014ifa,Bhatt:2016rek,Gaur:2017uaf}, ICAL can efficiently distinguish the upward-going and downward-going muon events. The resonance scattering and deep-inelastic scattering of neutrinos with iron nuclei result in the production of hadrons which produce showers inside the detector.

In the present work, the neutrino events are simulated using NUANCE Monte Carlo (MC) neutrino event generator~\cite{Casper:2002sd} using ICAL-geometry as a target. While simulating data, we use atmospheric neutrino flux at the INO site~\cite{Athar:2012it,Honda:2015fha} in the Theni district of Tamil Nadu, India, with a rock coverage of 1 km in all directions. Due to the rock coverage of about 1 km (3800 m water equivalent), the downward-going cosmic muon background gets reduced by $\sim 10^6$~\cite{Dash:2015blu}. The application of fiducial volume cut further vetoes these events. We incorporate the effect of solar modulation on atmospheric neutrinos flux by considering solar maxima for half exposure and solar minima for another half. Since we estimate the median sensitivities in our analysis, we simulate unoscillated neutrino events at ICAL for a large exposure of 1000-year MC to minimize the statistical fluctuation. The neutrino oscillation with LIV in the three-flavor framework in the presence of matter with the PREM profile~\cite{Dziewonski:1981xy} is incorporated using reweighting algorithm following Refs.~\cite{Ghosh:2012px,Thakore:2013xqa,Devi:2014yaa}.

The detector response to muons and hadrons is simulated using GEANT4 simulations of the ICAL detector as described in Refs~\cite{Chatterjee:2014vta,Devi:2013wxa}. The track-like events associated with muons are fitted with the Kalman filter technique to obtain information about the energy, direction, and charge of the reconstructed muons~\cite{Bhattacharya:2015bsp}. Although, the configuration of ICAL is mainly optimized to measure the four-momenta of muon, it can also measure the hadron energy in each event. In the resonance and deep-inelastic scattering processes, a significant fraction of neutrino energy is carried away by the hadrons and it is defined as $E'_\text{had} = E_\nu - E_\mu$. The hadron energy deposited in a shower is estimated using the total number of hits that are not part of the muon track. Since we can obtain information on the hadron energy apart from precisely measuring the four-momenta of muon on an event-by-event basis, we consider the binning in three separate observables: muon direction $\cos\theta_\mu$, muon energy $E_\mu$, and hadron energy $E'_\text{had}$. The use of both $E_\mu$ and $E'_{\rm had}$ as separate observables, allows us to indirectly probe the incoming neutrino energy. Note that if we add these two observables to reconstruct the neutrino energy, we may not be able to take the advantage of precise measurement of muon energy. This additional information on hadron energy is very important to improve the constraints on the LIV parameters, as we show in our result section. The reconstructed $\mu^-$ and $\mu^+$ events at ICAL detector with observables $\cos\theta_\mu^\text{rec}$, $E_\mu^\text{rec}$, and ${E'}_\text{had}^\text{rec}$ are obtained after folding with detector properties using the migration matrices~\cite{Chatterjee:2014vta,Devi:2013wxa} provided by ICAL collaboration following the procedure mentioned in  Refs~\cite{Ghosh:2012px,Thakore:2013xqa,Devi:2014yaa}. These reconstructed $\mu^-$ and $\mu^+$ events at 50 kt ICAL detector are scaled from 1000-yr MC to 10-yr MC for analysis. Now, we present the reconstructed $\mu^-$ and $\mu^+$ events expected for 500 kt$\cdot$yr exposure at ICAL for the case of SI and SI with LIV.

\subsection{Total event rates}
\label{sec:event_rates}

First of all, we would like to address the question: can we observe the signal of non-zero CPT-violating LIV parameter ($a_{\mu\tau}$, $a_{e\mu}$, and $a_{e\tau}$) in the total number of $\mu^-$ and $\mu^+$ events at 50 kt ICAL detector for 10-year exposure? To answer this question, we estimate total reconstructed $\mu^-$ and $\mu^+$ events at ICAL for 500 kt$\cdot$yr exposure for the cases of SI and SI with non-zero CPT-violating LIV parameters one at-a-time considering $a_{\mu\tau}=\pm\,1.0 \times 10^{-23}~\text{GeV}$, $a_{e\mu}=\pm\, 2.5 \times 10^{-23}~\text{GeV}$, and $a_{e\tau}=\pm\, 2.5 \times 10^{-23}~\text{GeV}$. We present these numbers in Table~\ref{tab:total_events} assuming NO as true mass ordering while using the values of benchmark oscillation parameters given in Table~\ref{tab:osc-param-value}. While calculating total number of events, we integrate over reconstructed muon energy $E_\mu^\text{rec}$ in the range 1 to 25 GeV, reconstructed muon direction $\cos\theta_\mu^\text{rec}$ in the range -1 to 1, and reconstructed hadron energy ${E'}_\text{had}^{\text{rec}}$ in the range 0 to 25 GeV.

\begin{table}
  \centering
  \begin{tabular}{|ccc*{6}{c|}}
    \hline
    \hline
    \multicolumn{3}{|c|}{\multirow{2}{*}{Observables}} & \multicolumn{3}{c|}{$\mu^{-}$} & \multicolumn{3}{c|}{$\mu^{+}$} \\ 
    \cline{4-9}
    \multicolumn{3}{|c|}{} & $P_{\mu\mu}$ & $P_{e\mu}$ & Total & $P_{\bar\mu\bar\mu}$ & $P_{\bar{e}\bar\mu}$ & Total \\
    \hline
    \hline
    \multicolumn{3}{|c|}{SI} & 4318 & 95 & 4413 & 2002 & 12 & 2014 \\
    \cline{1-9}
    \hline
    \hline
    \multicolumn{1}{|c|}{\multirow{2}{*}{$a_{\mu\tau}$ [GeV]}} & \multicolumn{2}{c|}{$+ 1.0 \times 10^{-23}$} & 4319 & 91 & 4410 & 2015 & 11 & 2026\\ \cline{2-9}
    \multicolumn{1}{|c|}{} &\multicolumn{2}{c|}{$- 1.0 \times 10^{-23}$} & 4352 & 98 & 4450 & 1999 & 12 & 2011 \\
    \cline{1-9}
    \hline
    \hline
    \multicolumn{1}{|c|}{\multirow{2}{*}{$a_{e\mu}$ [GeV]}} & \multicolumn{2}{c|}{$+ 2.5 \times 10^{-23}$} & 4295 & 136 & 4431 & 2011 & 8 & 2019 \\ \cline{2-9}
    \multicolumn{1}{|c|}{} & \multicolumn{2}{c|}{$- 2.5 \times 10^{-23}$} & 4343 & 68 & 4411 & 1961 & 33 & 1994 \\
    \cline{1-9}
    \hline
    \hline
    \multicolumn{1}{|c|}{\multirow{2}{*}{$a_{e\tau}$ [GeV]}} & \multicolumn{2}{c|}{$+ 2.5 \times 10^{-23}$} & 4233 & 150 & 4383 & 1979 & 24 & 2003\\  \cline{2-9}
    \multicolumn{1}{|c|}{} & \multicolumn{2}{c|}{$- 2.5 \times 10^{-23}$} & 4400 & 58 & 4458 & 2015 & 10 & 2025 \\
    \hline
    \hline
  \end{tabular}
  \mycaption{The reconstructed $\mu^-$ and $\mu^+$ events expected at ICAL for 500 kt$\cdot$yr exposure. We present the reconstructed $\mu^-$ and $\mu^+$ events for the cases of SI (first row), $a_{\mu\tau} = \pm\, 1.0 \times 10^{-23}$ (second and third rows), $a_{e\mu} = \pm\, 2.5 \times 10^{-23}$ (fourth and fifth rows), and $a_{e\tau} = \pm\, 2.5 \times 10^{-23}$ (sixth and seventh rows). Along with total $\mu^-$ events, we also show the events coming from $\nu_\mu \rightarrow \nu_\mu$ ($P_{\mu\mu}$) disappearance channel and $\nu_e \rightarrow \nu_\mu$ ($P_{e\mu}$) appearance channel separately. Similarly, for $\mu^+$ also, we show the contribution from $\bar{\nu}_\mu \rightarrow \bar{\nu}_\mu$ ($P_{\bar{\mu}\bar{\mu}}$) disappearance channel and $\bar{\nu}_e \rightarrow \bar{\nu}_\mu$ ($P_{\bar{e}\bar{\mu}}$) appearance channel separately. We assume $\sin^2\theta_{23} = 0.5$ and NO as true mass ordering. The values of other oscillation parameters are taken from Table~\ref{tab:osc-param-value}.}
  \label{tab:total_events}
\end{table}

In Table~\ref{tab:total_events}, we present total $\mu^-$ events along with the estimates of individual events contributed from $\nu_\mu \rightarrow \nu_\mu$ disappearance channel and $\nu_e \rightarrow \nu_\mu$ appearance channel. Similarly, we also show total $\mu^+$ events originating from $\bar{\nu}_\mu \rightarrow \bar{\nu}_\mu$ disappearance and $\bar{\nu}_e \rightarrow \bar{\nu}_\mu$ appearance channels. Here, we observe that about 2\% of total $\mu^-$ events at ICAL for SI are contributed by the appearance channel. The tau lepton may get produced during the interaction of $\nu_\tau$ inside the detector via $\nu_\mu \rightarrow \nu_\tau$ and $\nu_e \rightarrow \nu_\tau$ channels. The number of muon events produced during tau decay inside the detector is small (about 2\% of the total upward going muon events produced from $\nu_\mu$ interactions~\cite{Pal:2014tre}), and the energies of these muon events are lower and mostly below the 1 GeV energy threshold of the ICAL detector. In the present work, we do not consider this small contribution.

We would like to point out that the difference between total $\mu^-$ ($\mu^+$) events for the cases of SI and SI with non-zero CPT-violating LIV parameters  ($a_{\mu\tau}$, $a_{e\mu}$, and $a_{e\tau}$) one at-a-time is not large. But while presenting our results, we demonstrate that ICAL can place competitive bounds on LIV parameters using the information contained in the distributions of reconstructed $\mu^-$ and $\mu^+$ events as a function of $E_\mu^\text{rec}$, $\cos\theta_\mu^\text{rec}$, and ${E'}_\text{had}^\text{rec}$. To validate this claim, now we show the impact of non-zero LIV parameters one at-a-time on the spectral distributions of reconstructed $\mu^-$ and $\mu^+$ events as a function of $E_\mu^\text{rec}$, and  $\cos\theta_\mu^\text{rec}$ while integrating over ${E'}_\text{had}^\text{rec}$.

\subsection{Reconstructed event distributions }
\label{sec:event_dist}

In the present analysis, we use the binning scheme for reconstructed observables $E_\mu^\text{rec}$, $\cos\theta_\mu^\text{rec}$, and ${E'}_\text{had}^\text{rec}$ as shown in Table~\ref{tab:binning_scheme}. We have total 13 bins for $E_\mu^\text{rec}$ in the range of 1 to 25 GeV, 15 bin for $\cos\theta_\mu^\text{rec}$ in the range -1 to 1, and 4 bins for ${E'}_\text{had}^\text{rec}$ in the range 0 to 25 GeV. Although no significant oscillations are developed for downward-going events, these events are included in our analysis because they help in increasing the overall statistics and minimizing the normalization errors in the atmospheric neutrino events. In this way, we also take care of those upward-going (near horizon) neutrino events, which result in the downward-going reconstructed muon events due to angular smearing during the interaction of neutrinos as well as reconstruction. We have considered the same binning scheme for reconstructed $\mu^-$ as well as $\mu^+$ events. 

\begin{table}[t] 
  \centering 
  \begin{tabular}{|c| c| c| c|} 
    \hline 
    Observable & Range & Bin width & Total bins \\ 
    \hline
    $E_{\mu}^\text{rec}$ (GeV) & \makecell[c]{$[1, 11]$ \\ $[11, 21]$ \\ $[21, 25]$} & 
    \makecell[c]{1 \\ 5 \\ 4} & 
    $\left.\begin{tabular}{l}
    10 \\ 2 \\ 1
    \end{tabular}\right\}$  13 \\
    \hline
    $\cos\theta_\mu^\text{rec}$  & \makecell[c]{$[-1.0, 0.0]$ \\ $[0.0, 1.0]$} & 
    \makecell[c]{0.1 \\ 0.2} & 
    $\left.\begin{tabular}{l}
    10 \\ 5
    \end{tabular}\right\}$  15 \\
    \hline
    ${E'}_\text{had}^\text{rec}$ (GeV)  & \makecell[c]{$[0, 2]$ \\ $[2, 4]$ \\ $[4, 25]$} &
    \makecell[c]{1 \\ 2 \\ 21} & 
    $\left.\begin{tabular}{l}
    2 \\ 1 \\ 1
    \end{tabular}\right\}$  4\\ \hline
  \end{tabular}
  \mycaption{The binning scheme used in the present analysis for reconstructed observables $E_\mu^\text{rec}$, $\cos\theta_\mu^\text{rec}$, and ${E'}_\text{had}^\text{rec}$ for both reconstructed $\mu^-$ as well as $\mu^+$.}
  \label{tab:binning_scheme}
\end{table}

\begin{figure}[t]
  \centering
  \includegraphics[width=0.32\textwidth]{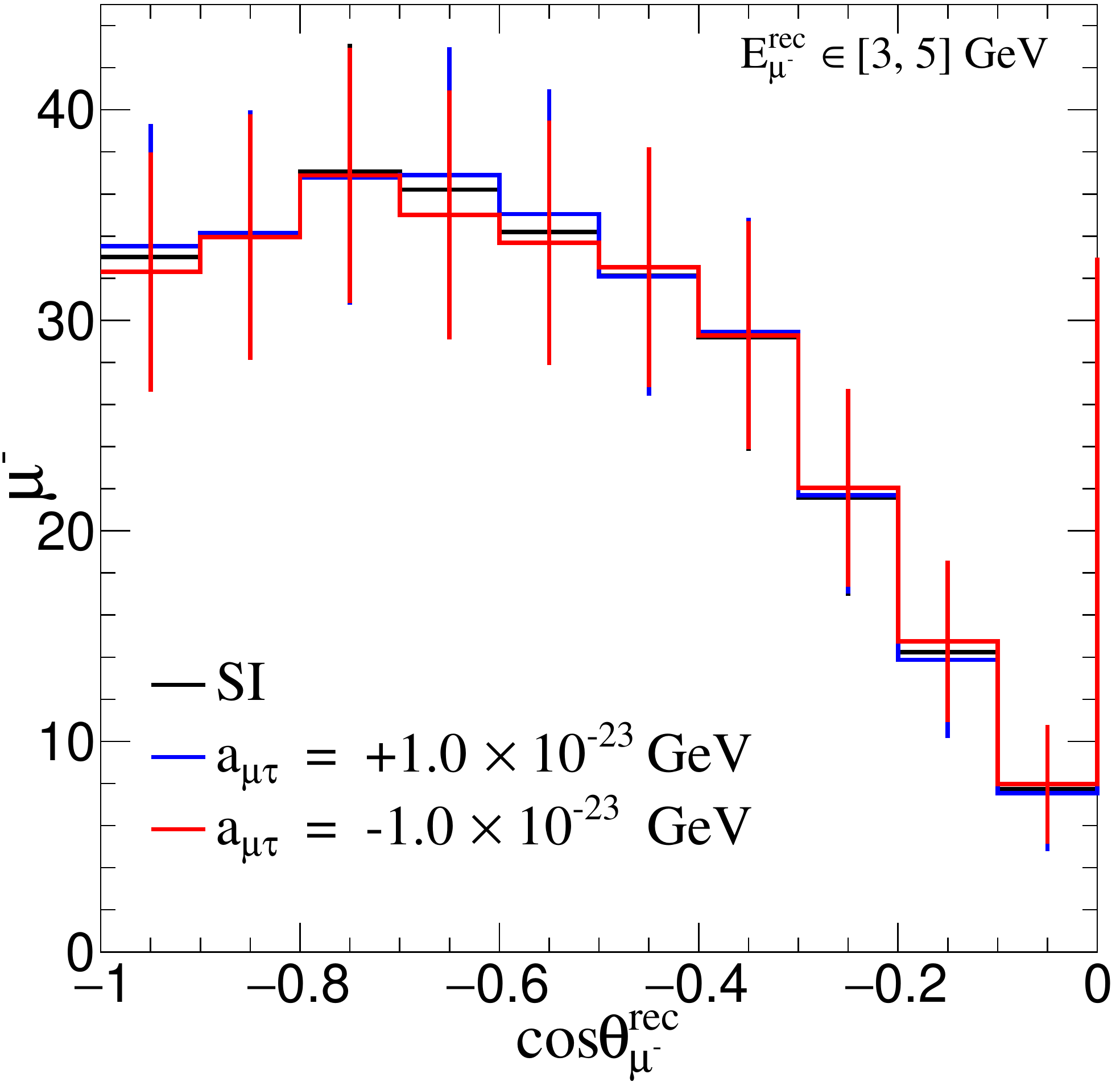}
  \includegraphics[width=0.32\textwidth]{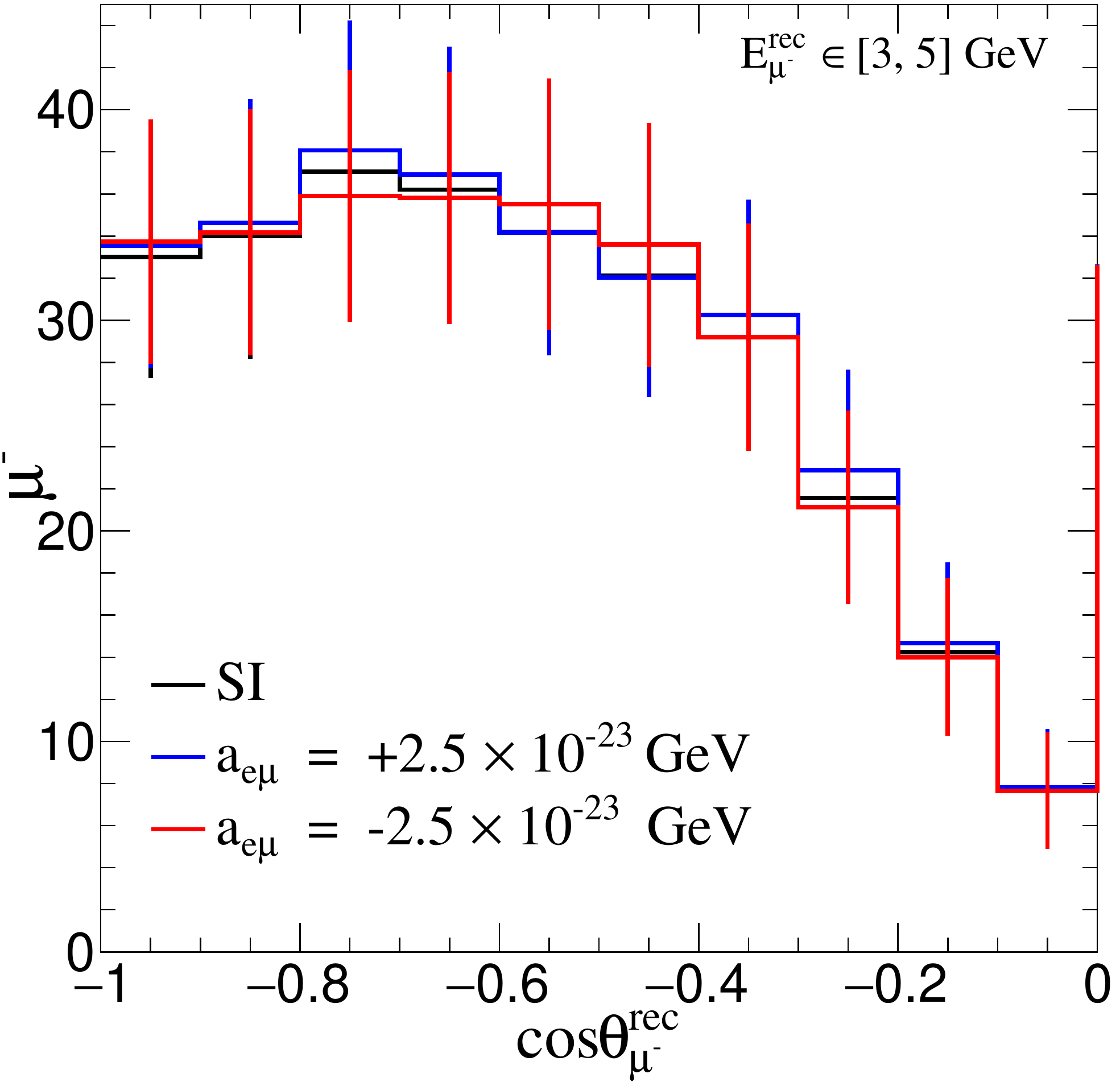}
  \includegraphics[width=0.32\textwidth]{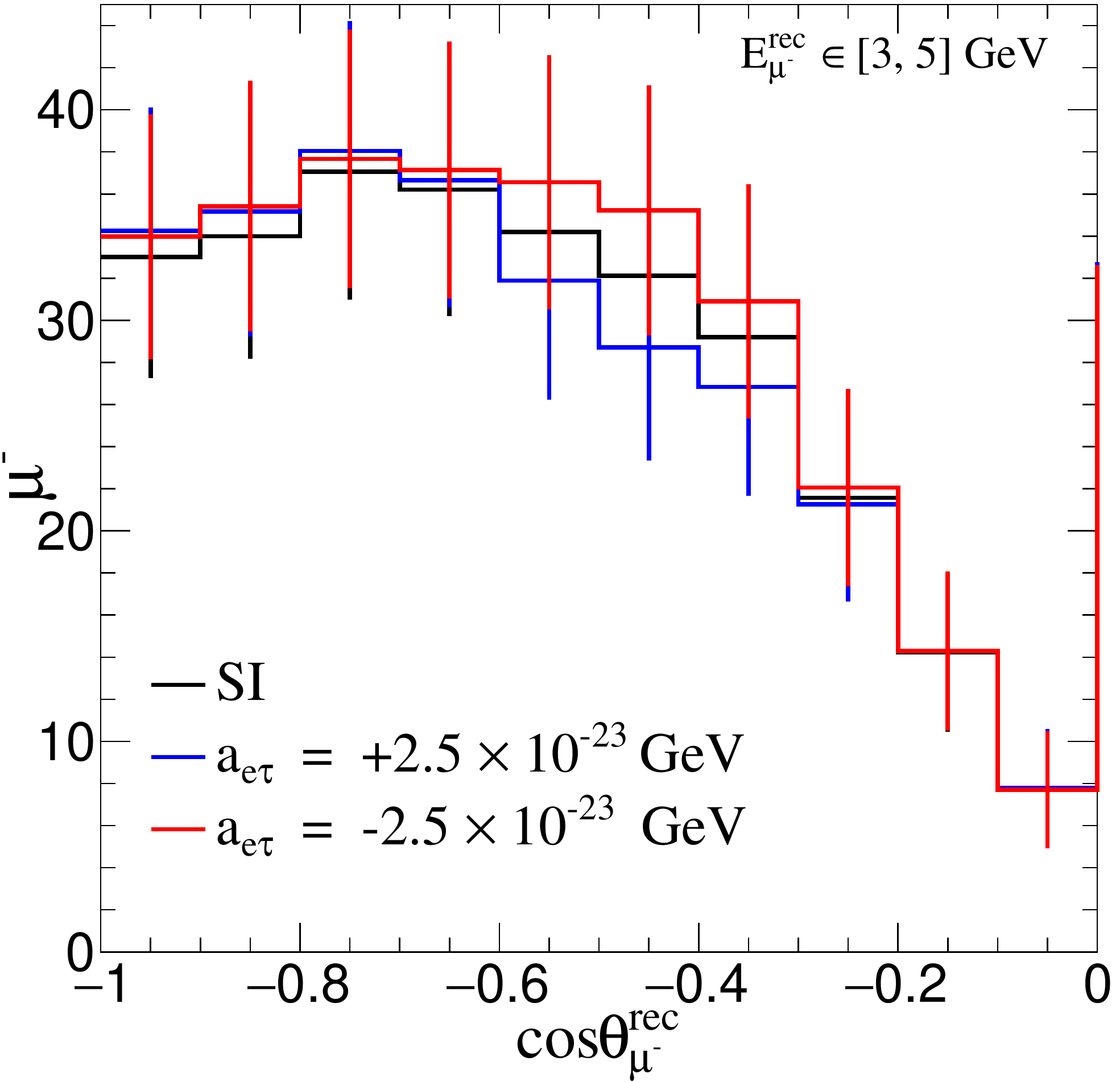}
  \includegraphics[width=0.32\textwidth]{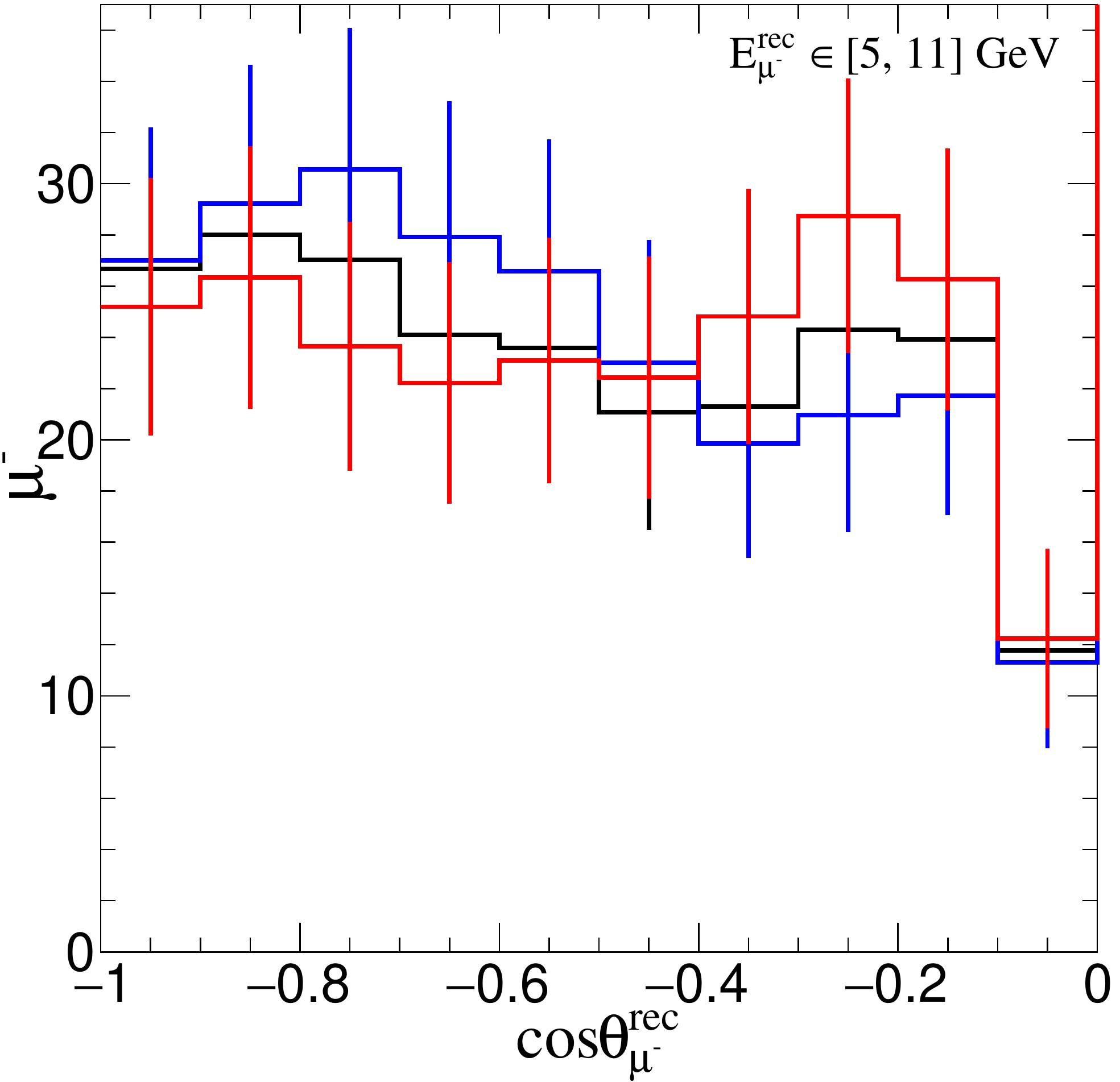}
  \includegraphics[width=0.32\textwidth]{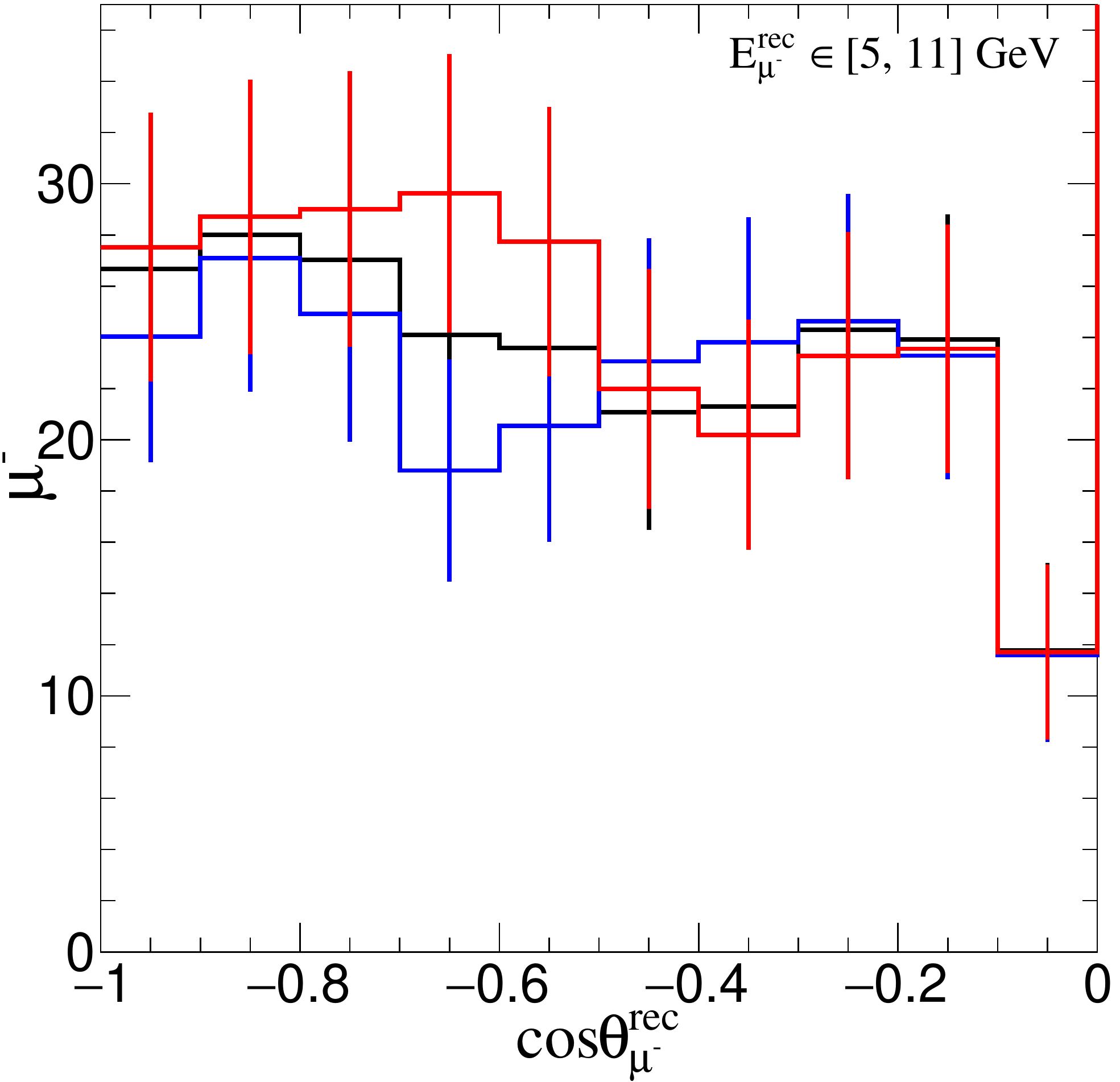}
  \includegraphics[width=0.32\textwidth]{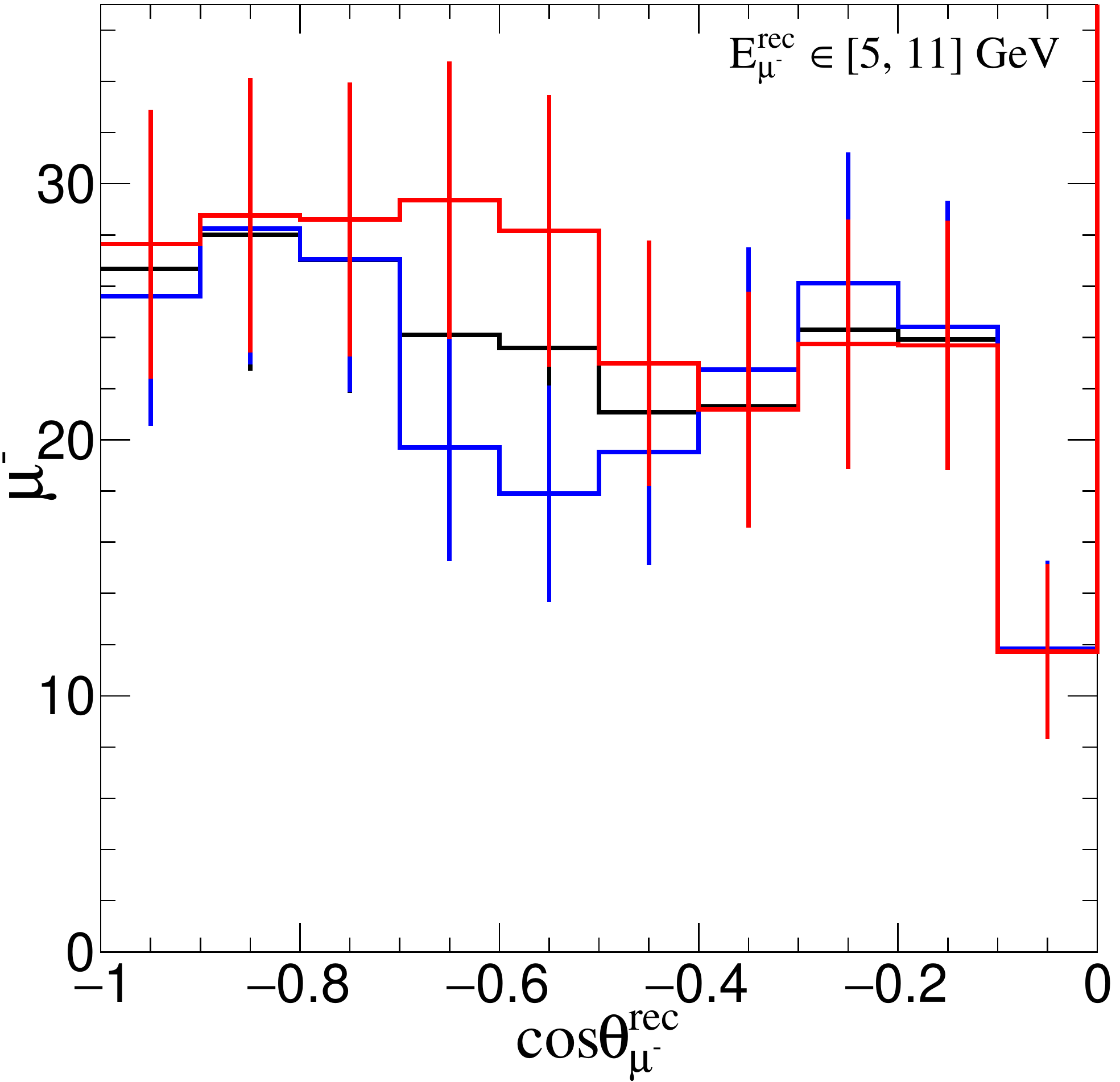}
  \includegraphics[width=0.32\textwidth]{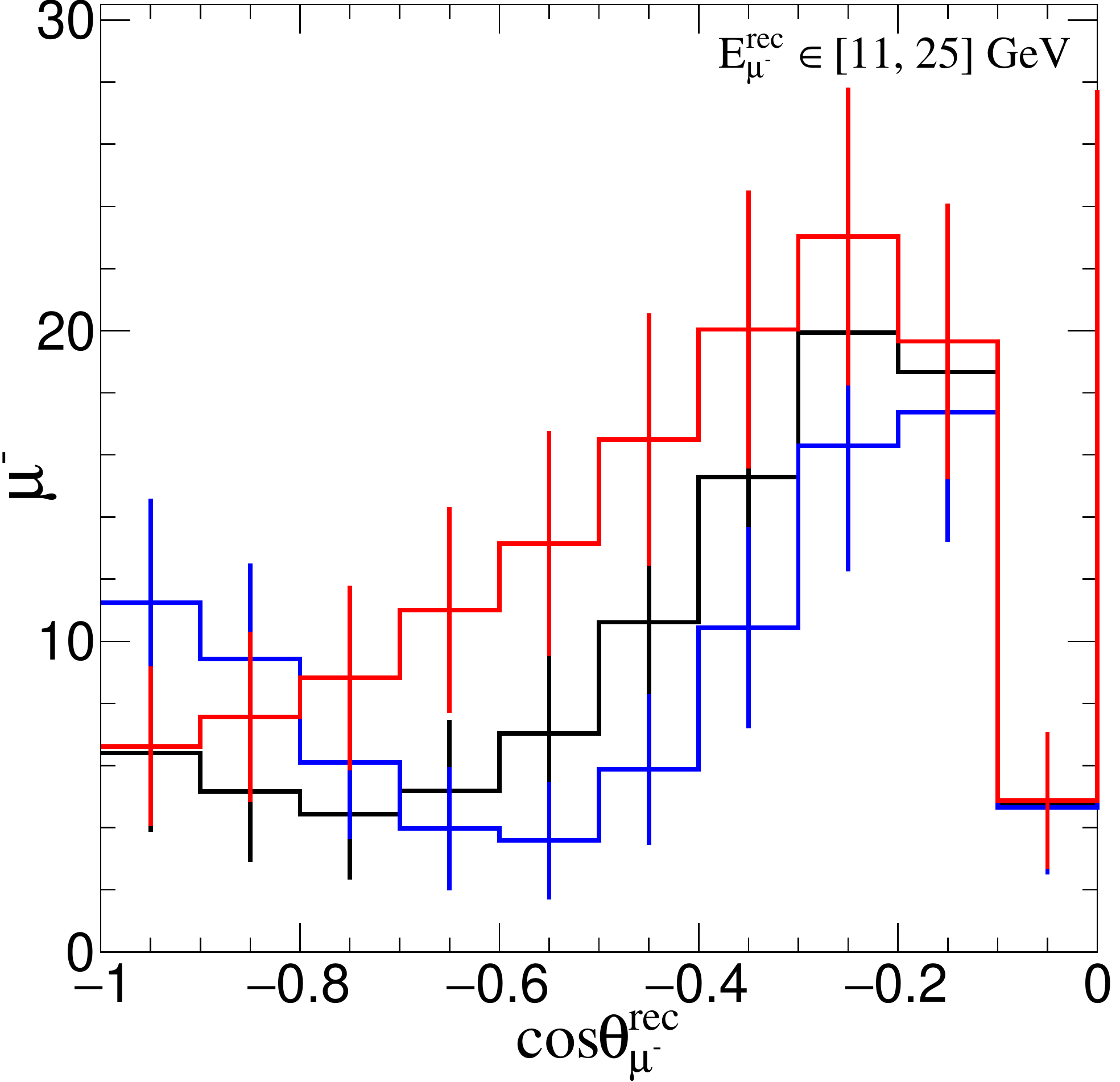}
  \includegraphics[width=0.32\textwidth]{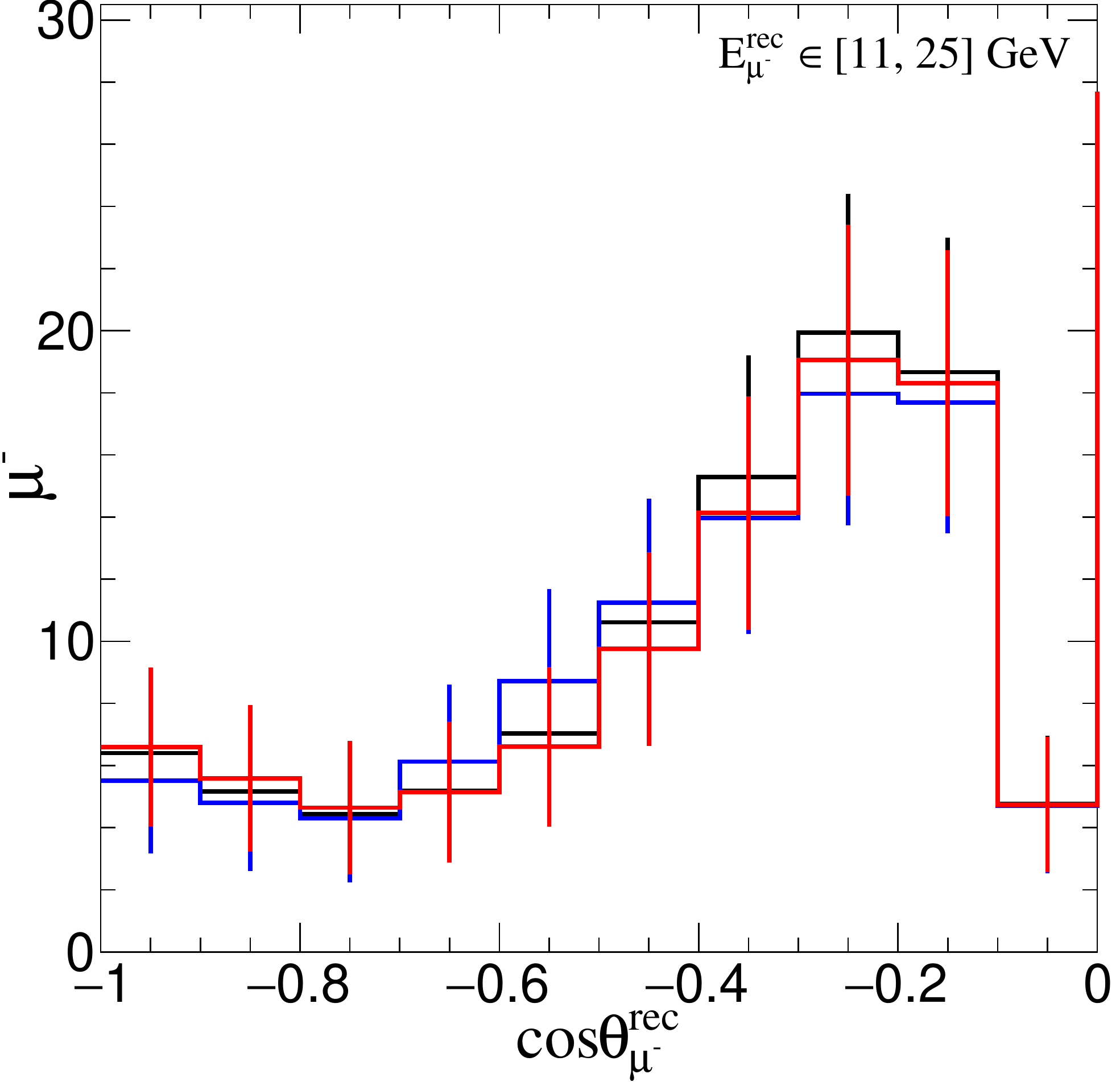}
  \includegraphics[width=0.32\textwidth]{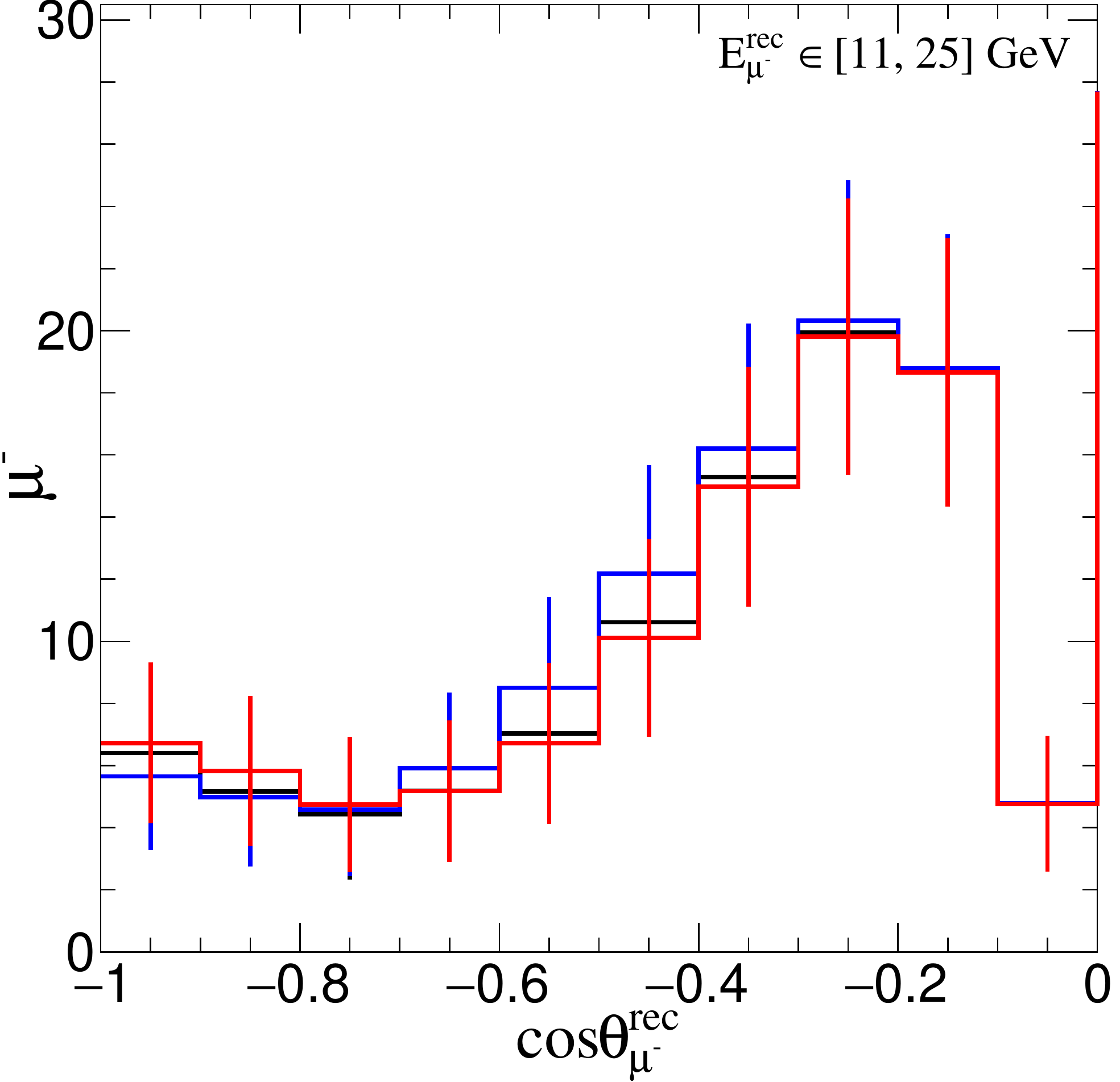}
  \mycaption{The distributions of reconstructed $\mu^-$ events at the ICAL detector for 500 kt$\cdot$yr exposure for three different range of $E_\mu^\text{rec}$ : [3, 5] GeV (top panels), [5, 11] GeV (middle panels) and [11, 25] GeV (bottom panels) where y-ranges are different among these rows. Here, error bars show the statistical uncertainties. Note that, although the events are binned in ($E_\mu^\text{rec}$, $\cos\theta_\mu^\text{rec}$, ${E'}_\text{had}^\text{rec}$) binning scheme given in Table~\ref{tab:binning_scheme}, the reconstructed events in the hadron energy bins are integrated in the range 0 to 25 GeV in these plots. The left, middle and right panels show the effect of $a_{\mu\tau}$, $a_{e\mu}$, and $a_{e\tau}$, respectively one parameter at-a-time. In the left panels, the red, black, and blue curves represent $a_{\mu\tau} = -1.0 \times 10^{-23}$ GeV, 0, and $1.0 \times 10^{-23}$ GeV, respectively. In the middle panels, the red, black, and blue curves refer to $a_{e\mu} = -2.5 \times 10^{-23}$ GeV, 0, and $2.5 \times 10^{-23}$ GeV, respectively. Similarly, in the right panels, the red, black, and blue curves stand for $a_{e\tau} = -2.5 \times 10^{-23}$ GeV, 0, and $2.5 \times 10^{-23}$ GeV, respectively. We assume $\sin^2\theta_{23} = 0.5$ and NO as true mass ordering. The values of other oscillation parameters are taken from Table~\ref{tab:osc-param-value}.}
  \label{fig:event_dist_mu-}
\end{figure}

\begin{figure}[t]
  \centering
  \includegraphics[width=0.32\textwidth]{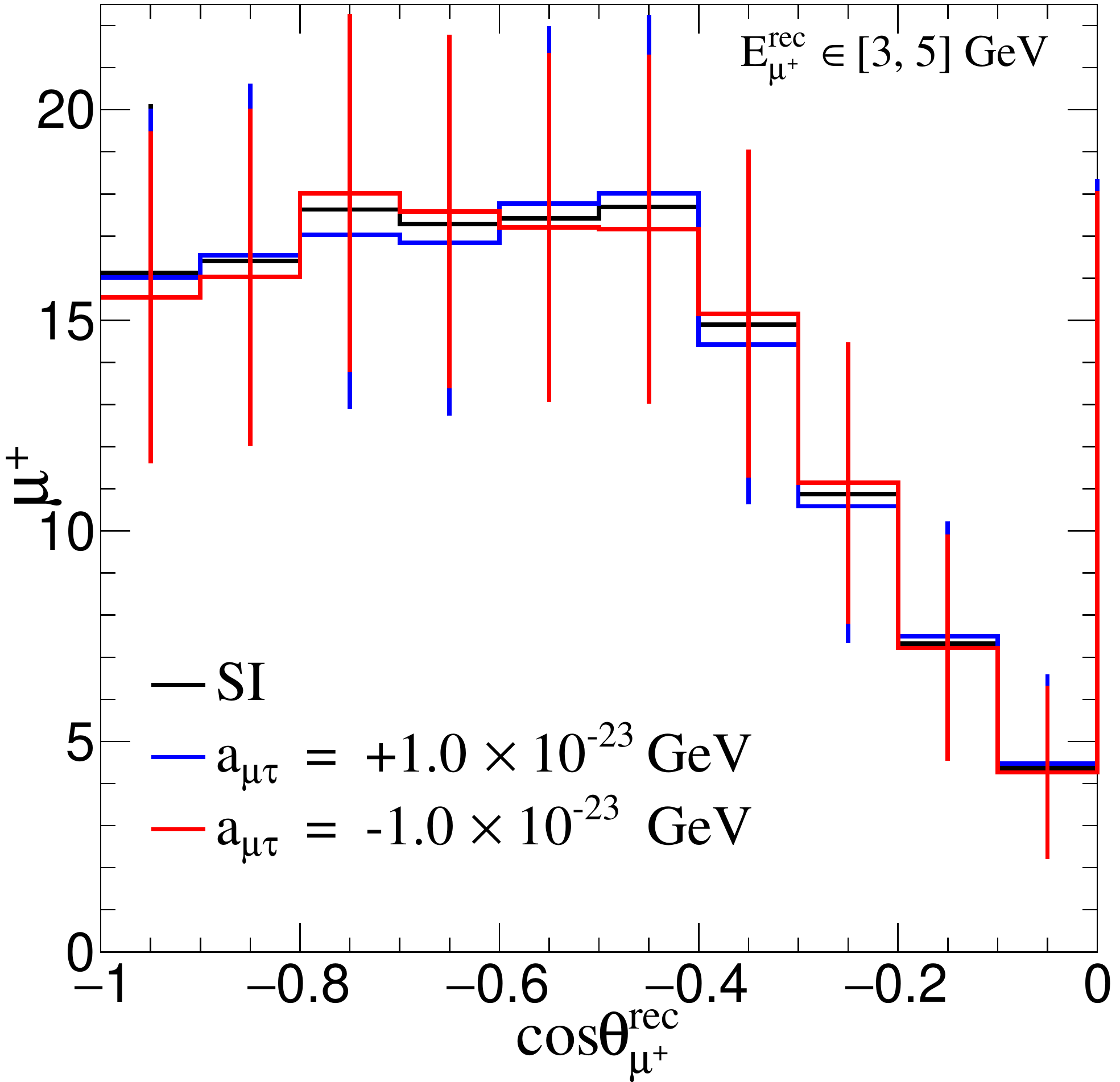}
  \includegraphics[width=0.32\textwidth]{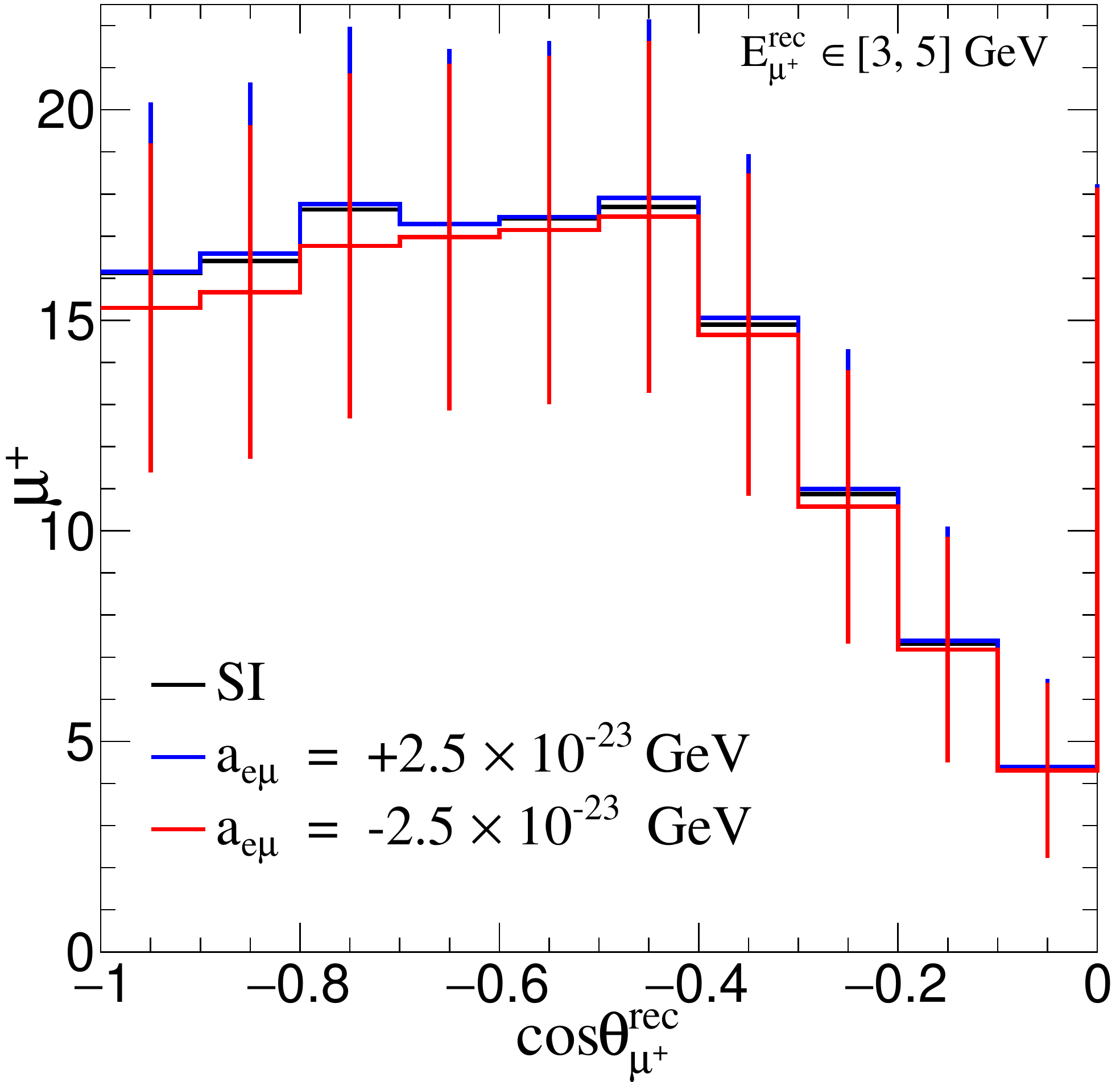}
  \includegraphics[width=0.32\textwidth]{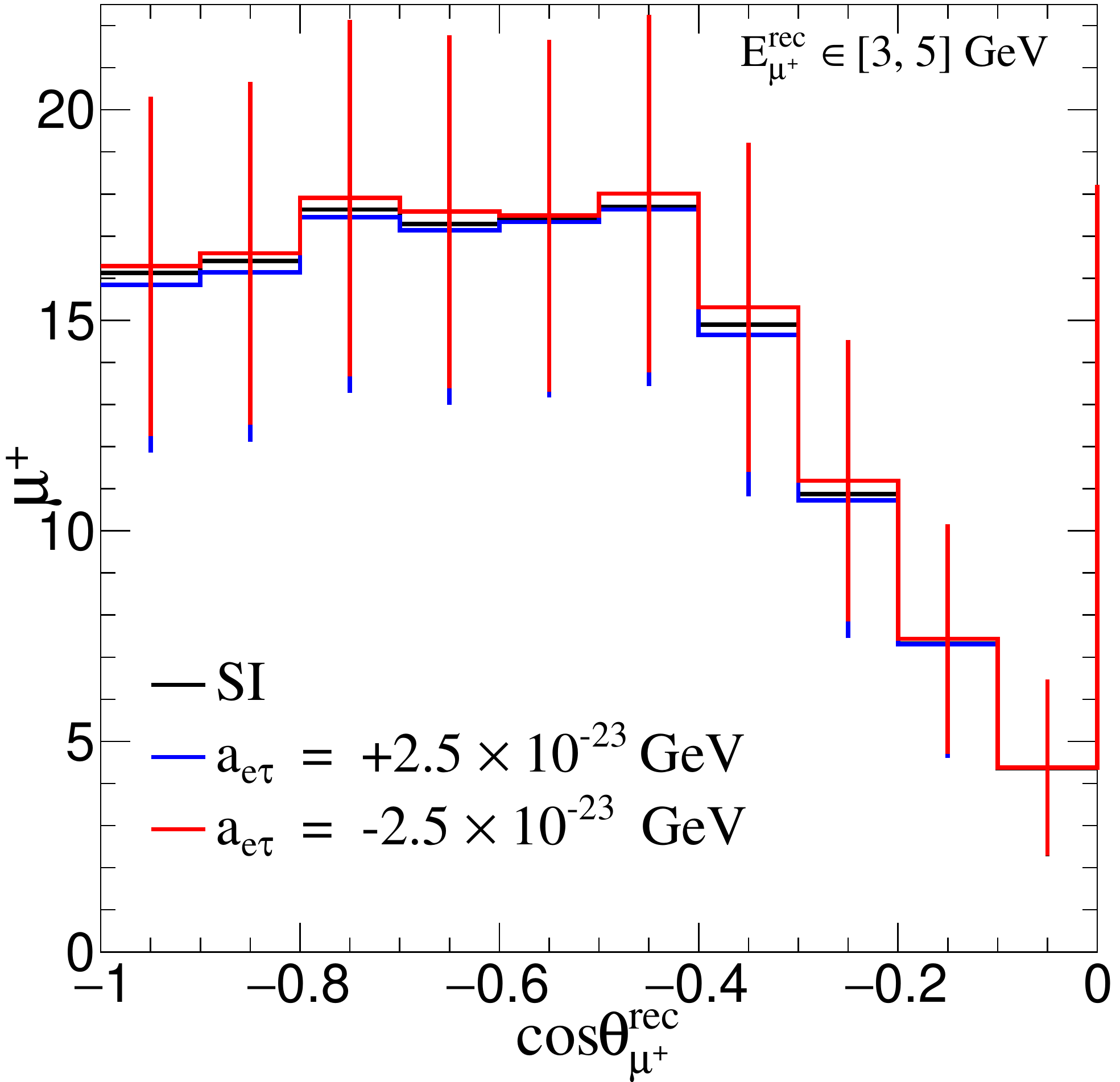}
  \includegraphics[width=0.32\textwidth]{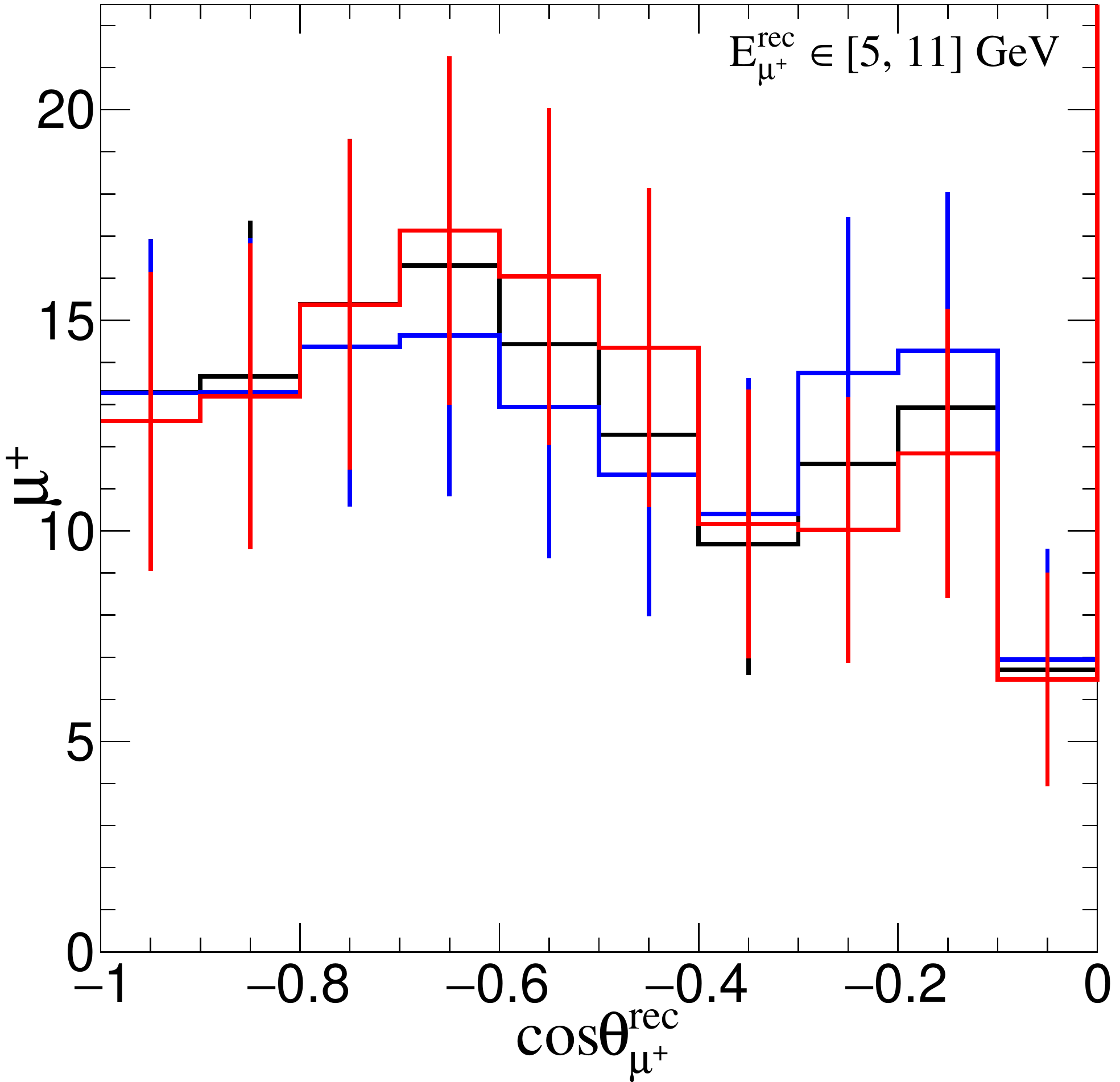}
  \includegraphics[width=0.32\textwidth]{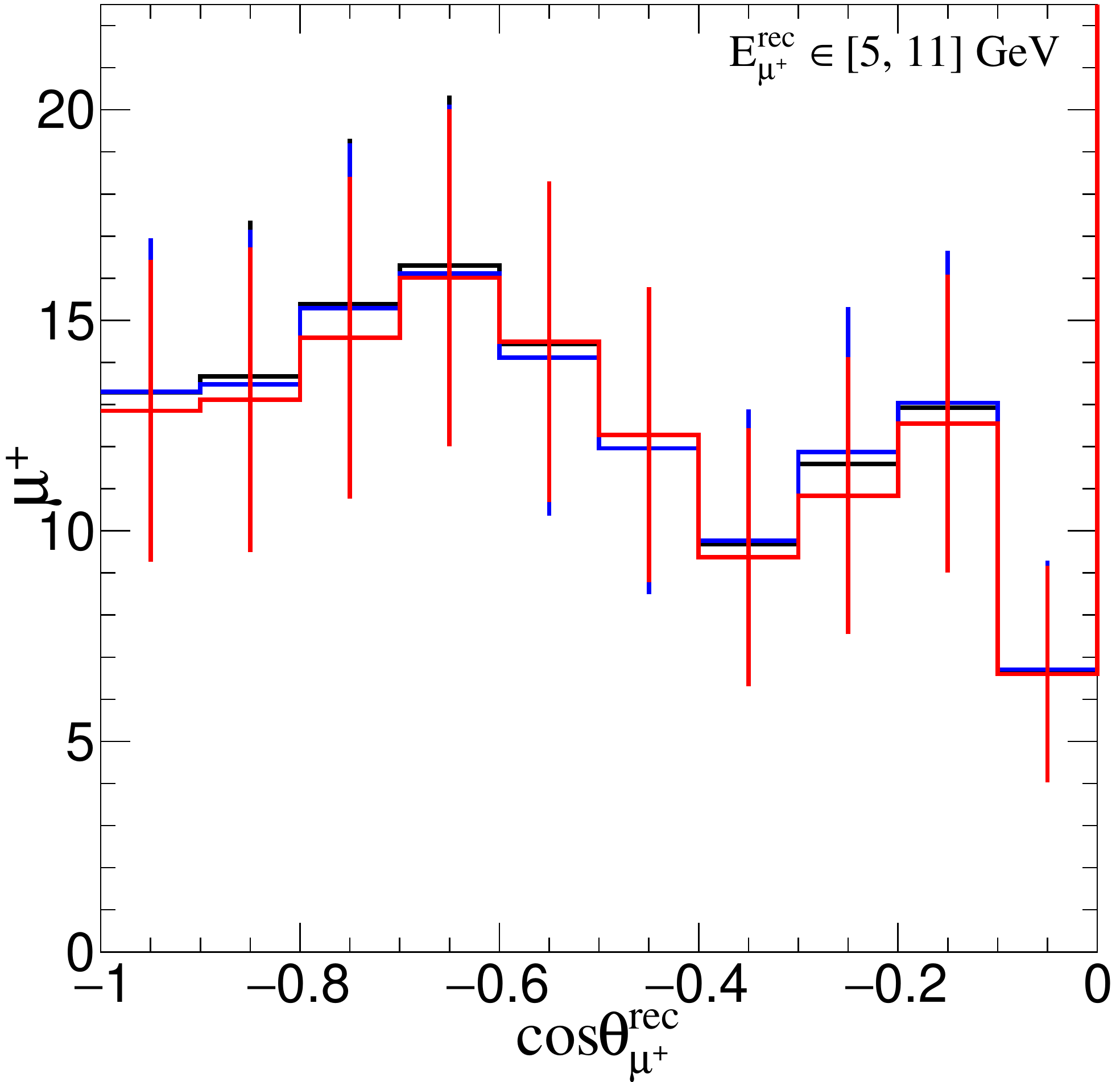}
  \includegraphics[width=0.32\textwidth]{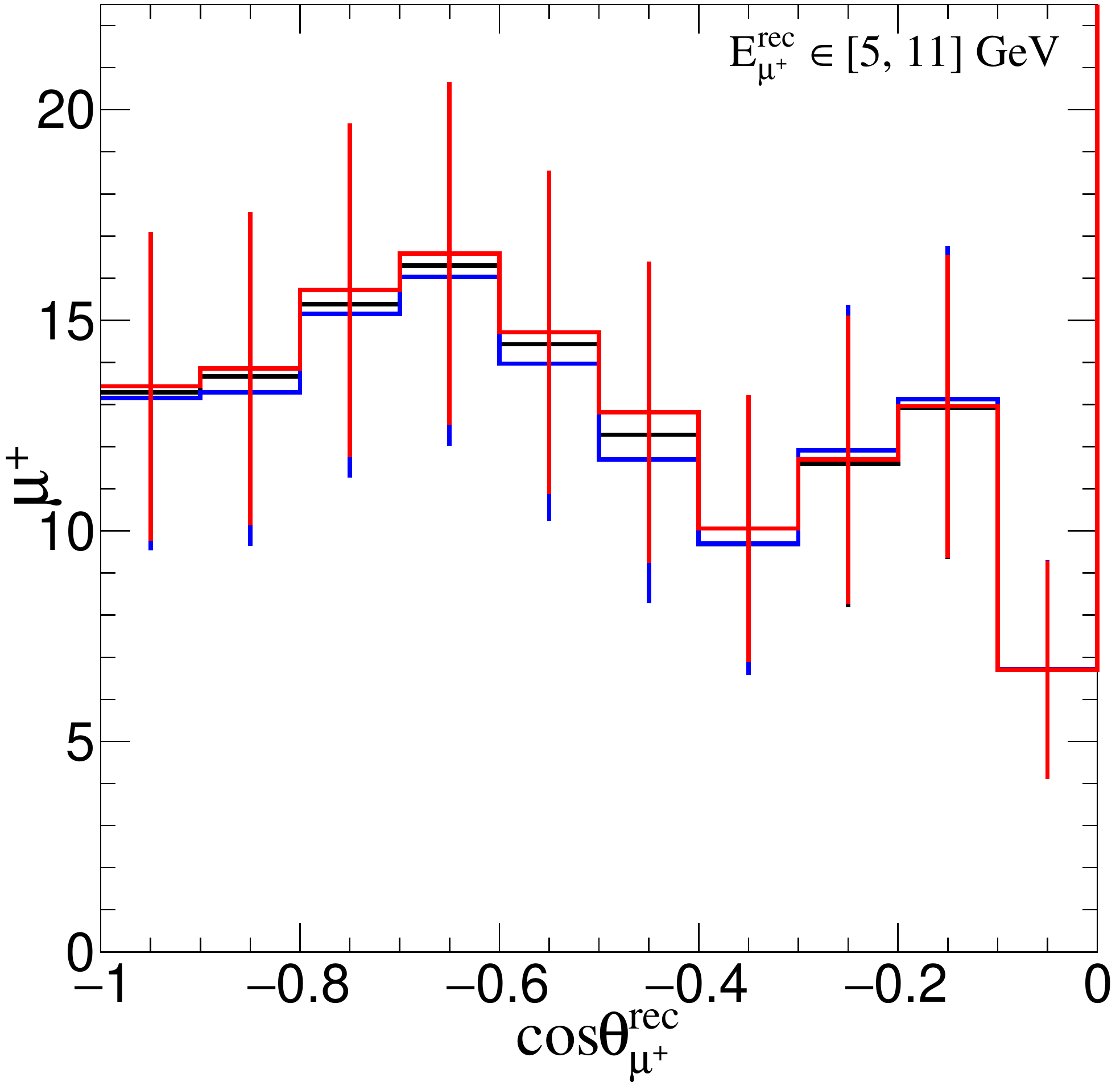}
  \includegraphics[width=0.32\textwidth]{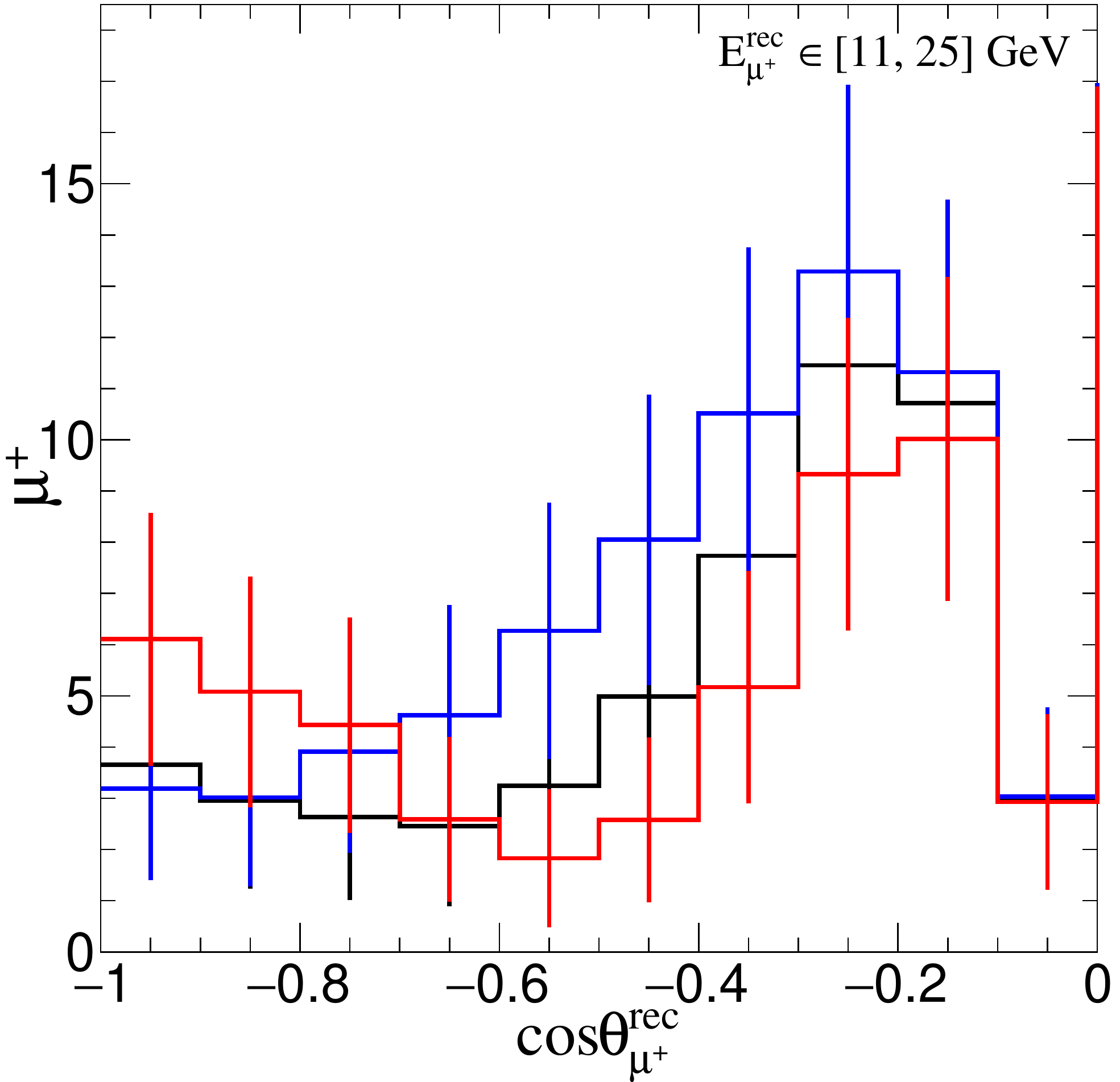}
  \includegraphics[width=0.32\textwidth]{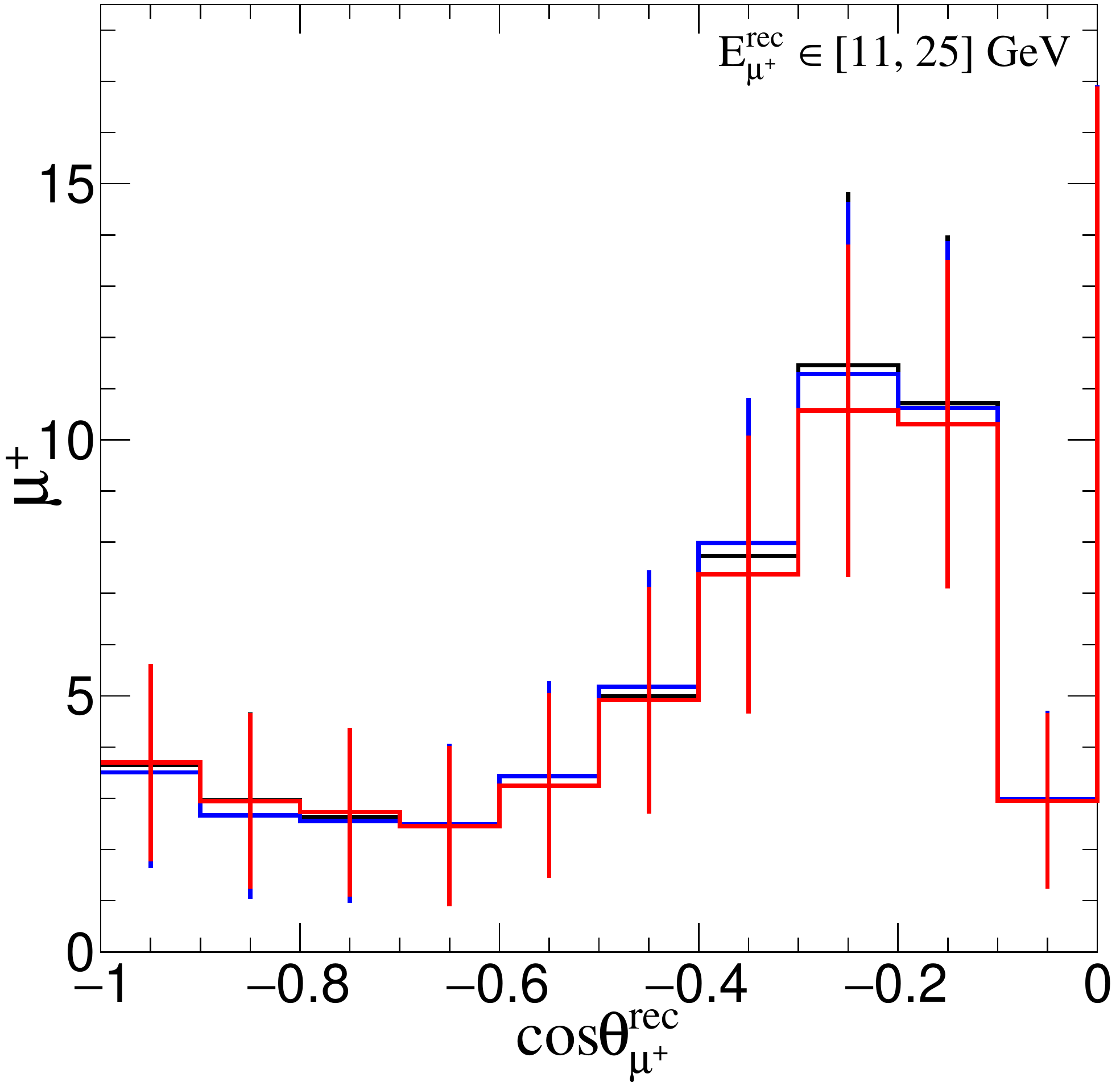}
  \includegraphics[width=0.32\textwidth]{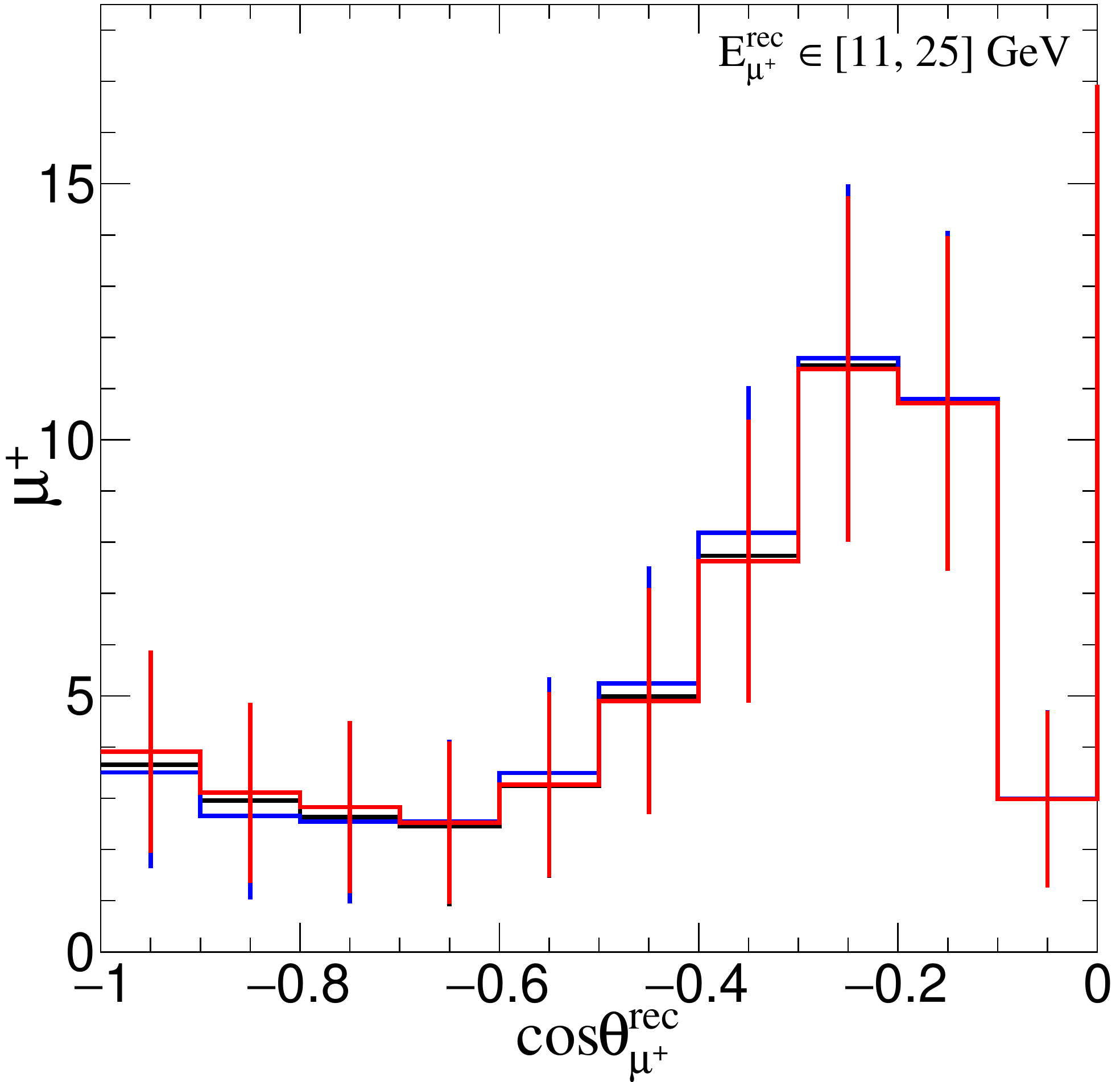}
  \mycaption{The distributions of reconstructed $\mu^+$ events at the ICAL detector for 500 kt$\cdot$yr exposure for three different range of $E_\mu^\text{rec}$ : [3, 5] GeV (top panels), [5, 11] GeV (middle panels) and [11, 25] GeV (bottom panels) where y-ranges are different among these rows. Here, error bars show the statistical uncertainties. Note that, although the events are binned in ($E_\mu^\text{rec}$, $\cos\theta_\mu^\text{rec}$, ${E'}_\text{had}^\text{rec}$) binning scheme given in Table~\ref{tab:binning_scheme}, the reconstructed events in the hadron energy bins are integrated in the range 0 to 25 GeV in these plots. The left, middle and right panels show the effect of $a_{\mu\tau}$, $a_{e\mu}$, and $a_{e\tau}$, respectively one parameter at-a-time. In the left panels, the red, black, and blue curves represent $a_{\mu\tau} = -1.0 \times 10^{-23}$ GeV, 0, and $1.0 \times 10^{-23}$ GeV, respectively. In the middle panels, the red, black, and blue curves refer to $a_{e\mu} = -2.5 \times 10^{-23}$ GeV, 0, and $2.5 \times 10^{-23}$ GeV, respectively. Similarly, in the right panels, the red, black, and blue curves stand for $a_{e\tau} = -2.5 \times 10^{-23}$ GeV, 0, and $2.5 \times 10^{-23}$ GeV, respectively. We assume $\sin^2\theta_{23} = 0.5$ and NO as true mass ordering. The values of other oscillation parameters are taken from Table~\ref{tab:osc-param-value}.}
  \label{fig:event_dist_mu+}
\end{figure}

In Fig.~\ref{fig:event_dist_mu-}, we present the event distributions of reconstructed $\mu^-$ events as a function of $\cos\theta_\mu^\text{rec}$ in the range -1 to 0 for three different ranges of reconstructed muon energies. For $E_\mu^\text{rec}$, we consider the energy range of [3, 5] GeV, [5, 11] GeV, and [11, 25] GeV as shown in the top, middle, and bottom panels, respectively. We integrate over ${E'}_\text{had}^\text{rec}$ in the range of 0 to 25 GeV. The error bars show the statistical uncertainties, which are obtained by taking the square root of the number of events. The left, middle and right panels show the impact of $a_{\mu\tau}$, $a_{e\mu}$, and $a_{e\tau}$, respectively, considering only one parameter at-a-time and remaining parameters are assumed to be zero. In the same fashion, we demonstrate the impact of $a_{\mu\tau}$, $a_{e\mu}$, and $a_{e\tau}$ on reconstructed $\mu^+$ events in Fig.~\ref{fig:event_dist_mu+}.

Before we discuss the impact of non-zero LIV parameters $a_{\mu\tau}$, $a_{e\mu}$, and $a_{e\tau}$,  we would like to mention few general features of the event spectra in Figs.~\ref{fig:event_dist_mu-} and \ref{fig:event_dist_mu+}. The number of reconstructed $\mu^-$ and $\mu^+$ events decrease as we go towards higher energies. This happens because of power law where the atmospheric neutrino flux has an energy dependence of $E_\nu^{-2.7}$. Though the neutrino and antineutrino fluxes along horizontal direction ($\cos\theta_\nu \sim 0$) are higher than the vertical direction ($\cos\theta_\nu \sim \pm1$), we observe lesser reconstructed $\mu^-$ and $\mu^+$ events around $\cos\theta_\mu^\text{rec} \in [-0.1, 0]$ due to the poor reconstruction efficiency of the ICAL detector for horizontal direction irrespective of the choice of $E_\mu^\text{rec}$ range. Note that the number of reconstructed $\mu^+$ events are smaller than the reconstructed $\mu^-$ events due to the lower interaction cross-section for antineutrinos.

In the atmospheric sector, the impact of $a_{\mu\tau}$ is largest among all the CPT-violating LIV parameters as shown in the left panels of  Fig.~\ref{fig:event_dist_mu-}. In left panels, we compare three cases corresponding to  $a_{\mu\tau} = -1.0 \times 10^{-23}$ GeV (red curves), 0 (black curves), and $1.0 \times 10^{-23}$ GeV (blue curves). The impact of $a_{\mu\tau}$ is significant at higher energies of $E_\mu^\text{rec} \in $ [5, 11] GeV and [11, 25] GeV. We also observe that the polarity of difference between SI and SI with non-zero $a_{\mu\tau}$ changes as move from $\cos\theta_\mu^\text{rec}$ of -1 to 0. Therefore, the total rate shows a diluted effect of non-zero  $a_{\mu\tau}$ and the good directional resolution of the ICAL detector plays an important role while constraining LIV parameter $a_{\mu\tau}$. The impact of the LIV parameter $a_{e\mu}$ is shown in the panels of the middle column where the red, black, and blue curves refers to $a_{e\mu} = -2.5 \times 10^{-23}$ GeV, 0, and $2.5 \times 10^{-23}$ GeV, respectively. We can observe that the difference between SI and SI with non-zero $a_{e\mu}$ is significant for $E_\mu^\text{rec} \in $ [5, 11] GeV. Similarly, we show the effect of non-zero $a_{e\tau}$ on the distribution of reconstructed $\mu^-$ events in the right panels of Fig.~\ref{fig:event_dist_mu-} where the red, black, and blue curves stand for $a_{e\tau} = -2.5 \times 10^{-23}$ GeV, 0, and $2.5 \times 10^{-23}$ GeV, respectively. The difference between SI and SI with non-zero $a_{e\tau}$ is largest for $E_\mu^\text{rec} \in $ [5, 11] GeV, but we also see significant difference for $E_\mu^\text{rec} \in $ [3, 5] GeV.

As far the case of $\mu^+$ is concerned, we observe that the impact of $a_{\mu\tau}$ is significant in the higher muon energy range of [5, 11] GeV and [11, 25] GeV as shown in the left panels in Fig.~\ref{fig:event_dist_mu+}. By comparing the left panels of Figs.~\ref{fig:event_dist_mu-} and \ref{fig:event_dist_mu+}, we observe that for a given value of $a_{\mu\tau}$, the effect is in the opposite direction for reconstructed $\mu^-$ and $\mu^+$ events. For example, the reconstructed $\mu^-$ events increase for the negative value of $a_{\mu\tau}$ and the red curve is higher than black curve in the bottom left panel of Fig.~\ref{fig:event_dist_mu-} for $E_\mu^\text{rec} \in $ [11, 25] GeV whereas the corresponding curve is lower in the case of $\mu^+$ as shown in Fig.~\ref{fig:event_dist_mu+}. The addition of reconstructed $\mu^-$ and $\mu^+$ events will dilute this feature. This observation indicates that the charge identification efficiency of the ICAL detector will play an important role while constraining LIV parameter $a_{\mu\tau}$. The effect of other LIV parameters $a_{e\mu}$ and $a_{e\tau}$ on the event spectra of reconstructed $\mu^+$ events is significantly lower than that for the case of $\mu^-$. This happens because the effect of $a_{e\mu}$ and $a_{e\tau}$ is driven by matter resonance term, which occurs for the case of neutrino for NO but not antineutrino as explained in the section~\ref{sec:oscillograms}. Next, we discuss the numerical method and analysis procedure that we use to obtain our final results. 

\section{Simulation method}
\label{sec:statistical_analysis}

To obtain the median sensitivity of the ICAL atmospheric neutrino experiment in the frequentist approach~\cite{Blennow:2013oma}, we define the following Poissonian $\chi_-^2$~\cite{Baker:1983tu} for reconstructed $\mu^{-}$ events in $E_\mu^\text{rec}$, $\cos\theta_\mu^\text{rec}$, and ${E'}_\text{had}^\text{rec}$ observables (the so-called ``3D'' analysis as considered in~\cite{Devi:2014yaa}):

\begin{equation}\label{eq:chisq_mu-}
\chi^2_- (3\text{D}) = \mathop{\text{min}}_{\xi_l} \sum_{i=1}^{N_{{E'}_\text{had}^\text{rec}}} \sum_{j=1}^{N_{E_{\mu}^\text{rec}}} \sum_{k=1}^{N_{\cos\theta_\mu^\text{rec}}} \left[2(N_{ijk}^\text{theory} - N_{ijk}^\text{data}) -2 N_{ijk}^\text{data} \ln\left(\frac{N_{ijk}^\text{theory} }{N_{ijk}^\text{data}}\right)\right] + \sum_{l = 1}^5 \xi_l^2
\end{equation}

where, 
\begin{equation}\label{eq:N_theory}
N_{ijk}^\text{theory} = N_{ijk}^0\left(1 + \sum_{l=1}^5 \pi^l_{ijk}\xi_l\right).
\end{equation}

In these equations, $N_{ijk}^\text{data}$ and $N_{ijk}^\text{theory}$ stand for the expected and observed number of reconstructed $\mu^-$ events in a given  $(E_\mu^\text{rec}, \cos\theta_\mu^\text{rec}, {E'}_\text{had}^\text{rec})$ bin, respectively. We used the binning scheme mentioned in Table~\ref{tab:binning_scheme} where $N_{E_{\mu}^\text{rec}} = 13$, $N_{\cos\theta_\mu^\text{rec}} = 15$, and $N_{{E'}_\text{had}^\text{rec}} = 4$. In Eq.~\ref{eq:N_theory}, $N_{ijk}^0$ corresponds to the number of expected events without considering systematic uncertainties. In this analysis, we consider five systematic uncertainties following Refs.~\cite{Kameda:2002fx, Ghosh:2012px, Thakore:2013xqa}: 20\% error in flux normalization, 10\% error in cross section, 5\% energy dependent tilt error in flux, 5\% zenith angle dependent tilt error in flux, and 5\% overall systematics. We use the well-known method of pulls~\cite{GonzalezGarcia:2004wg,Huber:2002mx,Fogli:2002pt} to incorporate these systematic uncertainties in our simulation. The pulls due to the systematic uncertainties are denoted by $\xi_l$ in Eqs.~\ref{eq:chisq_mu-} and \ref{eq:N_theory}.

When we do not incorporate the hadron energy information ${E'}_\text{had}^\text{rec}$ in our analysis and use only $E_\mu^\text{rec}$, and $\cos\theta_\mu^\text{rec}$ (the so-called ``2D'' analysis as considered in Ref.~\cite{Ghosh:2012px}), the Poissonian $\chi^2_-$ for $\mu^-$ events boils down to

\begin{equation}\label{eq:chisq_mu-2d}
\chi^2_- (2\text{D}) = \mathop{\text{min}}_{\xi_l}  \sum_{j=1}^{N_{E_{\mu}^\text{rec}}} \sum_{k=1}^{N_{\cos\theta_\mu^\text{rec}}} \left[2(N_{ijk}^\text{theory} - N_{jk}^\text{data}) -2 N_{jk}^\text{data} \ln\left(\frac{N_{jk}^\text{theory} }{N_{jk}^\text{data}}\right)\right] + \sum_{l = 1}^5 \xi_l^2
\end{equation}

with, 
\begin{equation}\label{eq:N_theory_2d}
N_{jk}^\text{theory} = N_{jk}^0\left(1 + \sum_{l=1}^5 \pi^l_{jk}\xi_l\right).
\end{equation}

In Eq.~\ref{eq:chisq_mu-2d}, $N_{jk}^\text{data}$ and $N_{jk}^\text{theory}$ correspond to the observed and expected number of reconstructed $\mu^-$ events in a given  $(E_\mu^\text{rec}, \cos\theta_\mu^\text{rec})$ bin, respectively. The expected number of events without systematic uncertainties are represented by $N_{jk}^0$ in Eq.~\ref{eq:N_theory_2d}. In case of binning scheme mentioned in Table~\ref{tab:binning_scheme}, $N_{E_{\mu}^\text{rec}} = 13$, and  $N_{\cos\theta_\mu^\text{rec}} = 15$.

Following the same procedure, as described above, the $\chi^2_+$ for reconstructed $\mu^+$ events is determined for both the ``2D'' and ``3D'' analyses. The total $\chi^2$ is estimated by adding the contribution from $\mu^-$ and $\mu^+$:

\begin{equation}\label{eq:chisq_total}
\chi^2_\text{ICAL} = \chi^2_{-} + \chi^2_+.
\end{equation}

In this analysis, we simulate the prospective data using the benchmark values of oscillation parameters given in Table~\ref{tab:osc-param-value}. We use Eq.~\ref{eq:eff_dmsq} to estimate the value of $\Delta m^2_{31}$ from $\Delta m^2_\text{eff}$ where $\Delta m^2_\text{eff}$ has the same magnitude for NO and IO with positive and negative signs, respectively. To incorporate the systematic uncertainties, the $\chi^2_\text{ICAL}$ is minimized with respect to pull parameters $\xi_l$ in the fit. After that the marginalization is performed over atmospheric mixing angle $\sin^2\theta_{23}$ in the range (0.36, 0.66), atmospheric mass square difference $|\Delta m^2_\text{eff}|$ in the range (2.1, 2.6) $\times 10^{-3}~\text{eV}^2$, and over both choices of mass ordering, NO and IO, in the theory. We do not consider any priors for $\sin^2\theta_{23}$ and $|\Delta m^2_\text{eff}|$ in our analysis. While performing fit, we keep the solar oscillation parameters $\sin^2 2\theta_{12}$ and $\Delta m^2_{21}$ fixed at their values as mentioned in Table~\ref{tab:osc-param-value}. Since the reactor mixing angle is already very well measured~\cite{Marrone:2021,NuFIT,Esteban:2020cvm,deSalas:2020pgw}, we use a fixed value of $\sin^2 2\theta_{13} = 0.0875$ both in data and theory. Throughout our analysis, we take $\delta_{\rm CP} = 0$ both in data and theory.

\section{Results}
\label{sec:results}

\subsection{Effective regions in $(E_\mu^\text{rec}, \cos\theta_\mu^\text{rec})$ plane to constrain LIV parameters}
\label{sec:effective_regions}

For simulating the prospective data for statistical analysis, we assume the oscillations with standard interactions only where LIV parameters are taken as zero. The statistical significance of the analysis for constraining the LIV parameters ($a_{\mu\tau}$, $a_{e\mu}$, $a_{e\tau}$) is quantified in the following fashion:

\begin{equation}\label{eq:chisq_LIV}
\Delta \chi^2_\text{ICAL-LIV} = \chi^2_\text{ICAL}~ (\text{SI} + a_{\alpha\beta} ) - \chi^2_\text{ICAL}~ (\text{SI}).
\end{equation}

Here, $\chi^2_\text{ICAL}~ (\text{SI})$ and $\chi^2_\text{ICAL}~ (\text{SI} + a_{\alpha\beta})$ are calculated by fitting the prospective data with only SI case (no LIV) and with SI in the presence of non-zero LIV parameter $a_{\alpha\beta}$, respectively. We consider only one LIV parameter at-a-time where $a_{\alpha\beta}$ can be $a_{\mu\tau}$, $a_{e\mu}$, or $a_{e\tau}$. Since, the statistical fluctuations are suppressed to estimate the median sensitivity, we have  $\chi^2_\text{ICAL}~ (\text{SI})\sim 0$.

As we will demonstrate in the later part of the result section that ICAL being an atmospheric neutrino experiment dominated by $\nu_\mu \rightarrow \nu_\mu$ survival channel places tightest constraint on $a_{\mu\tau}$ among all the LIV parameters. Here, we discuss what regions in ($E_\mu^\text{rec}$, $\cos\theta_\mu^\text{rec}$) plane contribute significantly towards the sensitivity of ICAL for $a_{\mu\tau}$. A similar discussion for other two off-diagonal LIV parameters $a_{e\mu}$ and $a_{e\tau}$ is provided in the appendix~\ref{app:effective_regions_emu_etau}. The sensitivity of ICAL towards the LIV parameter $a_{\mu\tau}$ stems from both $\mu^-$ and $\mu^+$ events for both the choices of mass ordering. While considering $a_{\mu\tau} = 1.0 \times 10^{-23}$ GeV ($-1.0 \times 10^{-23}$ GeV) in theory and $a_{\mu\tau} = 0$ in data, the contribution of $\mu^-$ and $\mu^+$ towards fixed-parameter\footnote{Please note that in the fixed-parameter scenario, we minimize only over the systematic uncertainties, but we keep the oscillation parameters fixed in both theory and data at their benchmark values as mentioned in Table~\ref{tab:osc-param-value}.} $\Delta\chi^2_\text{ICAL-LIV}$ for 500 kt$\cdot$yr exposure at ICAL for NO is 43.5 (43.9) and 25.0 (23.8), respectively.

\begin{figure}[t]
  \centering
  \includegraphics[width=0.45\textwidth]{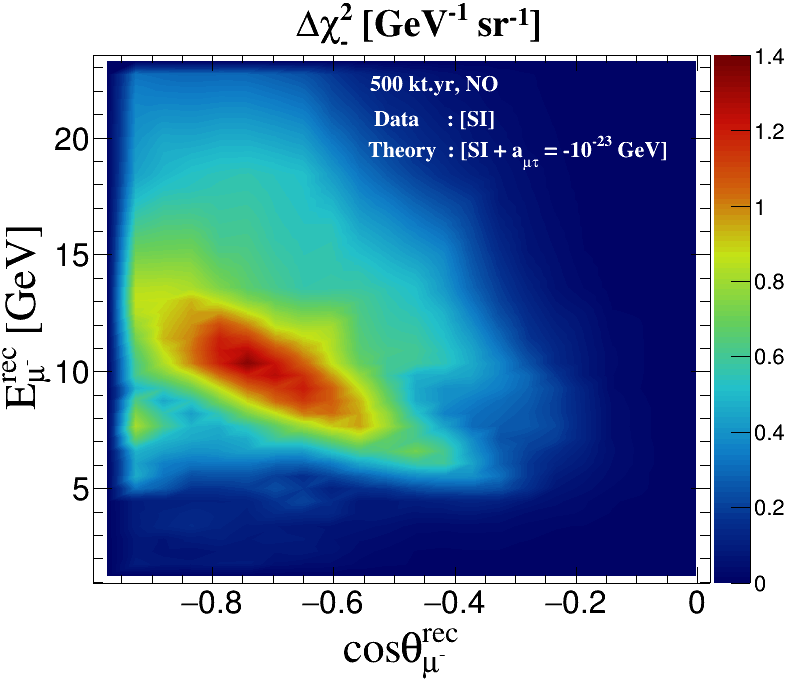}
  \includegraphics[width=0.45\textwidth]{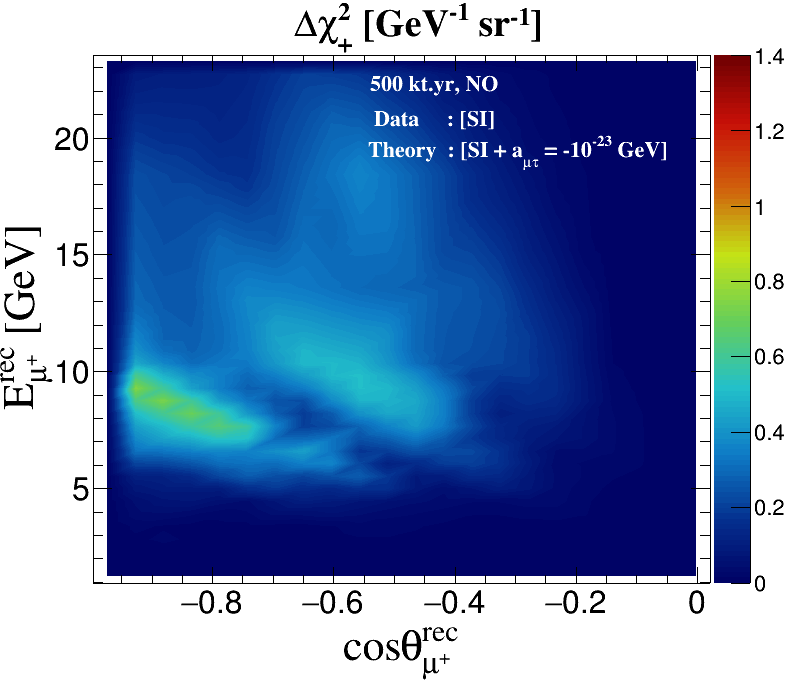}
  \includegraphics[width=0.45\textwidth]{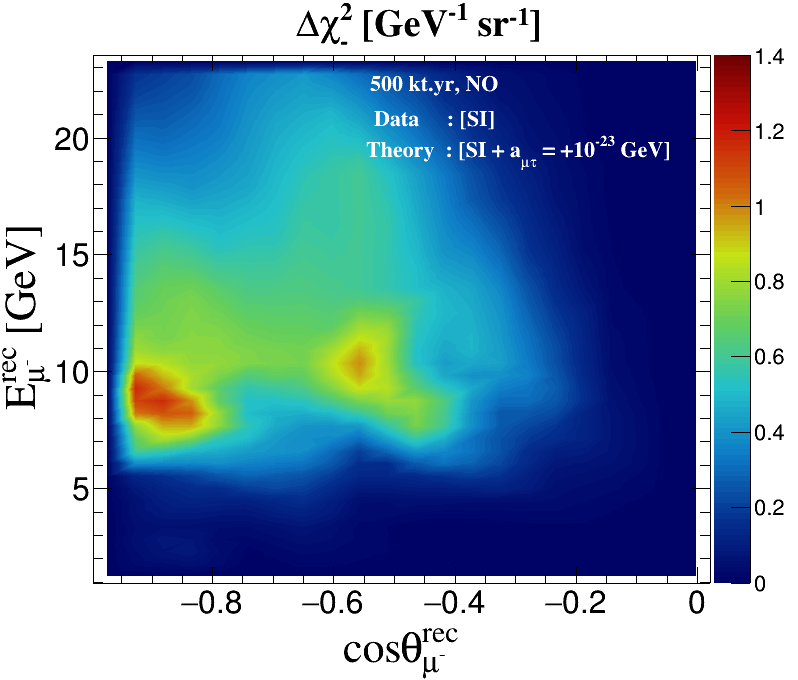}
  \includegraphics[width=0.45\textwidth]{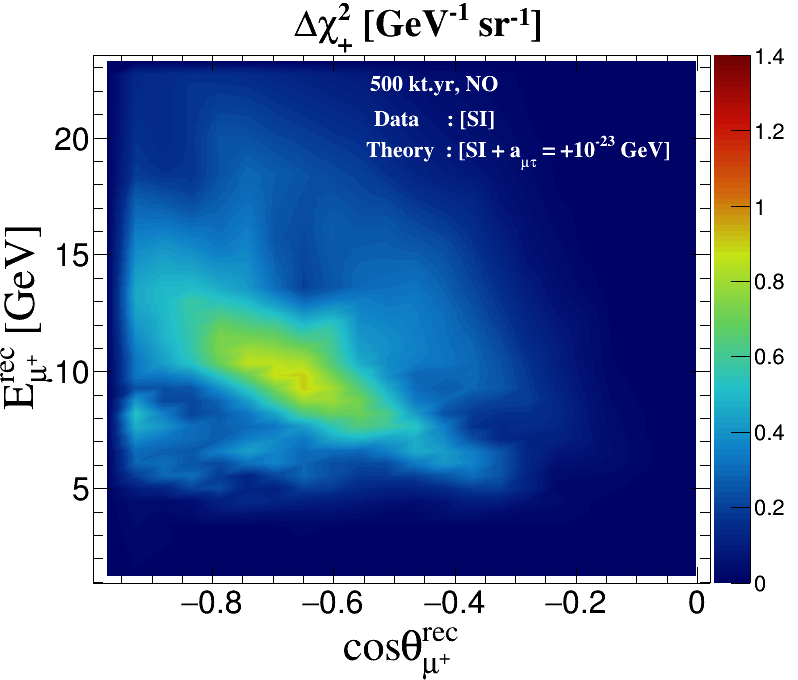}
  \mycaption{The distribution of fixed-parameter $\Delta\chi^2_-$ ($\Delta\chi^2_+$) in the plane of ($E_\mu^\text{rec}$, $\cos\theta_\mu^\text{rec}$) without pull penalty term  using 500 kt$\cdot$yr exposure of the ICAL detector as shown in the left (right) panels. Note that $\Delta\chi^2_-$ and $\Delta\chi^2_+$ are presented in the units of $\text{GeV}^{-1} \text{sr}^{-1}$ where we have divided them by 2$\pi ~ \times$ bin area. In data, $a_{\mu\tau} = 0$ with NO (true) using the benchmark value of oscillation parameters given in Table~\ref{tab:osc-param-value}. In theory, $a_{\mu\tau} = -1.0 \times 10^{-23} ~ \text{GeV}$ and $1.0 \times 10^{-23} ~ \text{GeV}$ in the top and bottom panels, respectively.
  }
  \label{fig:chisq_contour}
\end{figure}

Before we present our final results, let us first identify the effective regions in $(E_\mu^\text{rec}$, $\cos\theta_\mu^\text{rec})$ plane which contribute significantly towards $\Delta\chi^2_\text{ICAL-LIV}$. In Fig.~\ref{fig:chisq_contour}, we show the distribution of fixed-parameter $\Delta \chi^2_-$ ( $\Delta \chi^2_+$) without pull penalty term\footnote{We do not include pull penalty term  $\sum_{l = 1}^5 \xi_l^2$ (see Eq.~\ref{eq:chisq_mu-}) while calculating $\Delta \chi^2_{-}$ and $\Delta \chi^2_+$ to explore contributions from each bin in the plane of $E_\mu^\text{rec}$ and $\cos\theta_\mu^\text{rec}$ for $\mu^-$ and $\mu^+$ events, respectively.} from reconstructed $\mu^-$ ($\mu^+$) events in  $(E_\mu^\text{rec}$, $\cos\theta_\mu^\text{rec})$  plane for 500 kt$\cdot$yr exposure at ICAL assuming NO at true mass ordering. For demonstration purpose, we have added the $\Delta \chi^2$ contribution from all ${E'}_\text{had}^\text{rec}$ bins for each ($E_\mu^\text{rec}$, $\cos\theta_\mu^\text{rec}$) bin while using binning scheme mentioned in Table~\ref{tab:binning_scheme}. In the top (bottom) panels in Fig.~\ref{fig:chisq_contour}, we take non-zero LIV parameter $a_{\mu\tau} = 1.0 \times 10^{-23}$ GeV ($-1.0 \times 10^{-23}$ GeV) in theory while considering $a_{\mu\tau} = 0$ in the prospective data. The left and the right panels show the distribution of $\Delta \chi^2_-$ and $\Delta \chi^2_+$, respectively. In all the panels, we observe that a significant contribution is received from bins with $7 \text{ GeV} < E_\mu^\text{rec} < 17 \text{ GeV}$ and $\cos\theta_\mu^\text{rec} < -0.4$. These are the regions around the oscillation valley as observed in Fig.~\ref{fig:oscillograms_mutau}.

\subsection{Advantage of hadron energy information in constraining LIV parameters}
\label{sec:3D_impact}

\begin{figure}[t]
  \centering
  \includegraphics[width=0.32\textwidth]{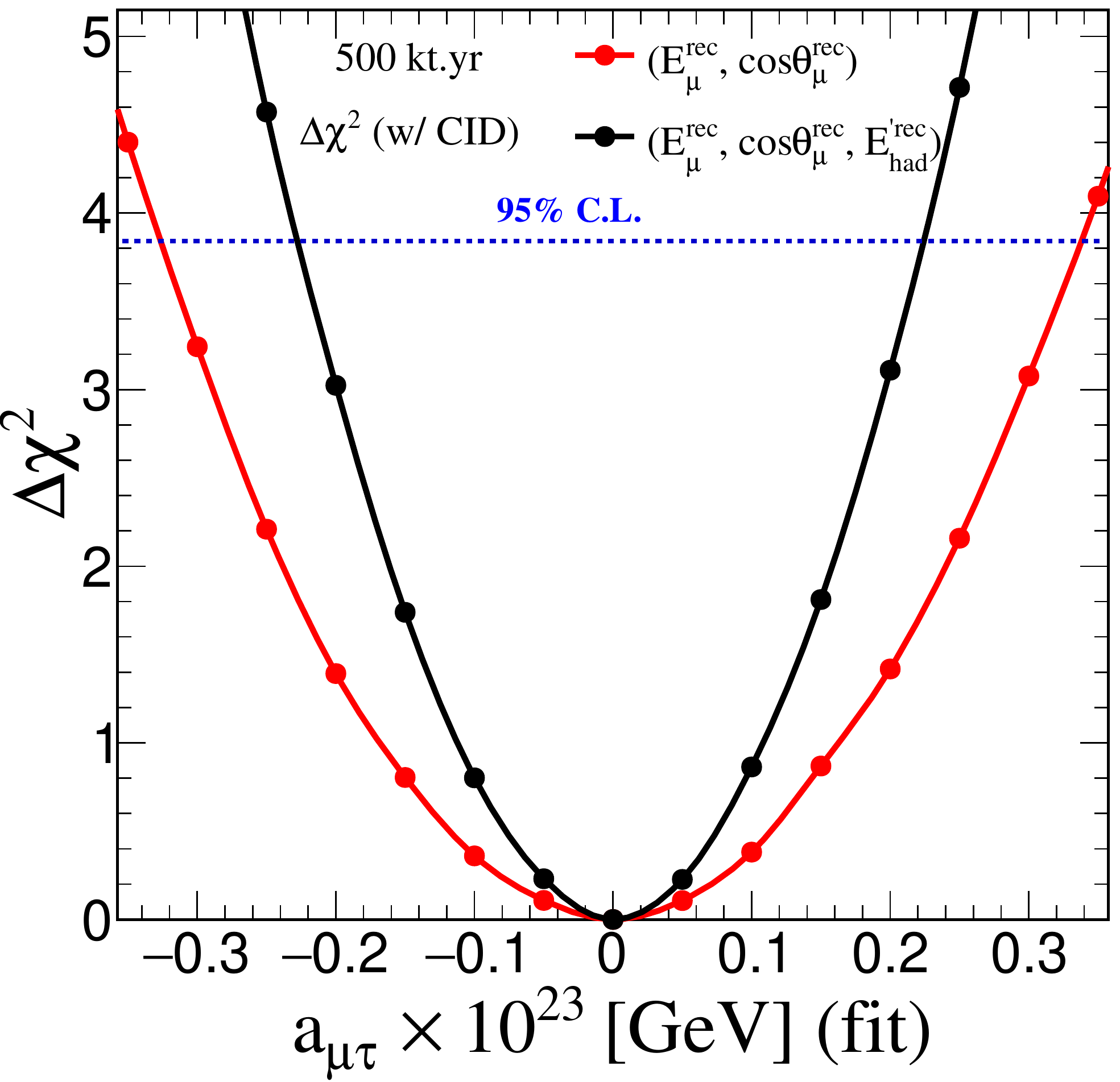} 
  \includegraphics[width=0.32\textwidth]{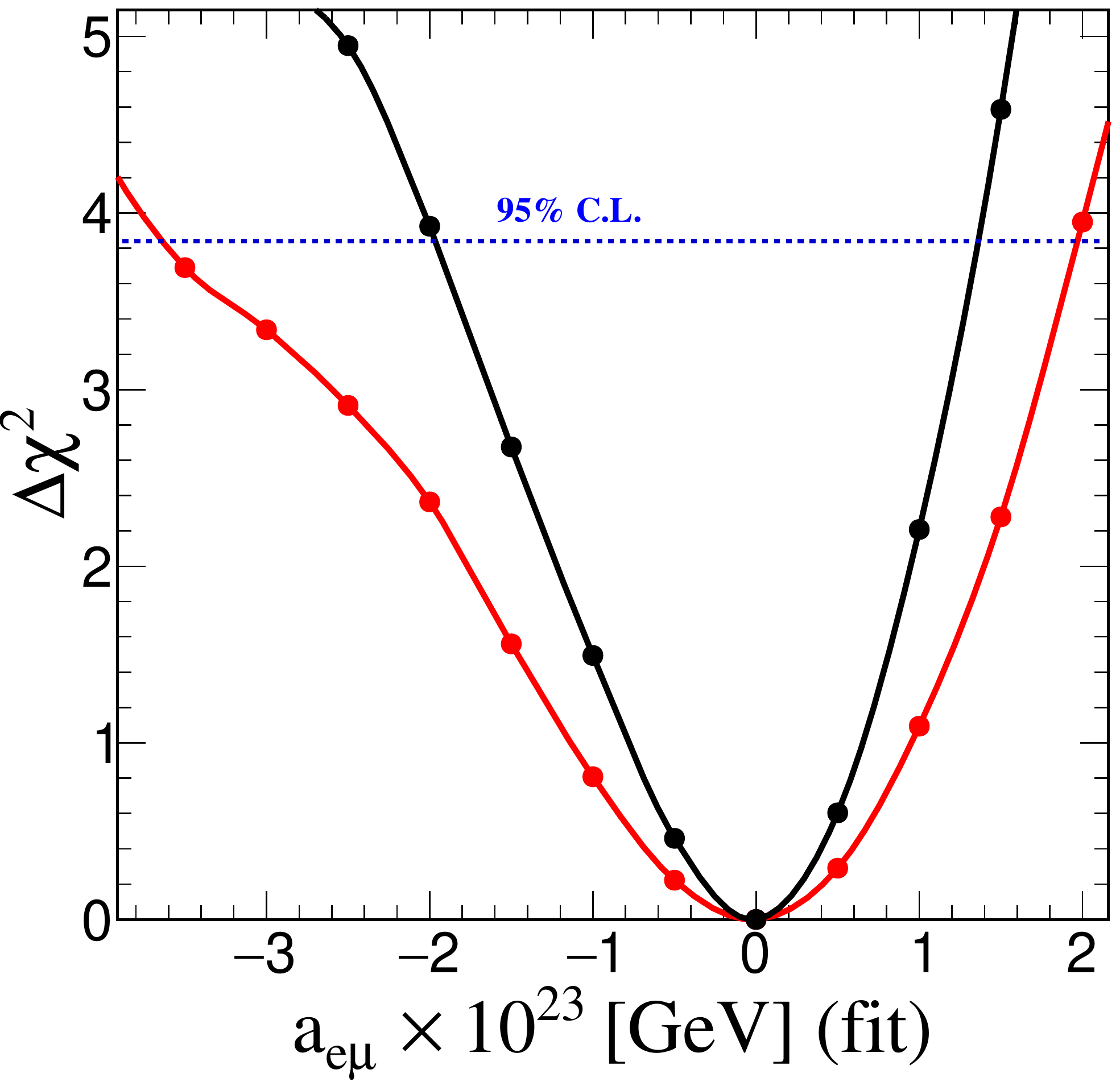}  
  \includegraphics[width=0.32\textwidth]{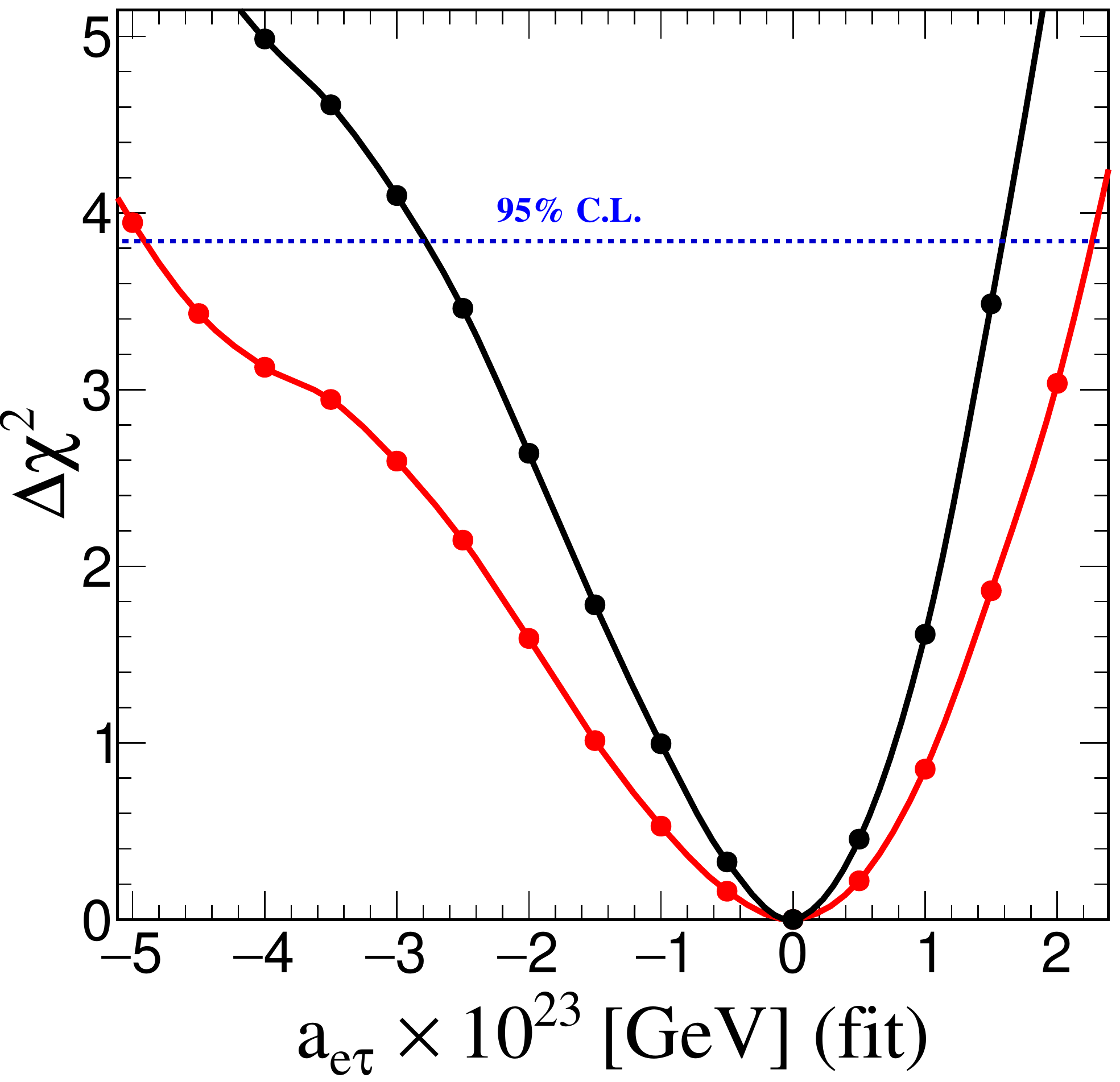}
  \mycaption{The sensitivities to constrain the LIV parameters $a_{\mu\tau}$, $a_{e\mu}$, and $a_{e\tau}$ using 500 kt$\cdot$yr exposure at the ICAL detector as shown in the left, middle, and right panels, respectively. In each panel, the red lines represent the 2D analysis using reconstructed observables ($E_\mu^\text{rec}$, $\cos\theta_\mu^\text{rec}$) whereas the black lines refers to the 3D analysis using reconstructed observables ($E_\mu^\text{rec}$, $\cos\theta_\mu^\text{rec}$, ${E'}_\text{had}^\text{rec}$). We have used the benchmark value of oscillation parameters given in Table~\ref{tab:osc-param-value}. In theory, we have marginalized over oscillation parameters $\sin^2\theta_{23}$, $|\Delta m^2_\text{eff}|$, and both choices of mass orderings.} 
  \label{fig:chisq-LIVhad}
\end{figure}

In Fig.~\ref{fig:chisq-LIVhad}, we constrain the LIV parameters $a_{\mu\tau}$, $a_{e\mu}$, and $a_{e\tau}$ using 500 kt$\cdot$yr exposure at the ICAL detector and describe the advantage of incorporating the hadron energy information. In the prospective data, we assume the case of SI where the values of all LIV parameters are considered to be zero, whereas in theory, we consider SI + LIV case where the value of the LIV parameters $a_{\mu\tau}$, $a_{e\mu}$, and $a_{e\tau}$ are varied one-at-a-time. In the left, middle and right panels, we constrain the LIV parameters $a_{\mu\tau}$, $a_{e\mu}$, and $a_{e\tau}$, respectively. In all the panel, the red curves represent the 2D case where we use only $E_\mu^\text{rec}$ and $\cos\theta_\mu^\text{rec}$ variables without considering any hadron energy information. For reconstructed variables of $E_\mu^\text{rec}$ and $\cos\theta_\mu^\text{rec}$, we use binning scheme mentioned in Table~\ref{tab:binning_scheme}, where the events are integrated over ${E'}_\text{had}^\text{rec}$ bins in the hadron energy range of 0 to 25 GeV. On the other hand, the black curves represent the 3D case considering reconstructed variables of $E_\mu^\text{rec}$, $\cos\theta_\mu^\text{rec}$, and  ${E'}_\text{had}^\text{rec}$ where we use the hadron energy information available at the ICAL detector. For 3D case, we used the binning scheme given in Table~\ref{tab:binning_scheme} where ${E'}_\text{had}^\text{rec}$ is divided into 4 bins in the hadron energy range of 0 to 25 GeV. 

We can observe in Fig.~\ref{fig:chisq-LIVhad} that the incorporation of hadron energy information results in the improved constraints for all three cases of LIV parameters $a_{\mu\tau}$, $a_{e\mu}$, and $a_{e\tau}$. The ICAL detector places the tightest constraint on LIV parameter $a_{\mu\tau}$ among all the three off-diagonal LIV parameters. For the case of $a_{e\mu}$ and $a_{e\tau}$, we observe that the bounds are asymmetric with stronger constraints for the positive values than that for the negative values. These observations are consistent with the effect of $a_{e\mu}$ and $a_{e\tau}$ on $\nu_\mu$ survival probability oscillograms in Figs.~\ref{fig:oscillograms_emu} and \ref{fig:oscillograms_etau}, respectively. In section~\ref{sec:oscillograms}, we use Eq.~\ref{eq:Pmue-mat-LIV} to explain the reason behind stronger effects of $a_{e\mu}$ and $a_{e\tau}$ for positive values. In the next section, we explore the impact of CID capability of the ICAL detector in constraining the LIV parameters.

\subsection{Advantage of charge identification capability to constrain LIV parameters}
\label{sec:cid_impact}

\begin{figure}[t]
  \centering
  \includegraphics[width=0.32\textwidth]{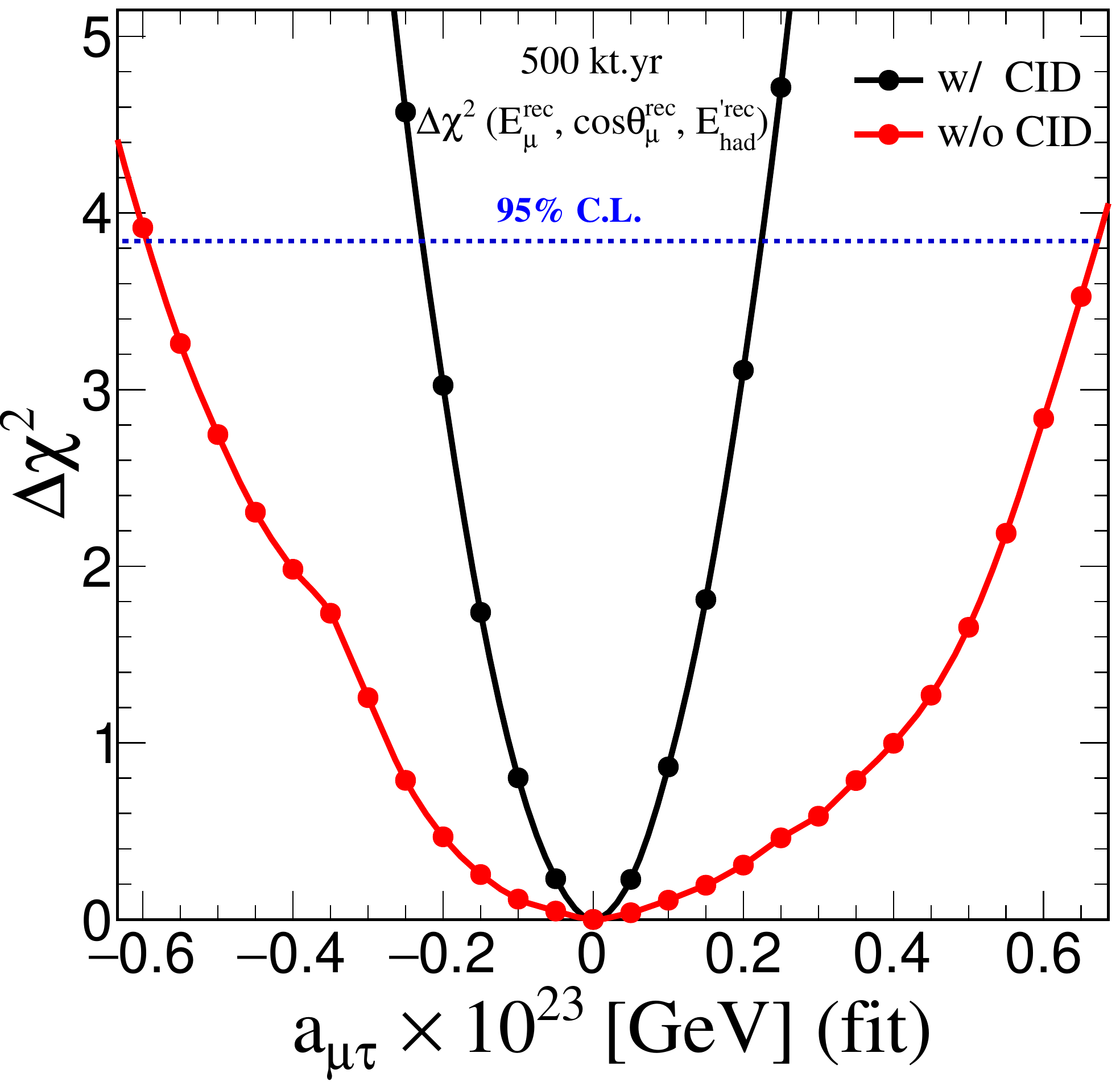} 
  \includegraphics[width=0.32\textwidth]{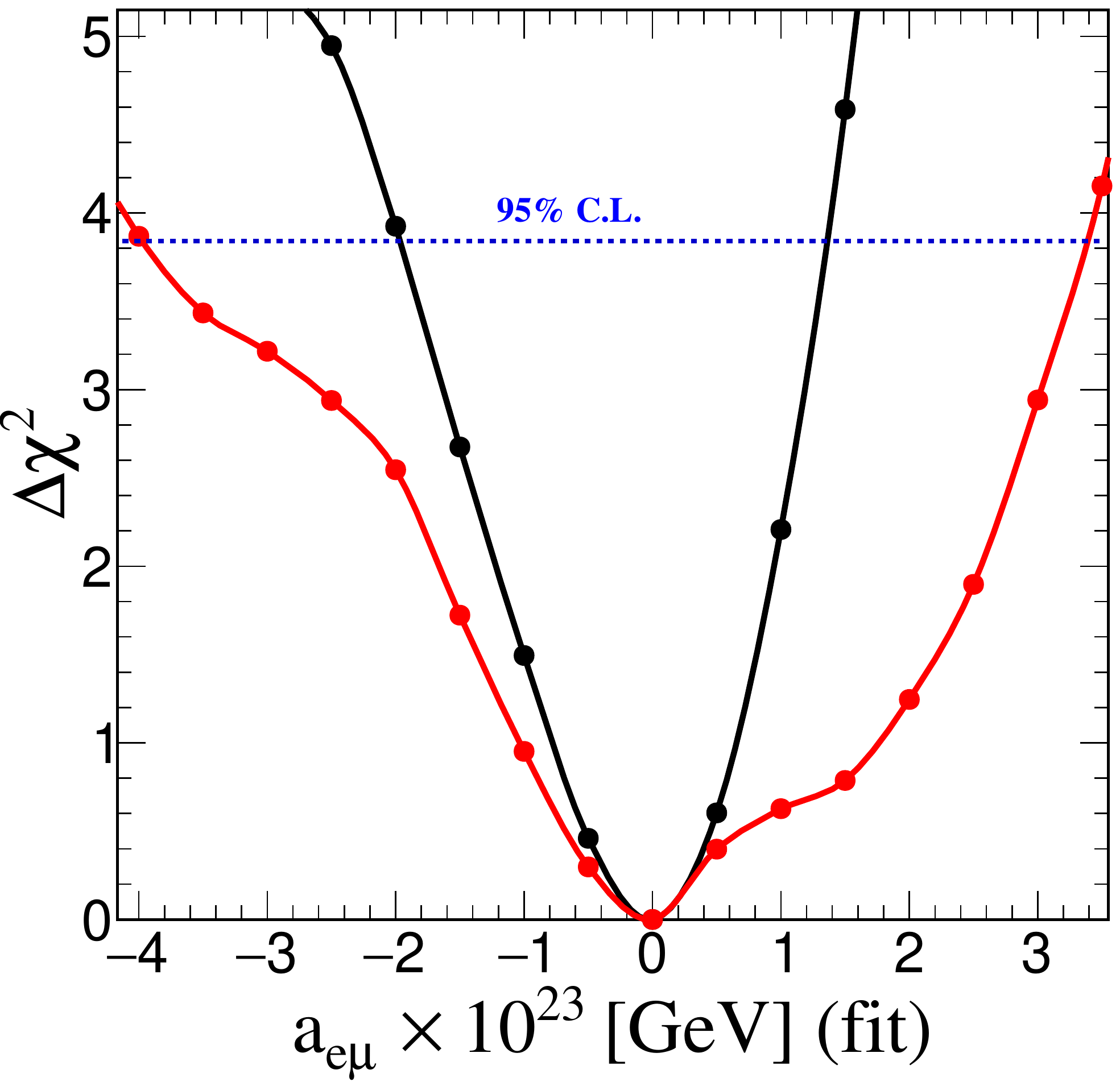}
  \includegraphics[width=0.32\textwidth]{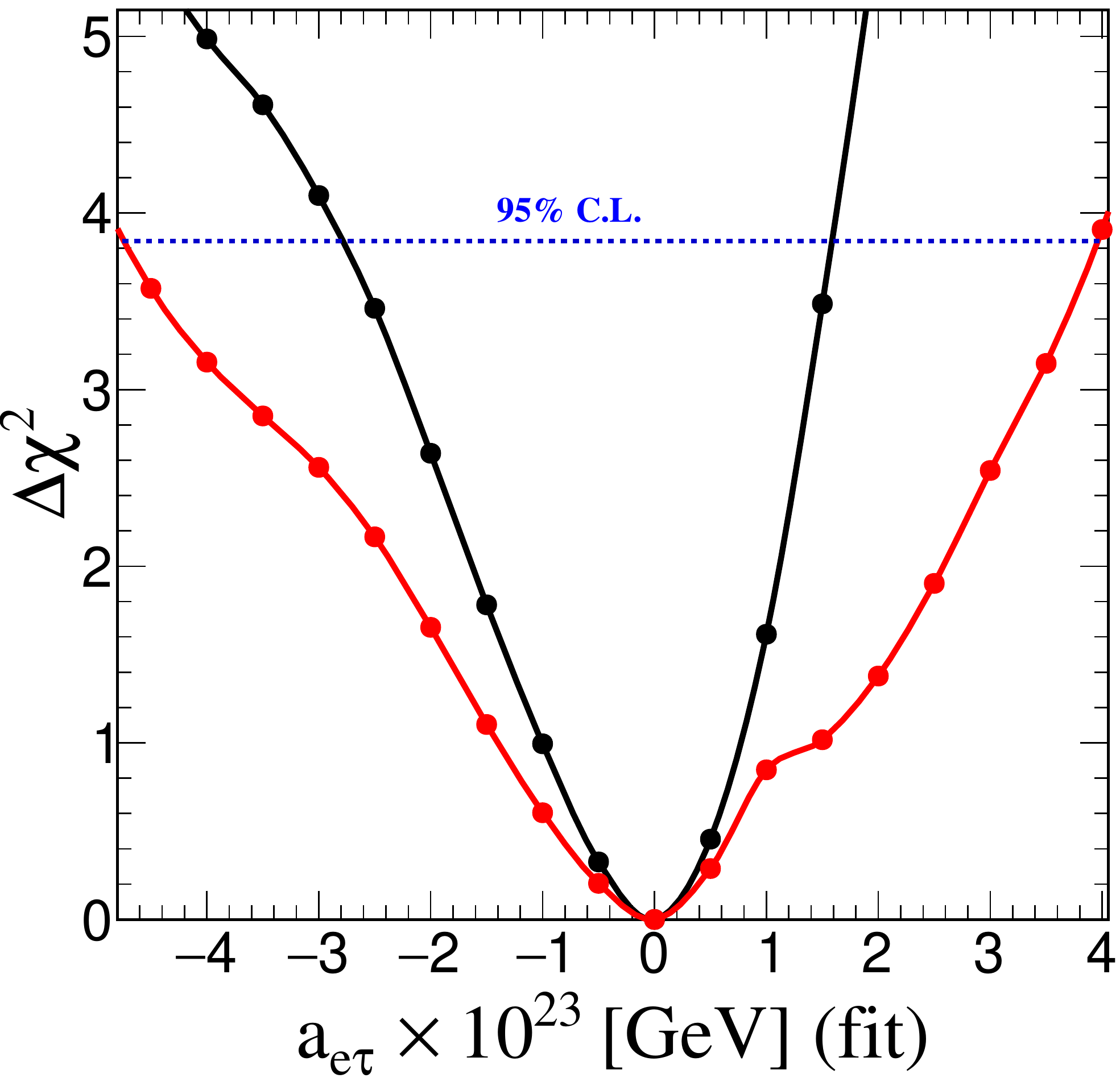} 
  \mycaption{The sensitivities to constrain the LIV parameters $a_{\mu\tau}$, $a_{e\mu}$, and $a_{e\tau}$ using 500 kt$\cdot$yr exposure at the ICAL detector as shown in the left, middle, and right panels, respectively. In each panel, the black lines represent the case when the charge identification capability of the ICAL detector is used while estimating the sensitivities, whereas the red lines refer to the case when the charge identification capability of ICAL is absent. We have used the benchmark value of oscillation parameters given in Table~\ref{tab:osc-param-value}. In theory, we have marginalized over oscillation parameters $\sin^2\theta_{23}$, $|\Delta m^2_\text{eff}|$, and both choices of mass orderings.
  } 
  \label{fig:chisq-LIV3Dwo}
\end{figure}

The presence of a 1.5 T magnetic field enables the ICAL detector to distinguish $\mu^-$ and $\mu^+$ events, which leads to the capability of the ICAL detector to separately identify their parent particles neutrinos and antineutrinos, respectively. In Fig.~\ref{fig:chisq-LIV3Dwo}, we discuss the advantage of charge identification capability of the ICAL detector while constraining the CPT-violating LIV parameters $a_{\mu\tau}$, $a_{e\mu}$, and $a_{e\tau}$ using 500 kt$\cdot$yr exposure. In each panel of Fig.~\ref{fig:chisq-LIV3Dwo}, black curves show the sensitivity of the ICAL detector where CID capability is exploited, and two separate sets of bins for reconstructed $\mu^-$ and $\mu^+$ events are used. On the other hand, the red curves represent the case where the CID capability of the ICAL detector is not utilized, and a single set of combined bins for both reconstructed $\mu^-$ and $\mu^+$ events is considered.

We can observe in Fig.~\ref{fig:chisq-LIV3Dwo} that the incorporation of CID capability of the ICAL detector significantly improves the sensitivity to constrain the LIV parameters $a_{\mu\tau}$, $a_{e\mu}$, and $a_{e\tau}$. In section~\ref{sec:oscillograms}, we discussed that for a given value of $a_{\mu\tau}$, the oscillation valley bends in the opposite directions for neutrinos and antineutrinos. The combined binning of both reconstructed $\mu^-$ and $\mu^+$ events in a single set of bins will lead to a dilution of these features, which are caused due to the presence of non-zero LIV parameter. Thus, it is important to separately identify reconstructed $\mu^-$ and $\mu^+$ events to preserve the information about LIV. All these observations validate the fact that the charge identification capability of the ICAL detector will play a crucial role in constraining LIV parameters.

In the subsections~\ref{sec:3D_impact} and \ref{sec:cid_impact}, we discussed the advantage of incorporating hadron energy information and presence of CID while constraining CPT-violating LIV parameters $a_{\mu\tau}$, $a_{e\mu}$, and $a_{e\tau}$.  Now, Table~\ref{tab:results_bounds} nicely summarizes the findings from these two studies in tabular form, and at the same time, we compare the performance of ICAL with other experiments. In Table~\ref{tab:results_bounds}, we present the constraints on LIV parameters $a_{\mu\tau}$, $a_{e\mu}$, and $a_{e\tau}$ at 95\% C.L. using 500 kt$\cdot$yr exposure at the ICAL detector. By comparing the results shown in the second row with respect to that in the first row, we can infer that the incorporation of hadron energy information improves the bounds on all of these LIV parameters. The improvement due to the presence of CID capability can be observed by comparing the second row with respect to the third row. While doing comparison with existing constraints, we have mentioned the results from Super-K analysis~\cite{Super-Kamiokande:2014exs} for LIV parameters $a_{\mu\tau}$, $a_{e\mu}$, and $a_{e\tau}$ at 95\% C.L. with real and imaginary parts separately. The IceCube collaboration~\cite{IceCube:2017qyp} has performed analysis only for $a_{\mu\tau}$. We show constraints for the real and imaginary parts of $a_{\mu\tau}$ separately at 99\% C.L. as provided by the IceCube collaboration.

\begin{table}[t]
  \centering
  \begin{center}
    \begin{tabular}{|c| c| c| c|}
      \hline \hline
      \multicolumn{4}{|c|}{Constraints on CPT-violating LIV parameters at 95\% C.L. using ICAL} \\ \hline \hline
      Observables & $a_{\mu\tau} [\rm 10^{-23} ~ GeV]$ &  $a_{e\mu} [\rm 10^{-23} ~ GeV]$ & $a_{e\tau} [\rm 10^{-23} ~GeV]$ \\ \hline \hline
      ($E_{\mu}^\text{rec}$, $\cos\theta_\mu^\text{rec}$) w/ CID  & [$-0.33, 0.34 $]& [$-3.61, 1.97$] & [$-4.90, 2.25$] \\ 
      ($E_{\mu}^\text{rec}$, $\cos\theta_\mu^\text{rec}$, ${E'}_{\rm had}^\text{rec}$) w/ CID & [$-0.23, 0.22$] & [$-1.97, 1.34$] & [$-2.80, 1.58$] \\ 
      ($E_{\mu}^\text{rec}$, $\cos\theta_\mu^\text{rec}$, ${E'}_{\rm had}^\text{rec}$) w/o CID & [$-0.59, 0.67$] & [$-3.97, 3.37 $] & [$-4.71, 3.96 $] \\
      \hline \hline
      \multicolumn{4}{|c|}{} \\ \hline \hline
      \multicolumn{4}{|c|}{Existing constraints on CPT-violating LIV parameters} \\ \hline \hline
      Experiments & $a_{\mu\tau} [\rm 10^{-23} ~ GeV]$ &  $a_{e\mu} [\rm 10^{-23} ~ GeV]$ & $a_{e\tau} [\rm 10^{-23} ~GeV]$ \\ \hline \hline
      \multirow{2}{*}{Super-K\tablefootnote{In Super-K analysis~\cite{Super-Kamiokande:2014exs}, the authors have scanned the parameters on a logarithmic scale where the only positive values of parameters are considered assuming $10^{-28}$ GeV to be the minimum value which is equivalent to the case of no LIV.} (95\% C.L.)~\cite{Super-Kamiokande:2014exs}} & $\mathrm{Re}(a_{\mu\tau}) < 0.65$  & $\mathrm{Re}(a_{e\mu}) < 1.8$ & $\mathrm{Re}(a_{e\tau}) < 4.1$ \\
      & $\mathrm{Im}(a_{\mu\tau}) < 0.51$  & $\mathrm{Im}(a_{e\mu}) < 1.8$ & $\mathrm{Im}(a_{e\tau}) < 2.8$ \\ \hline
      \multirow{2}{*}{IceCube\tablefootnote{To compare with our results, we mention the dimension-three results for $a_{\mu\tau}$ as given by IceCube~\cite{IceCube:2017qyp}.} (99\% C.L.)~\cite{IceCube:2017qyp}} &$|\mathrm{Re}(a_{\mu\tau})| < 0.29$ & \multirow{2}{*}{--} & \multirow{2}{*}{--} \\ 
      &$|\mathrm{Im}(a_{\mu\tau})| < 0.29$ & & \\ \hline \hline
      
    \end{tabular}
  \end{center}
  \mycaption{Constraints on the CPT-violating LIV parameters $a_{\mu\tau}$, $a_{e\mu}$, and $a_{e\tau}$ at 95\% C.L. using 500 kt$\cdot$yr exposure at the ICAL detector. The second row shows results using 2D variables ($E_{\mu}^\text{rec}$, $\cos\theta_\mu^\text{rec}$) with CID, whereas the third row present the results for the case of 3D variables ($E_{\mu}^\text{rec}$, $\cos\theta_\mu^\text{rec}$, ${E'}_{\rm had}^\text{rec}$) with CID. The results in the third row use 3D variables but do not use the CID capability of the ICAL detector. We have used the benchmark value of oscillation parameters given in Table~\ref{tab:osc-param-value}. Note that we have marginalized over oscillation parameters $\sin^2\theta_{23}$, $|\Delta m^2_\text{eff}|$, and both choices of mass orderings in theory. For comparison, the last two rows mention the existing constraints on LIV parameters obtained from Super-K and IceCube experiments. All these results consider only the time component assuming isotropic nature.}
  \label{tab:results_bounds}
\end{table}

\subsection{Impact of marginalization on constraining LIV parameters}
\label{sec:margin_impact}

\begin{figure}[t]
  \centering
  \includegraphics[width=0.32\textwidth]{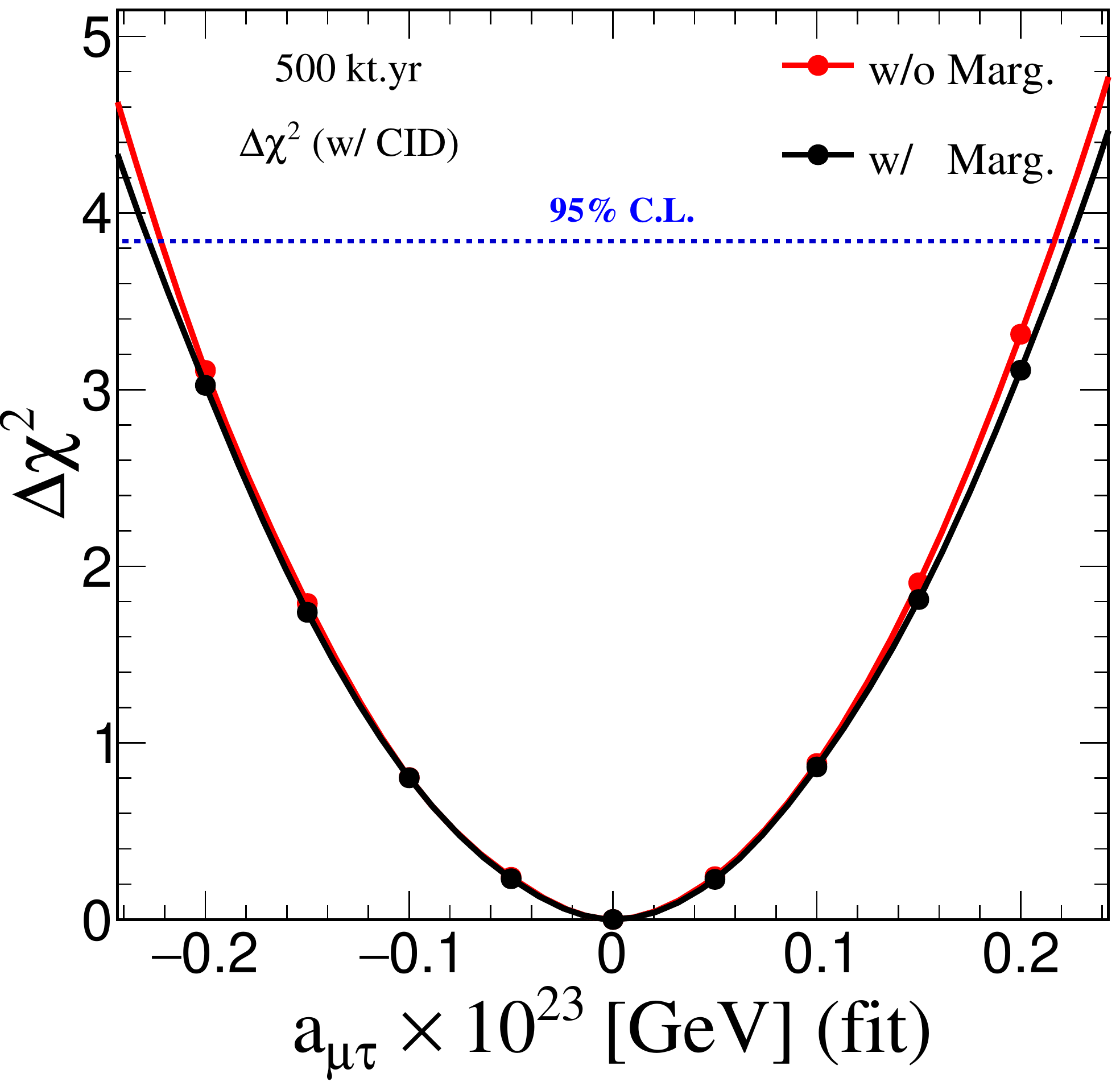} 
  \includegraphics[width=0.32\textwidth]{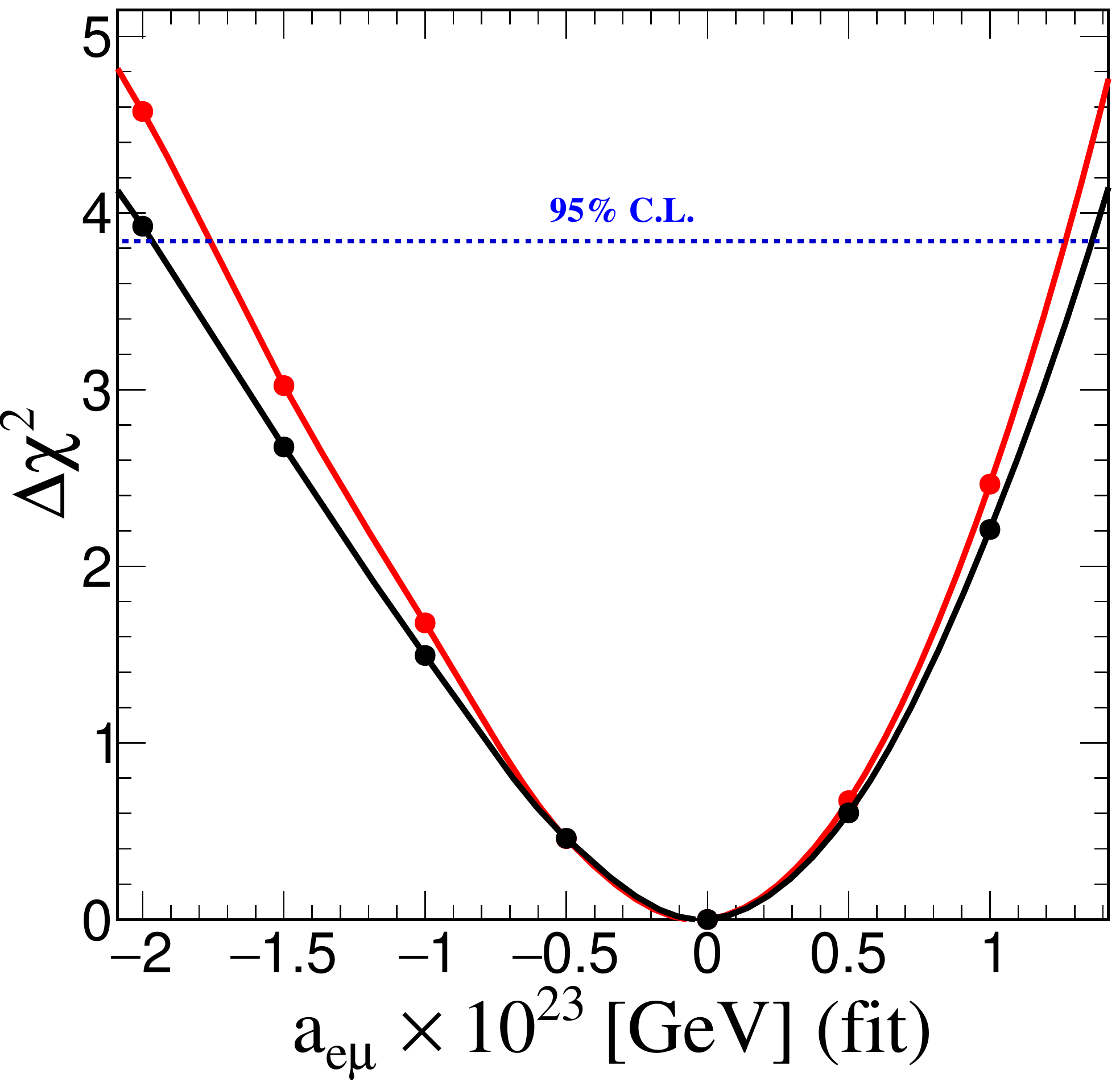}
  \includegraphics[width=0.32\textwidth]{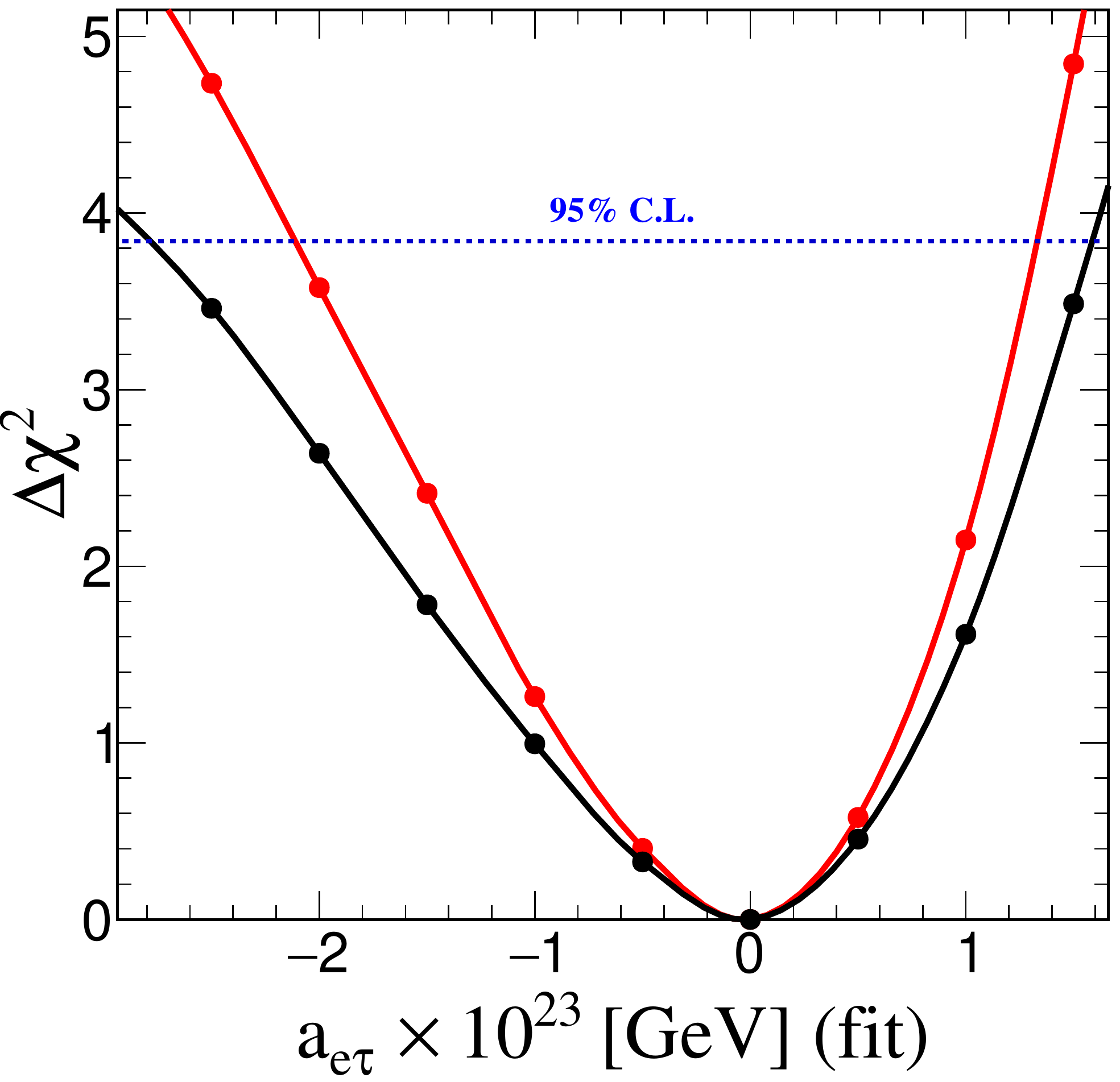}
  \mycaption{The sensitivities to constrain the LIV parameters $a_{\mu\tau}$, $a_{e\mu}$, and $a_{e\tau}$ using 500 kt$\cdot$yr exposure at the ICAL detector as shown in the left, middle, and right panels, respectively. In each panel, the black lines represent the case where we have marginalized over oscillation parameters $\sin^2\theta_{23}$, $|\Delta m^2_\text{eff}|$, and both choices of mass orderings. On the other hand, the red curves in each panel portray the case where the marginalization over oscillation parameter is not performed in theory. We have used the benchmark value of oscillation parameters given in Table~\ref{tab:osc-param-value}.} 
  \label{fig:chisq-LIVmargin}
\end{figure}

In the previous sections, we have performed complete marginalization over the uncertain neutrino oscillation parameters in their 3$\sigma$ ranges given by the present global fit study, but we expect in the coming decades, the precisions of neutrino oscillation parameters are going to improve because of the currently running and upcoming experiments. To understand the impact of uncertainties in oscillation parameters on our results, in this section, we marginalize over oscillation parameters while constraining the CPT-violating LIV parameters $a_{\mu\tau}$, $a_{e\mu}$, and $a_{e\tau}$ using the ICAL detector with 500 kt$\cdot$yr exposure and compared this with the fixed-parameter scenario as shown in Fig.~\ref{fig:chisq-LIVmargin}. The red curves in Fig.~\ref{fig:chisq-LIVmargin} represent the fixed-parameter case where the oscillation parameters are kept fixed in theory. Please note that in the fixed-parameter scenario, we do minimize the systematic errors using the pull method. The black curves show the case where we marginalize over oscillation parameters $\sin^2\theta_{23}$, $|\Delta m^2_\text{eff}|$, and both choices of mass orderings in theory as explained in section~\ref{sec:statistical_analysis}.

We can observe in the left panel in Fig.~\ref{fig:chisq-LIVmargin} that the marginalization over oscillation parameters do not affect the constraints by ICAL on LIV parameter $a_{\mu\tau}$ by a large amount. The constraints on LIV parameters $a_{e\mu}$ and $a_{e\tau}$ show some deterioration after marginalization over oscillation parameters. The largest impact of uncertainties of oscillation parameters can be observed for the case of $a_{e\tau}$ as shown in the right panel of Fig.~\ref{fig:chisq-LIVmargin}. Thus, we can infer that the more precise determination of the oscillation parameters in the future will improve the constraints on LIV parameters $a_{e\mu}$ and $a_{e\tau}$ using 500 kt$\cdot$yr exposure at ICAL.

\subsection{Impact of true values of $\sin^2\theta_{23}$ on constraining LIV parameters}
\label{sec:th23_impact}

\begin{figure}[t]
  \centering
  \includegraphics[width=0.32\linewidth]{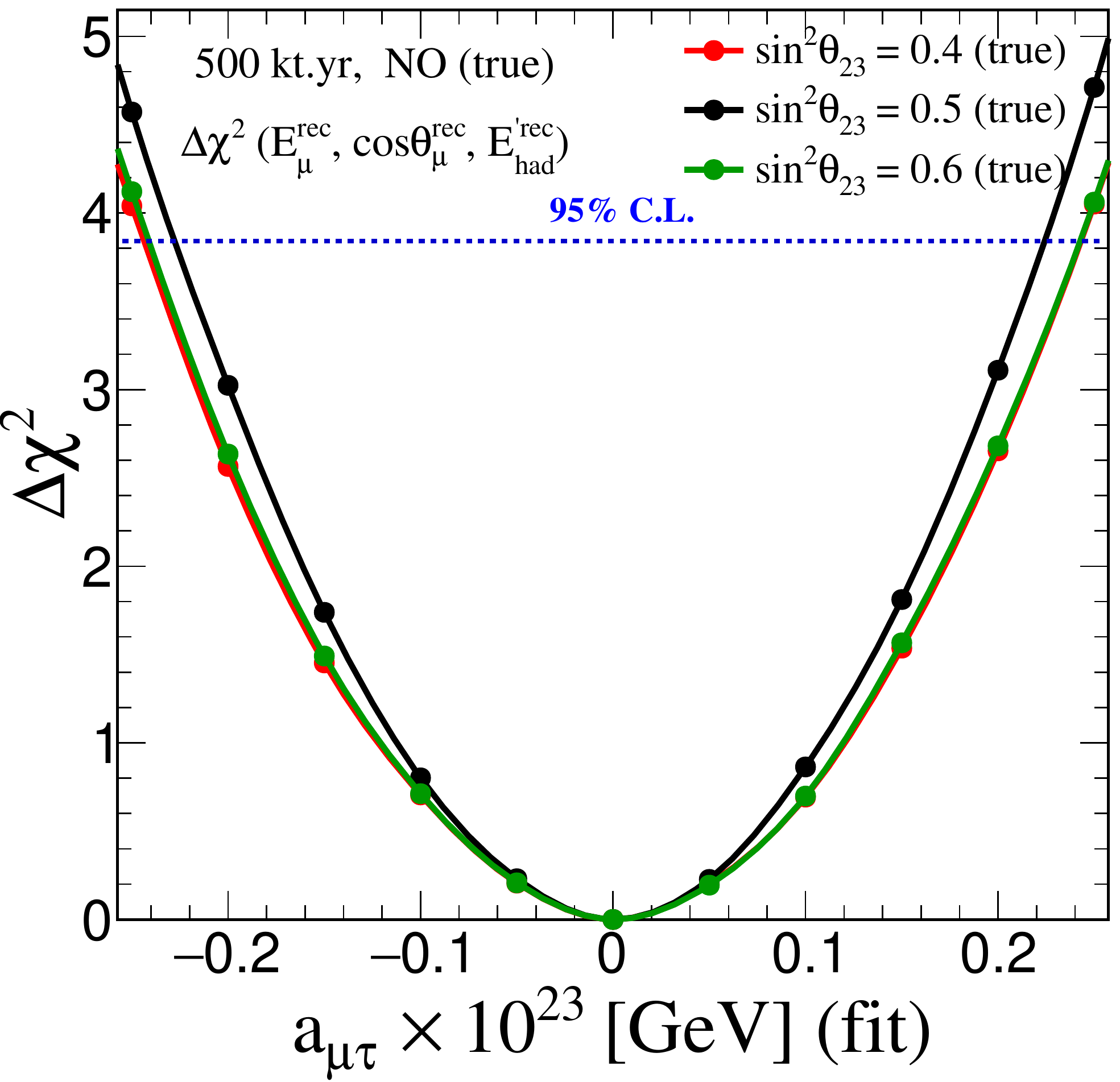}
  \includegraphics[width=0.32\linewidth]{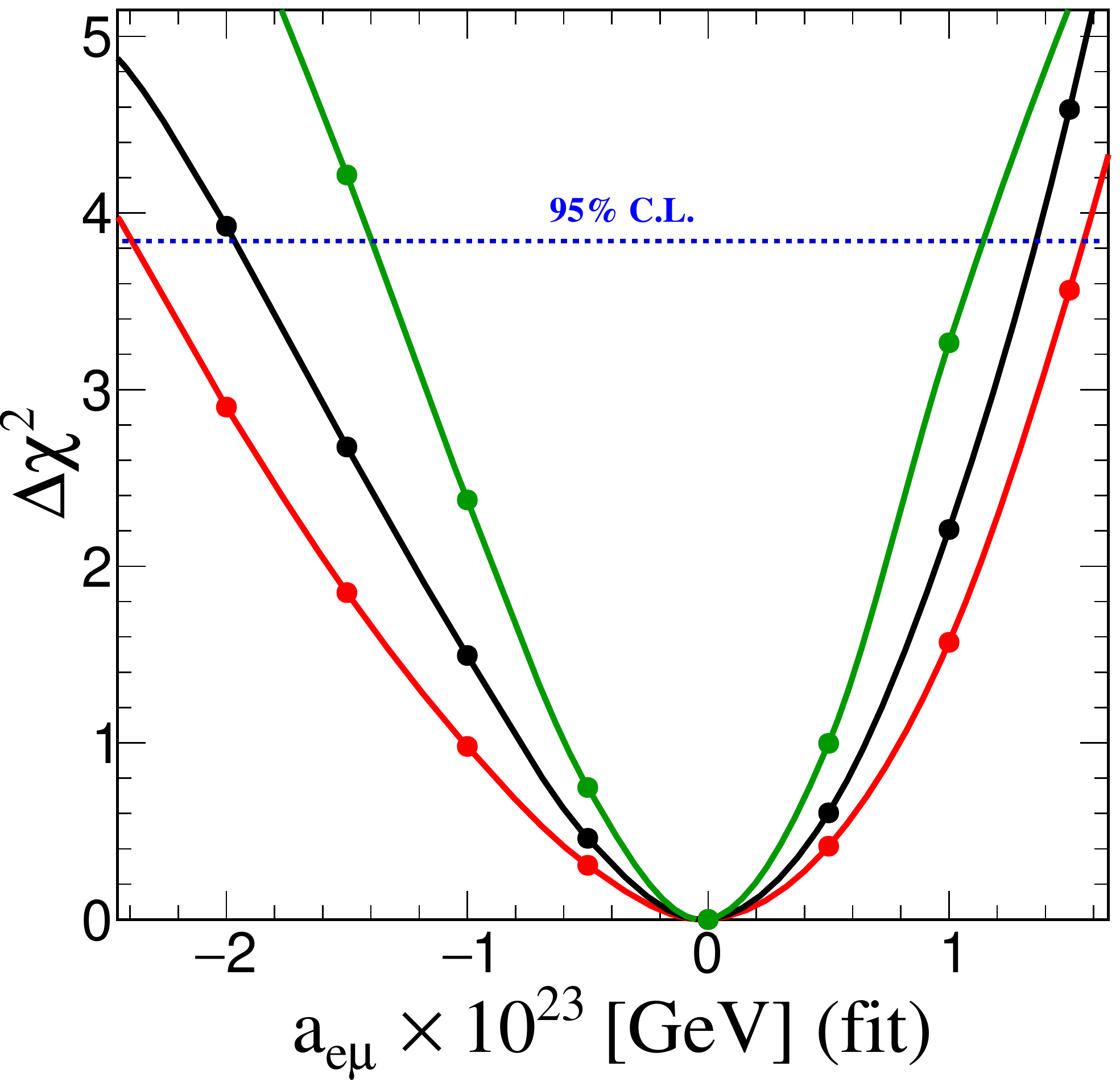}
  \includegraphics[width=0.32\linewidth]{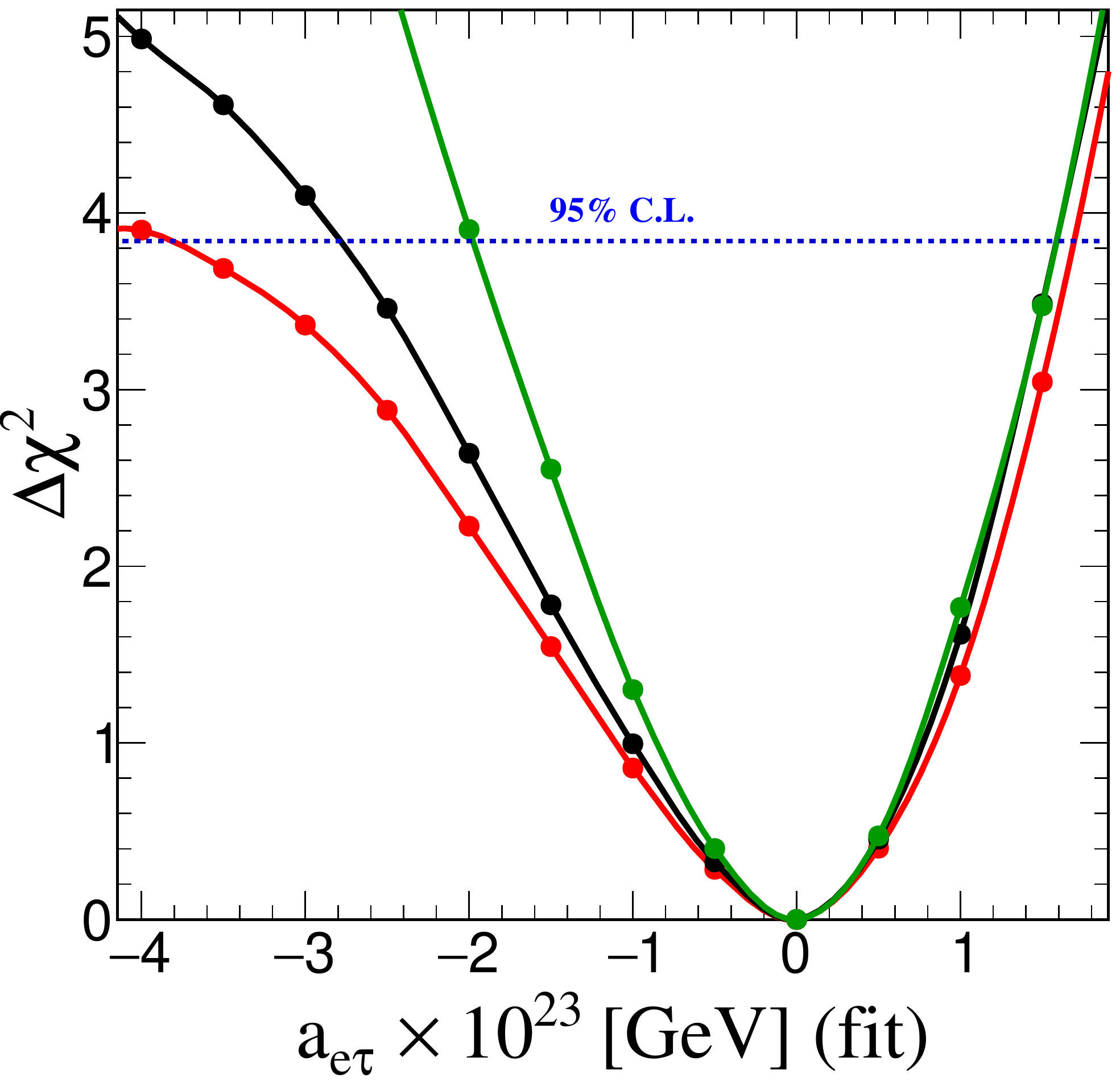}
  \mycaption{The sensitivities to constrain the LIV parameters $a_{\mu\tau}$, $a_{e\mu}$, and $a_{e\tau}$ using 500 kt$\cdot$yr exposure at the ICAL detector as shown in the left, middle, and right panels, respectively. In each panel, the red, black, and green curves represent the cases when the true value of $\sin^2\theta_{23}$ is taken as 0.4, 0.5, and 0.6, respectively. We have used the benchmark value of other oscillation parameters given in Table~\ref{tab:osc-param-value}. In theory, we have marginalized over oscillation parameters $\sin^2\theta_{23}$, $|\Delta m^2_\text{eff}|$, and both choices of mass orderings.}
  \label{fig:th23_impact}
\end{figure}

In the results presented so far, we have assumed $\sin^2\theta_{23}=0.5$ which corresponds to the maximal mixing , i.e. $\theta_{23} = 45^\circ$. The current global fit of neutrino data indicates that $\theta_{23}$ may be non-maximal where $\theta_{23}$ can be found in lower octant for $\theta_{23}<45^\circ$  or higher octant for $\theta_{23}>45^\circ$. At present, $\theta_{23}$ is the most uncertain oscillation parameter apart from $\delta_{\rm CP}$. Thus, it is an important question to ask how the constraints on CPT-violating LIV parameters  $a_{\mu\tau}$, $a_{e\mu}$, and $a_{e\tau}$ change, if $\theta_{23}$(true) is found to be non-maximal in nature. To answer this question, we present Fig.~\ref{fig:th23_impact} where we show the impact of non-maximal $\theta_{23}$ on constraining CPT-violating LIV parameters  $a_{\mu\tau}$, $a_{e\mu}$, and $a_{e\tau}$ as shown in the left, middle, and right panels, respectively. In each panel, we demonstrate constraints for three different true values of $\sin^2\theta_{23}=0.4$ (red curves), 0.5 (black curves), and 0.6 (blue curves).

The left panel of Fig.~\ref{fig:th23_impact} shows that the constraints on LIV parameter $a_{\mu\tau}$ deteriorates for non-maximal values of $\theta_{23}$(true) where  $\sin^2\theta_{23}{\rm (true)} =0.4$ and 0.6. We can also observe that the constraints on $a_{\mu\tau}$ are the same for $\sin^2\theta_{23}{\rm (true)} = 0.4$ and 0.6 which is consistent with the Eq.~\ref{eq:pmumu-nsi-mutau-omsd} where the term containing  $a_{\mu\tau}$ in $P(\nu_\mu \rightarrow \nu_\mu)$ is proportional to $\sin^2 2\theta_{23}$. The middle and right panels in Fig.~\ref{fig:th23_impact} illustrate that the constraints on both $a_{e\mu}$, and $a_{e\tau}$ improves for $\sin^2\theta_{23}{\rm (true)} = 0.6$ and deteriorates for $\sin^2\theta_{23}{\rm (true)} = 0.4$. This feature can be explained using Eq.~\ref{eq:Pmue-mat-LIV} for appearance channel which can also be translated to the effect on $\nu_\mu$ survival channel following the similar arguments as in section~\ref{sec:oscillograms}. The fifth term in Eq.~\ref{eq:Pmue-mat-LIV} with significant contribution to the effect of $a_{e\mu}$ is proportional to $\sin^3\theta_{23}$ . In a similar fashion, the seventh term in Eq.~\ref{eq:Pmue-mat-LIV} having dominant effect for $a_{e\tau}$ is proportional to $\sin^2\theta_{23}\cos\theta_{23}$. From all these observations in Fig.~\ref{fig:th23_impact}, we can conclude that if $\theta_{23}$ is found to be lying in the higher octant in nature then the sensitivity of ICAL for constraining $a_{e\mu}$, and $a_{e\tau}$ will enhance whereas it will reduce for $a_{\mu\tau}$.

\subsection{Impact of non-zero LIV parameters on mass ordering determination}
\label{sec:MO_impact}

\begin{table}[t]
  \centering
  \begin{tabular}{|c|c|c|c|c|}
    \hline
    \multirow{2}{*}{Cases}& \multicolumn{2}{c|}{NO (true)} & \multicolumn{2}{c|}{IO (true)} \\ \cline{2-5}
    & $\Delta \chi^2_{\rm ICAL-MO}$ & Deterioration & $\Delta \chi^2_{\rm ICAL-MO}$ & Deterioration \\ \hline
    SI & 7.55 & -- & 7.48 & -- \\
    SI + $a_{\mu\tau}$ & 6.27 & 16.8 \% & 6.34  & 15.2 \% \\
    SI + $a_{e\mu}$ & 5.08 & 32.7 \% & 3.90 & 47.9 \%\\
    SI + $a_{e\tau}$ & 5.23 & 30.7 \% & 4.24 & 15.2 \% \\ \hline
  \end{tabular}
  \mycaption{The sensitivity of the ICAL detector to determine mass ordering with 500 kt$\cdot$yr exposure. For the SI case (first row), we do not consider LIV in data and theory. For the cases of LIV parameters, we introduce $a_{\mu\tau}$ (second row), $a_{e\mu}$ (third row), and $a_{e\tau}$ (fourth row) in the theory one-at-a-time and marginalize over them along with oscillation parameters $\sin^2\theta_{23}$ and $\Delta m^2_{\rm eff}$ while assuming SI case in data. In the third and fifth columns, we show how much the mass ordering sensitivity deteriorates due to the presence of LIV parameters compared to the SI case.  We present our results assuming NO (IO) in data as given in the second and third (fourth and fifth) columns. We have used the benchmark value of oscillation parameters given in Table~\ref{tab:osc-param-value}.}
  \label{tab:MO_impact}
\end{table}

In this section, we are going to study the impact of non-zero LIV parameters $a_{\mu\tau}$, $a_{e\mu}$, and $a_{e\tau}$ on the sensitivity of the ICAL detector to determine the neutrino mass ordering. For statistical analysis, we simulate the prospective data assuming a given mass ordering. The sensitivity of ICAL to rule out the wrong mass ordering is calculated in the following fashion:
\begin{equation}
\Delta \chi^2_{\rm ICAL-MO} = \chi^2_{\rm ICAL} \text{(false MO)} - \chi^2_{\rm ICAL} \text{(true MO)}, 
\end{equation}
where, we calculate $\chi^2_{\rm ICAL} \text{(true MO)}$ and $\chi^2_{\rm ICAL} \text{(false MO)}$ by fitting the prospective data assuming true and false mass ordering, respectively. Since, the statistical fluctuations are suppressed while calculating the median sensitivity, we have $\chi^2_\text{ICAL} \text{(true MO)} \sim 0$. We calculate the sensitivity of ICAL to measure the neutrino mass ordering with 500 kt$\cdot$yr exposure using neutrino flux at the INO site following the procedure mentioned in Ref.~\cite{Devi:2014yaa}. The sensitivity towards the neutrino mass ordering is found to be 7.55 (7.48) for true NO (IO) as mentioned in the first row of Table~\ref{tab:MO_impact}.

In order to estimate the impact of non-zero LIV parameters, we generate the prospective data with a given mass ordering assuming no LIV where $a_{\mu\tau} = 0$, $a_{e\mu} =0$, and $a_{e\tau}=0$. Then, we fit the prospective data with opposite mass ordering assuming non-zero LIV parameter $a_{\mu\tau}$ in the theory and perform marginalization over $a_{\mu\tau}$ in the range\footnote{\label{fnote:LIV_ranges}The marginalization ranges of $a_{\mu\tau}$, $a_{e\mu}$, and $a_{e\tau}$ are guided by their 95\% C.L. bounds obtained in this work for ICAL as given in Table~\ref{tab:results_bounds}.} $[-0.23, 0.23] \times 10^{-23}$ GeV along with the oscillation parameters $\sin^2\theta_{23}$ and $\Delta m^2_{\rm eff}$ in the ranges as mentioned in section~\ref{sec:statistical_analysis}. The mass-ordering sensitivity with 500 kt$\cdot$yr exposure at ICAL for the cases of $a_{\mu\tau}$ is mentioned in the second row of Table~\ref{tab:MO_impact}. Similarly, the third (fourth) row in Table~\ref{tab:MO_impact} shows the sensitivity of ICAL towards neutrino mass ordering in the presence of $a_{e\mu}$ ($a_{e\tau}$) while perfoming marginalization over the range of $[-2.0, 1.5] \times 10^{-23}$ GeV ($[-2.8, 1.6] \times 10^{-23}$ GeV). In the third and fifth columns, we show the deterioration in the sensitivity for determining mass ordering due to the presence of non-zero LIV parameters with respect to the SI case. We observe in Table~\ref{tab:MO_impact} that depending upon true mass ordering, the results deteriorate by 15 to 50 \% due to the presence of non-zero LIV parameters when considered one parameter at-a-time.

\subsection{Allowed regions in ($|\Delta m^2_{32}|$ -- $\sin^2\theta_{23}$) plane with non-zero LIV parameters} 
\label{sec:precision_impact}

\begin{figure}[t]
  \centering
  \includegraphics[width=0.6\linewidth]{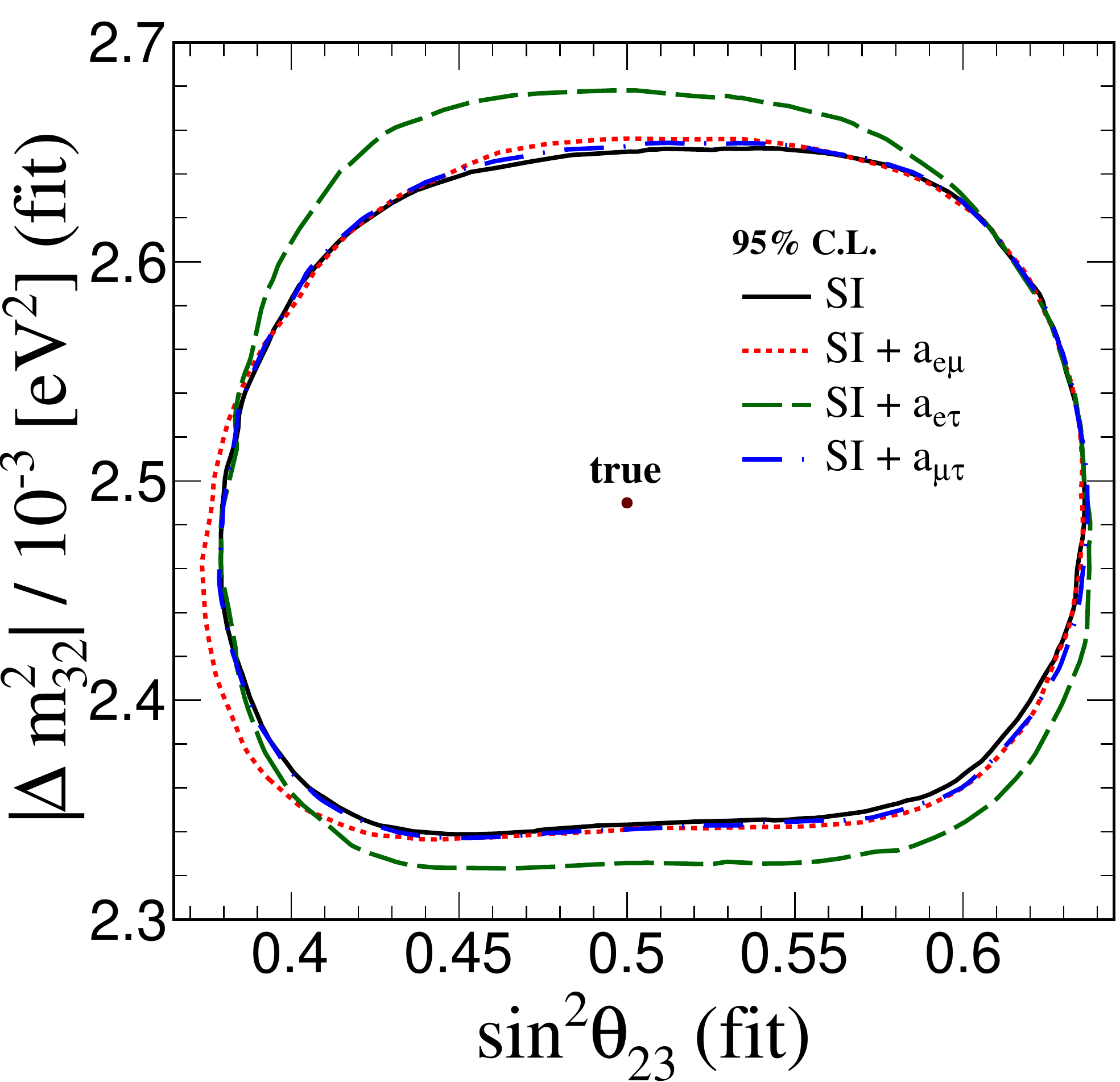}
  \mycaption{The allowed regions in ($|\Delta m^2_{32}|$ -- $\sin^2\theta_{23}$) plane at 95\% C.L. (2 d.o.f.) using 500 kt$\cdot$yr exposure at the ICAL detector assuming NO (true). The brown dot represents true choice, i.e. $\sin^2 \theta_{23}{\rm (true)} = 0.5$ and $|\Delta m^2_{32}|{\rm (true)} = 2.46 \times 10^{-3} {\rm~eV}^2$. The solid black curve shows the results for the SI case where we do consider LIV in data and fit. The dash-dotted blue, dotted red, and dashed green curves illustrate the results for the cases of non-zero LIV parameters $a_{\mu\tau}$, $a_{e\mu}$, and $a_{e\tau}$, respectively one-at-a-time in the fit and marginalization over them.}
  \label{fig:precisionmeasurement}
\end{figure}

Now, we explore the impact of CPT-violating LIV parameters $a_{\mu\tau}$, $a_{e\mu}$, and $a_{e\tau}$ on the precision measurement of atmospheric oscillation parameters $\sin^2 \theta_{23}$ and $|\Delta m^2_{32}|$ using the ICAL detector with 500 kt$\cdot$yr exposure. First of all, we simulate the prospective data assuming $\sin^2 \theta_{23}$ (true) = 0.5 and $|\Delta m^2_{32}| {\rm (true)} = 2.46 \times 10^{-3} ~{\rm eV}^2$ for case of SI with no LIV. The statistical significance for this is quantified using the following expression:
\begin{equation}
  \Delta \chi^2_{\rm ICAL-PM} (\sin^2\theta_{23},\,|\Delta m^2_{32}|) = \chi^2_{\rm ICAL} (\sin^2\theta_{23},\,|\Delta m^2_{32}|) - \chi^2_0,
\end{equation}
where, $\chi^2_{\rm ICAL} (\sin^2\theta_{23},\,|\Delta m^2_{32}|)$ is estimated by fitting the prospective data with theory for a given value of $\sin^2\theta_{23}$, and $|\Delta m^2_{32}|$. $\chi^2_0$ is the minimum value of $\chi^2_{\rm ICAL} (\sin^2\theta_{23},\,|\Delta m^2_{32}|)$ in the allowed range of oscillation parameters $\sin^2 \theta_{23}$ and $|\Delta m^2_{32}|$. Here, $\chi^2_0 \sim 0$ because the statistical fluctuations are suppressed. We estimate the allowed regions in ($|\Delta m^2_{32}|$ -- $\sin^2\theta_{23}$) plane for the case of SI at 95\% C.L. (2 d.o.f.) as shown by black line in Fig.~\ref{fig:precisionmeasurement}.

Now, we discuss the impact of non-zero LIV parameters $a_{\mu\tau}$, $a_{e\mu}$, and $a_{e\tau}$ one-at-a-time on the precision measurement of atmospheric oscillation parameters. We simulate the prospective data considering the true values of $\sin^2 \theta_{23}$ and $|\Delta m^2_{32}|$ as mentioned above without LIV. Then, we estimate allowed regions in ($|\Delta m^2_{32}|$ -- $\sin^2\theta_{23}$) plane with non-zero LIV parameters $a_{\mu\tau}$, $a_{e\mu}$, and $a_{e\tau}$ one-at-a-time in the fit. Further, we marginalize over $a_{\mu\tau}$ in the range $[-0.23, 0.23] \times 10^{-23}$ GeV, $a_{e\mu}$ in the range $[-2.0, 1.5] \times 10^{-23}$ GeV, and  $a_{e\tau}$ in the range $[-2.8, 1.6] \times 10^{-23}$ GeV one-at-a-time in the fit (see footnote~\ref{fnote:LIV_ranges}). We show these results for the cases of non-zero LIV parameters $a_{\mu\tau}$, $a_{e\mu}$, and $a_{e\tau}$ at 95 \% C.L. (2 d.o.f.) using dotted red, dashed green, and dash-dotted blue lines, respectively, in Fig.~\ref{fig:precisionmeasurement}. We do not observe a significant change in the allowed regions at 95\% C.L. when we introduce $a_{\mu\tau}$ in fit and marginalize over it. It indicates that the precision measurement of atmospheric oscillation parameters using the ICAL detector is quite robust against the presence of $a_{\mu\tau}$ in the fit. However, the presence of non-zero LIV parameter $a_{e\mu}$ in fit deteriorate the allowed region in the direction of $\sin^2\theta_{23}$. As far as the presence of LIV parameter $a_{e\tau}$ in fit is concerned, it deteriorates allowed region in the direction of $|\Delta m^2_{32}|$.

\subsection{Correlation between off-diagonal LIV parameters}
\label{sec:correlation}

\begin{figure}
  \centering
  \includegraphics[width=0.32\linewidth]{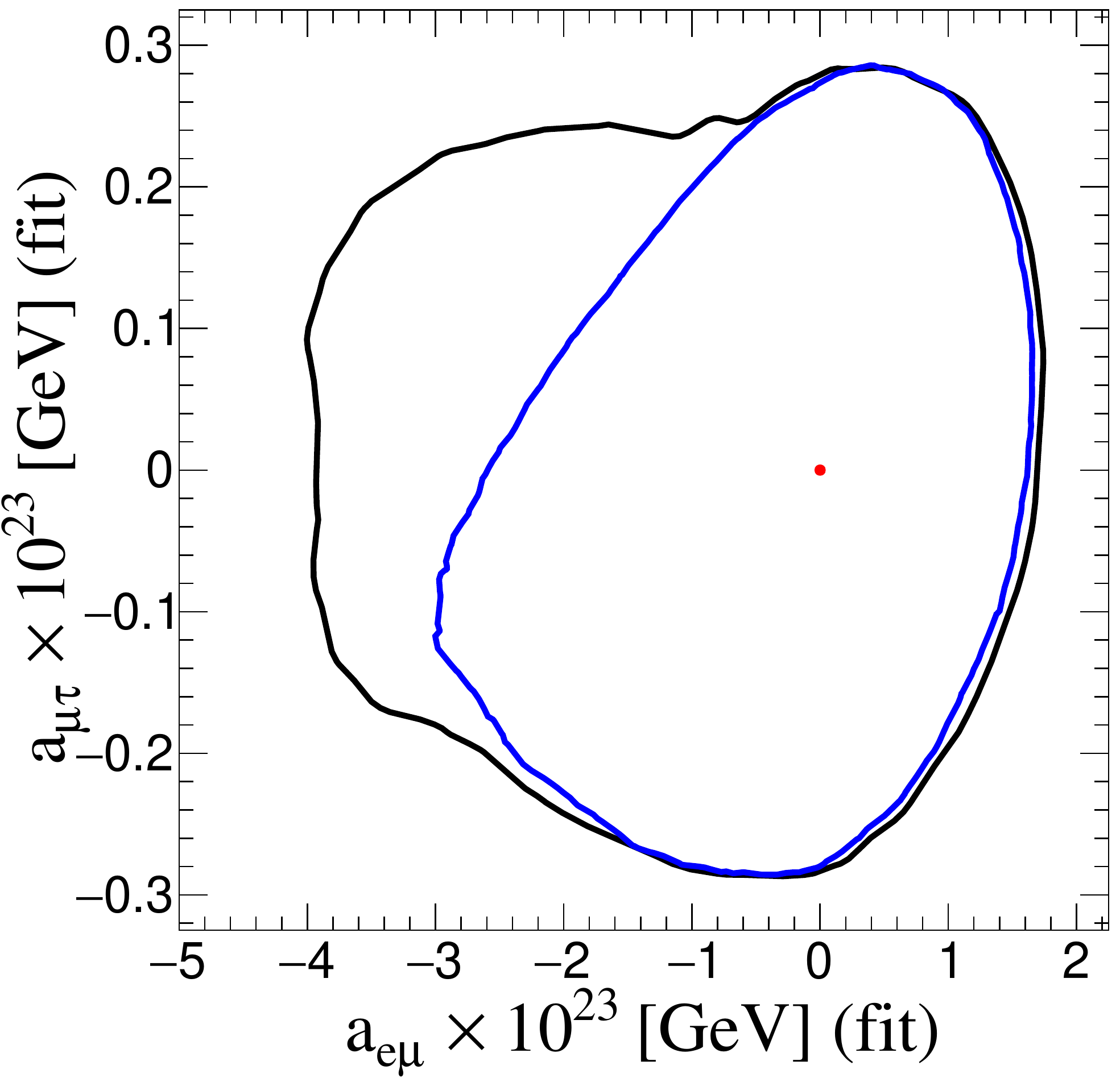}
  \includegraphics[width=0.32\linewidth]{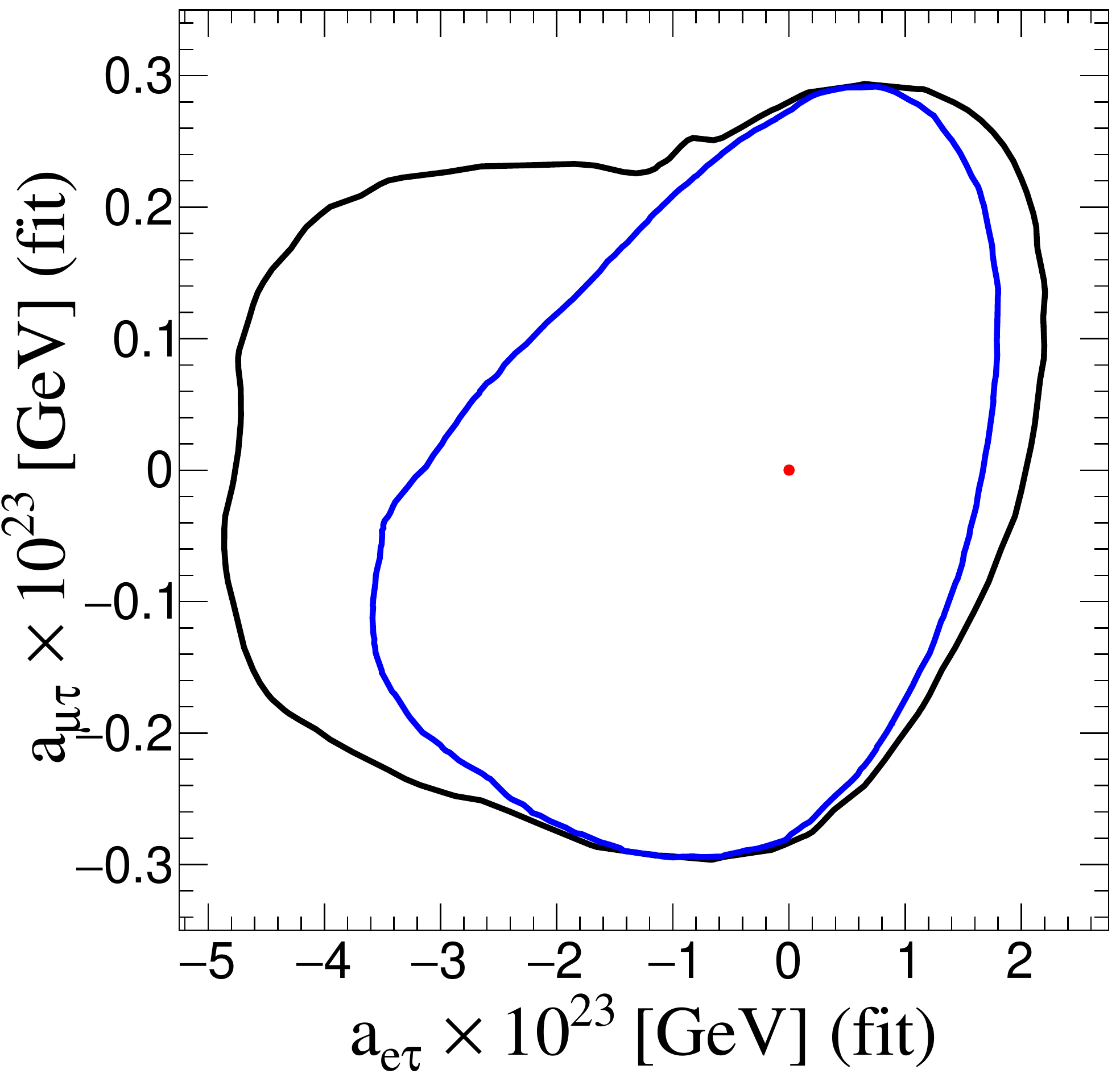}
  \includegraphics[width=0.32\linewidth]{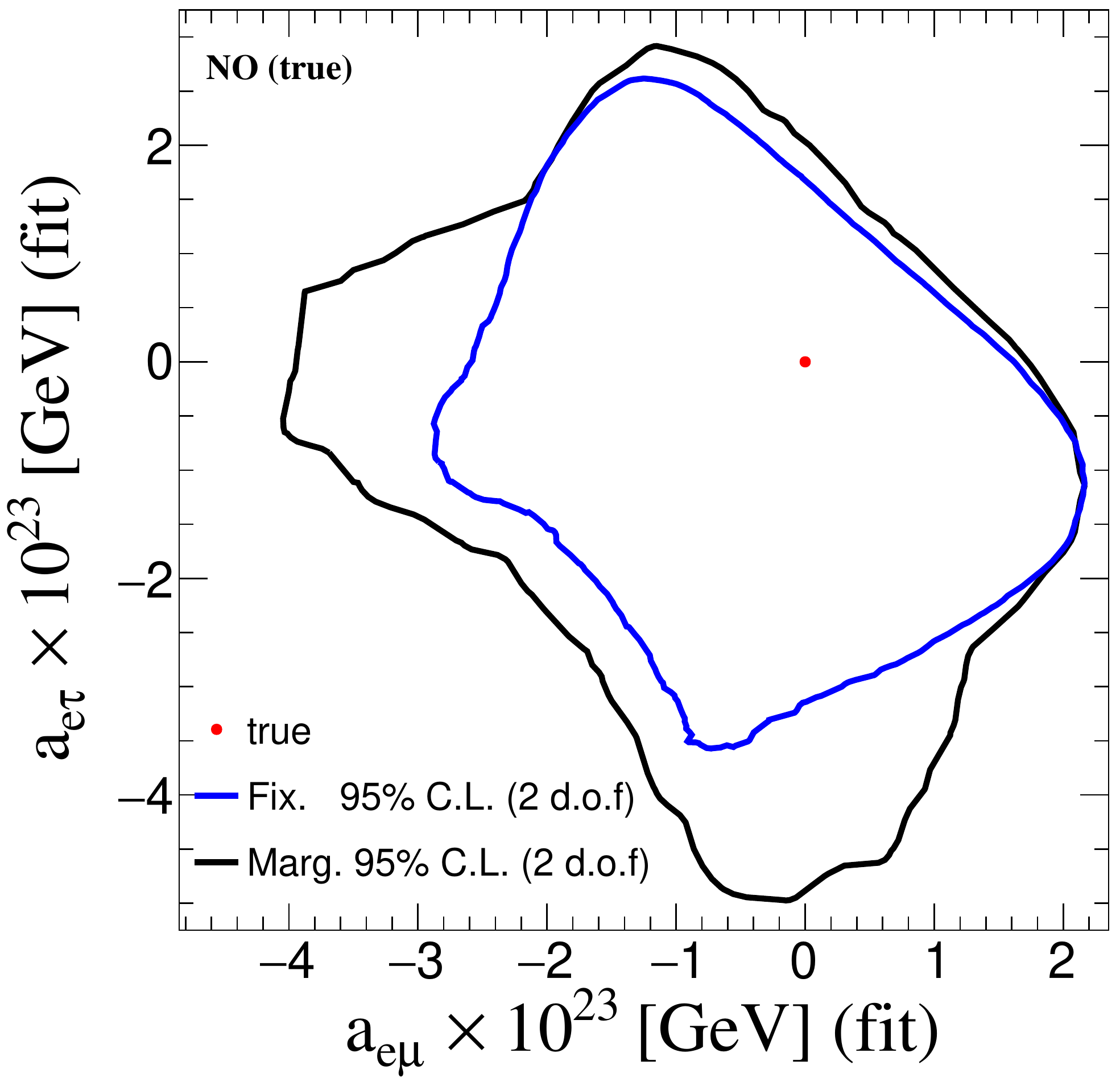}
  \mycaption{The correlated constraints in the plane of CPT-violating LIV parameters $(a_{\mu\tau}, a_{e\mu})$ (left panel), $(a_{\mu\tau}, a_{e\tau})$ (middle panel), and $(a_{e\tau}, a_{e\mu})$ (right panel) at 95\% C.L.  (2 d.o.f.) with 500 kt$\cdot$yr exposure at the ICAL detector. We generate prospective data considering no LIV (red dots) with NO (true) and benchmark value of oscillation parameters mentioned in Table~\ref{tab:osc-param-value}. In theory, we vary two LIV parameters at-a-time with  marginalization over the oscillation parameters $\sin^2\theta_{23}$, $\Delta m^2_{\rm eff}$, and both choices of mass ordering as shown by black curves. The blue curves represent the fixed-parameter scenario where we do not marginalize over any oscillation parameter in theory, but we do marginalize over systematic uncertainties.}
  \label{fig:correlated_bound}
\end{figure}

So far, we have studied the neutrino oscillations in the presence of only one CPT-violating LIV parameter at-a-time. In this section, we explore the correlation between different  CPT-violating LIV parameters. To begin with, we generate prospective data with 500 kt$\cdot$yr exposure at the ICAL detector considering no LIV with NO as true mass ordering and benchmark value of oscillation parameters mentioned in Table~\ref{tab:osc-param-value}. While constraining two LIV parameters at-a-time, we quantify the statistical significance using the following expression:
\begin{equation}
\Delta \chi^2_{\rm ICAL-LIV} (a_{\mu\tau}, a_{e\mu}) = \chi^2_{\rm ICAL} (\text{SI} + (a_{\mu\tau}, a_{e\mu})) - \chi^2_{\rm ICAL} {\rm (SI)},
\end{equation}
where, $\chi^2_{\rm ICAL} {\rm (SI)}$ and $\chi^2_{\rm ICAL} (\text{SI} + (a_{\mu\tau}, a_{e\mu}))$ are calculated by fitting the prospective data with only SI case (no LIV) and with SI + LIV $(a_{\mu\tau}, a_{e\mu})$ case, respectively. Since, the statistical fluctuations are suppressed, we have $\chi^2_{\rm ICAL} {\rm (SI)}\sim 0$. Then, we estimate the allowed regions in the plane of CPT-violating LIV parameters $(a_{\mu\tau}, a_{e\mu})$ at 95\% C.L. (2 d.o.f.) as shown in the left panel of Fig.~\ref{fig:correlated_bound}. The true choices of LIV parameters $a_{\mu\tau}=0$, $a_{e\mu}=0$, and $a_{e\tau}=0$ in data are depicted as red dots in Fig.~\ref{fig:correlated_bound}. In theory, we vary two LIV parameters at-a-time with (without) marginalization over the oscillation parameters $\sin^2\theta_{23}$, $\Delta m^2_{\rm eff}$, and both choices of mass ordering as shown by black (blue) curve. Similarly, we present the allowed regions in the plane of LIV parameters $(a_{\mu\tau}, a_{e\tau})$ and $(a_{e\tau}, a_{e\mu})$ in the middle and left panels, respectively, in Fig.~\ref{fig:correlated_bound}.    

\section{Summary and concluding remarks}
\label{sec:conclusion}

In this paper, for the first time, we explore the Lorentz Invariance Violation (LIV) through the mass-induced flavor oscillations of atmospheric neutrino in the multi-GeV range of energies over a wide range of baselines using the proposed ICAL detector at INO. In the present analysis, we focus our attention on the isotropic CPT-violating LIV parameters $a_{\mu\tau}$, $a_{e\mu}$, and $a_{e\tau}$ where we consider their real values only with both positive and negative signs.

We perform a detailed analysis by demonstrating the effect of non-zero CPT-violating LIV parameters ($a_{\mu\tau}$, $a_{e\mu}$, and $a_{e\tau}$) one-at-a-time on probability oscillograms of the disappearance ($\nu_\mu \rightarrow \nu_\mu$) channel, which has a dominant contribution (more than 98\%) to the reconstructed muon events at the ICAL detector. We observe that the vacuum oscillation valley bends due to the presence of non-zero LIV parameter $a_{\mu\tau}$ where the curvature of valley depends on the sign of $a_{\mu\tau}$. For a given $a_{\mu\tau}$, the curvatures of bendings are opposite for neutrino and antineutrino. This fact indicates that the charged identification capability of the ICAL detector is a crucial property while probing LIV parameter $a_{\mu\tau}$. If we turn our attention to the other off-diagonal LIV parameters $a_{e\mu}$ and $a_{e\tau}$, we observe that they do not result in any bending of oscillation valley, but they do disturb the vacuum oscillation valley as well as matter effect regions.

Next, we present the impact of LIV on the reconstructed event distributions at the ICAL detector for 500 kt$\cdot$yr exposure. We observe that $a_{\mu\tau}$ affects the event distributions by the largest amount among all off-diagonal CPT-violating LIV parameters. Since the effects of LIV parameters depend on energy and direction, the good energy and direction resolution of the ICAL detector is going to play an important role.

After describing the modification in reconstructed event distributions due to LIV, we calculate statistical significance for the presence of LIV using the ICAL detector at the $\chi^2$ level. In order to calculate $\Delta\chi^2$, we simulate the prospective data at ICAL with 500 kt$\cdot$yr exposure for SI case with no LIV and perform fitting assuming SI+LIV case in theory where we consider the CPT-violating LIV parameters $a_{\mu\tau}$, $a_{e\mu}$, and $a_{e\tau}$ one-at-a-time. First of all, we identify the energy and direction bins that contribute significantly to $\Delta\chi^2$. Further, we constrain the LIV parameters $a_{\mu\tau}$, $a_{e\mu}$, and $a_{e\tau}$ at 95\% C.L. using 500 kt$\cdot$yr exposure at the ICAL detector.

We show the advantage of incorporating hadron energy information and the CID capability of ICAL while constraining LIV parameters. We also demonstrate that the constraints on $a_{\mu\tau}$ are robust against the uncertainties in oscillation parameters where the estimated bounds on $a_{e\mu}$ and $a_{e\tau}$ are expected to improve with more precise determination of oscillation parameters. As far as the non-maximal value of $\theta_{23}$ is concerned, it deteriorates the constraints on $a_{\mu\tau}$ whereas constraints on $a_{e\mu}$ and $a_{e\tau}$ improve (worsen) if $\theta_{23}$ is found to be lying in the higher (lower) octant.

The aim of the ICAL detector at INO is to determine the mass ordering and the precise determination of atmospheric oscillation parameters $\theta_{23}$ and $|\Delta m^2_{32}|$. In the present analysis, we also explore how the sensitivity to measure mass ordering gets affected due to the presence of non-zero LIV parameters considered one-at-a-time. We observe that depending upon true mass ordering, the results get deteriorated by about 15 to 50\% due to the presence of LIV. If we look at the effect of LIV on precision measurement of atmospheric oscillation parameters $\theta_{23}$ and $|\Delta m^2_{32}|$, we find that the precision measurement is quite robust against the marginalization over $a_{\mu\tau}$ in the fit. However, the presence of $a_{e\mu}$ ($a_{e\tau}$) worsens the allowed region in the direction of $\sin^2 \theta_{23}$ ($|\Delta m^2_{32}|$).

The above-mentioned analyses consider one LIV parameter at-a-time. In order to explore the correlation among various off-diagonal CPT-violating LIV parameters, we considered two LIV parameter at-a-time and constrain (2 d.o.f.) them using 500 kt$\cdot$yr exposure at the ICAL detector. We have obtained allowed regions in three different planes of $(a_{\mu\tau}, a_{e\mu})$, $(a_{\mu\tau}, a_{e\tau})$ and $(a_{e\tau}, a_{e\mu})$ while marginalizing over oscillation parameters. 

Before we conclude, we would like to highlight the fact that the present analysis has been performed in the multi-GeV range of energies which is complementary to energy ranges used in the analyses of Super-K and IceCube.  We would also like to emphasize that the LIV bounds obtained in the present analysis using the ICAL detector are quite competitive compared to the existing bounds from Super-K and IceCube due to the charge identification capability of ICAL as well as the good resolutions in reconstructed muon energy and direction. In the absence of CID, the constraints on LIV parameters deteriorate significantly due to the dilution of LIV features while combining $\mu^-$ and $\mu^+$ events. This unique capability of CID enables ICAL to probe LIV separately in neutrino and antineutrino modes which may not be possible at any other existing or planned neutrino detector based on water, ice, or argon.  

\subsubsection*{Acknowledgements}

This work is performed by the members of the INO-ICAL collaboration to estimate the constraints on the CPT-violating off-diagonal LIV parameters using 500 kt$\cdot$yr exposure of the ICAL-INO atmospheric neutrino detector. We thank M. Masud, S. Das, and A. Khatun for useful communications. We thank A. Dighe, S. Goswami, J. F. Libby, A. K. Nayak, A. M. Srivastava, V. A. Kosteleck$\acute{\rm y}$, and J. S. Diaz for their useful suggestions and constructive comments on our work. We sincerely thank the INO internal referees, M. V. N. Murthy and S. Uma Sankar for their careful reading of the manuscript and for providing useful suggestions. S. Sahoo would like to thank the organizers of the XXIV DAE-BRNS High Energy Physics Online Symposium at NISER, Bhubaneswar, India, from 14th to 18th December 2020, for providing him an opportunity to present a poster based on this work. We get benefited from the valuable discussions during the Fourth IUCSS Summer School and Workshop on the Lorentz and CPT-violating Standard-Model Extension during May-2021. We acknowledge the support of the Department of Atomic Energy (DAE), Govt. of India. S.K.A. is supported by the DST/INSPIRE Research Grant [IFA-PH-12] from the Department of Science and Technology (DST), Govt. of India, and the Young Scientist Project [INSA/SP/YSP/144/2017/1578] from the Indian National Science Academy (INSA). S.K.A. acknowledges the financial support from the Swarnajayanti Fellowship Research Grant (No. DST/SJF/PSA-05/2019-20) provided by the Department of Science and Technology (DST), Govt. of India, and the Research Grant (File no. SB/SJF/2020-21/21) provided by the Science and Engineering Research Board (SERB) under the Swarnajayanti Fellowship by the DST, Govt. of India. We acknowledge CCHPC-19 and Sim01: High-Performance Computing facilities at Tata Institute of Fundamental Research, Mumbai, and SAMKHYA: High-Performance Computing facility at Institute of Physics, Bhubaneswar, for performing the numerical simulations.

\begin{appendix}

\section{Some properties of gauge invariant LIV parameters}
\label{app:L_CPTV}

In this section, we describe the properties of CPT-violating and conserving LIV parameters. Let us consider a set of Dirac spinors representing the neutrino field, $\psi = \{\nu_e, \nu_\mu, \nu_\tau\}$ and corresponding charge conjugated field, $\psi^c = \{\nu^c_e, \nu^c_\mu, \nu^c_\tau\}$. Here, the charge conjugated field transforms as $\psi^c = C\bar\psi^T$, and $C$ is the charge conjugation operator. Now, lets define a Langragian density $\mathcal{L} = \mathcal{L}_{o} - \mathcal{L}^{'}$, where $\mathcal{L}_{o}$ represents the weak interactions in the Standard Model and $\mathcal{L}^{'}$ represents the new physics interactions induced due to the violation of Lorentz symmetry. In order to address the new physics interactions, we are adopting the spontaneous Lorentz violation mechanism that is proposed in the string theory. At the Planck scale ($M_p$), spontaneous breaking of the Lorentz symmetry can be expected when a tensor field of non-perturbative vacuum in the proposed model acquires the non-zero vacuum expectation values (VEVs). These VEVs effectively can act as a fixed background to a given observer's frame of reference. The interactions induced due to this background can have the boost dependency which breaks the Lorentz symmetry\footnote{ Note that there are also proposed models where Lorentz violation can occur via explicit breaking mechanism.}.

At low energy, the induction strength of Lorentz violation is expected to be suppressed by an order of $1/M_p$~\cite{Colladay:1996iz, Kostelecky:2000mm, Kostelecky:2003cr, Bluhm:2005uj}. In the low-energy effective field theory, the LIV interaction can have many coupling coefficients which effectively maintain the power-counting renormalizability as follows~\cite{Kostelecky:1995qk}:
\begin {equation}
\mathcal{L}^{'} \supseteq \frac{\lambda}{(M_p)^k}\left\langle T \right\rangle \overline{\psi}\Gamma\left( i\partial\right)^k \psi + h.c.,
\end {equation}
where, $\lambda$ stands for a dimensionless coupling constant, $M_p $ denotes the Planck mass scale, $\left\langle T \right\rangle$ are the non-zero tensor VEVs, $\Gamma$ represent some gamma-matrix structure, and $k$ is an integer power where the dominant terms with $k \leq 1$ are renormalizable. Considering only the gauge-invariant terms, the CPT-violating and CPT-conserving parameters at low energies are obtained by choosing the  mass dimension to be zero and one, respectively. For $k = 0$, $\left\langle T \right\rangle \sim \left(\frac{m^2}{M_p}\right)$ and for $k = 1$, $\left\langle T \right\rangle \sim m$, where $m$ is the mass of fermionic field, $\psi$. The CPT-violating and CPT-conserving terms appear in the following fashion~\cite{Colladay:1998fq}:

\begin{align}
  \mathcal{L}^{'}_{\rm CPT-violating}  & = a_{\mu}\overline{\psi}\gamma^{\mu}\psi + b_{\mu}\overline{\psi}\gamma_{5}\gamma^{\mu}\psi + h.c.\,, \\
  \mathcal{L}^{'}_{\rm CPT-conserving} & = i c_{\mu\nu}\overline{\psi}\gamma^{\mu}\partial^\nu\psi + i d_{\mu\nu}\overline{\psi}\gamma_{5}\gamma^{\mu}\partial^\nu\psi + h.c. \\
  & = \frac{1}{4} c_{\mu\nu}\overline{\psi}\gamma^{\mu}\gamma^{\nu}(i\slashed{\partial})\psi + \frac{1}{4} d_{\mu\nu}\overline{\psi}\gamma_{5}\gamma^{\mu}\gamma^{\nu}(i\slashed{\partial})\psi + h.c.\,.
\end{align}
Here, the quantities $a_\mu$, $b_\mu$, $c_{\mu\nu}$, and $d_{\mu\nu}$ are real because of their mode of origin via spontaneous symmetry breaking mechanism followed by the hermitian nature of the underlying theory~\cite{Colladay:1996iz}. Note that the running of these Lorentz-violating couplings between the Planck scale and the low-energy scale is discussed in Ref.~\cite{Kostelecky:2001jc}.
 
Now, by interchanging the space-time indices $\mu\leftrightarrow\nu$, we get for the CPT-conserving parameters
  \begin{align}
  \mathcal{L}^{'}_{\rm CPT-conserving} & = \frac{1}{4} c_{\nu\mu}\overline{\psi}\gamma^{\nu}\gamma^{\mu}(i\slashed{\partial})\psi + \frac{1}{4} d_{\nu\mu}\overline{\psi}\gamma_{5}\gamma^{\nu}\gamma^{\mu}(i\slashed{\partial})\psi + h.c.\, \\
  & = -\frac{1}{4} c_{\nu\mu}\overline{\psi}\gamma^{\mu}\gamma^{\nu}(i\slashed{\partial})\psi - \frac{1}{4} d_{\nu\mu}\overline{\psi}\gamma_{5}\gamma^{\mu}\gamma^{\nu}(i\slashed{\partial})\psi + h.c.\quad ({\rm for~} \mu \neq \nu)\,.
  \end{align}
Note, $\mathcal{L}^{'}_{\rm CPT-conserving}$ must be invariant under the interchange of space-time indices ($\mu\leftrightarrow\nu$), which implies that $c_{\mu\nu}$ and $d_{\mu\nu}$ are the antisymmetric tensors, \textit{i.e.},
  \begin{align}
  c_{\mu\nu} & = - c_{\nu\mu} = - c^{\rm T}_{\mu\nu}\,, \label{eq:antisymm_c}\\
  d_{\mu\nu} & = - d_{\nu\mu} = - d^{\rm T}_{\mu\nu}\,. \label{eq:antisymm_d}
  \end{align}
Thus, $\mathcal{L}^{'}$ is defined as~\cite{Kostelecky:2011gq,KumarAgarwalla:2019gdj,Antonelli:2020nhn},
  \begin{align}
  \mathcal{L}^{'} & = \mathcal{L}^{'}_{\rm CPT-violating} + \mathcal{L}^{'}_{\rm CPT-conserving} \\
   & = \frac{1}{2}\left[a_{\mu}\overline{\psi}\gamma^{\mu}\psi + b_{\mu}\overline{\psi}\gamma_{5}\gamma^{\mu}\psi - i c_{\mu\nu}\overline{\psi}\gamma^{\mu}\partial^\nu\psi - i d_{\mu\nu}\overline{\psi}\gamma_{5}\gamma^{\mu}\partial^\nu\psi\right] + h.c.\,, \label{eq:LIV-Lagrangian-Density}
  \end{align}  
where, $a_\mu$ and $b_\mu$ are the CPT-violating LIV parameters, whereas $c_{\mu\nu}$ and $d_{\mu\nu}$ are the CPT-conserving LIV parameters. The $1/2$ factor in Eq.~\ref{eq:LIV-Lagrangian-Density} appears due to the cannonical renormalization of the neutrino field. 

A single field $\chi$ consisting of the $\psi$ and $\psi^c$ fields can be defined as
  \begin{equation}
  \chi = \begin{bmatrix} \psi \\ \psi^c \end{bmatrix} \rm and \,\,\,\chi^c = \mathcal{C}\chi, \rm \,\,\,where \,\,\,\,\mathcal{C} = \left[\begin{array}{cc} 0 & 1 \\ 1 & 0 \end{array}\right].
  \end{equation}
  The  Lagrangian $\mathscr{L}$ is defined in terms of field $\chi$ as \cite{Kostelecky:2003cr}
  \begin{align}
  \mathscr{L}  = \frac{1}{2}\bar\chi\Gamma^\mu\left(i\partial_\mu\right)\chi - \bar\chi\rm M \chi + \rm h.c. \, ,
  \end{align}
  where,
  \begin{align}
  \Gamma^\mu  & = \gamma^\mu + c^{\mu\nu}\gamma_\nu + d^{\mu\nu}\gamma_5\gamma_\nu\,,\\
  \rm M       & = \left(m + i m_5\gamma_5\right) + a^\mu\gamma_\mu + b^\mu\gamma_5\gamma_\mu\,.
  \end{align}
  The hermiticity demands that $\Gamma_\mu$ = $\gamma^0 \Gamma^\dagger_\mu \gamma^0$, and $\rm M$ = $\gamma^0 \rm M^\dagger \gamma^0$. It is expected that at low energies, the LIV observables don't have the significant effects on the SM weak interactions. Hence, the active neutrino can be treated as left-handed and the antineutrino as right-handed. The structure of the equation of motion should remain unaltered under the charge conjugation operation, which is denoted by the operator, $C = i\gamma^2\gamma^0$. Under the charge conjugation operation, $\Gamma_\mu$ = -$C (\Gamma_\mu)^{\rm T} C^{-1}$ and $\rm M$ = $C {\rm M^T} C^{-1}$, which further implies
\begin{subequations}
    \begin{align}
    a_\mu\gamma^\mu              & =  C \left(a_\mu\gamma^\mu\right)^{\rm T} C^{-1} , \label{eq:a}\\
    b_\mu\gamma^5\gamma^\mu      & =  C \left(b_\mu\gamma^5\gamma^\mu\right)^{\rm T} C^{-1} , \label{eq:b}\\
    c_{\mu\nu}\gamma^\mu         & = -C \left(c_{\mu\nu}\gamma^\mu\right)^{\rm T} C^{-1}, \label{eq:c}\\
    d_{\mu\nu}\gamma^5\gamma^\mu & = -C \left(d_{\mu\nu}\gamma^5\gamma^\mu\right)^{\rm T} C^{-1} . \label{eq:d}
    \end{align}
  \end{subequations}
  Now, expanding \ref{eq:a},
  \begin{subequations}
    \begin{align}
    a_\mu\gamma^\mu   & = C \left(a_\mu\gamma^\mu\right)^{\rm T}C^{-1} \\
    & = C \left(\gamma^\mu\right)^{\rm T} a^{\rm T}_\mu C^{-1}\\
    & = C \left(\gamma^\mu\right)^{\rm T} C^{-1}\,C a^{\rm T}_\mu C^{-1}\,\,[{\rm using}\,\,C^{-1}\,C = I]\\
    & = -\gamma^\mu C\,a^{\rm T}_\mu\,C^{-1}\,\,[{\rm using}\,\,C \left(\gamma^\mu\right)^{\rm T} C^{-1} = -\gamma^\mu]\\
    \Rightarrow a_\mu & = -C\,a^{\rm T}_\mu\,C^{-1}\, . \label{eq:aL}
    \end{align}
  \end{subequations}
  Similarly, expanding \ref{eq:b},
  \begin{subequations}
    \begin{align}
    b_\mu\gamma^5\gamma^\mu & = C \left(b_\mu\gamma^5\gamma^\mu\right)^{\rm T} C^{-1} \\
    & = C \left(\gamma^5\gamma^\mu\right)^{\rm T} b^{\rm T}_\mu C^{-1}\\
    & = C (\gamma^\mu)^{\rm T}C^{-1}\,C(\gamma^5)^{\rm T} C^{-1}\,C b^{\rm T}_\mu C^{-1}\,\,[{\rm using}\,\,C^{-1}\,C = I]\\
    & = -\gamma^\mu\gamma^5 C\,b^{\rm T}_{\mu}\,C^{-1}[{\rm using}\,\,C \left(\gamma^\mu\right)^{\rm T} C^{-1} = -\gamma^\mu,\,\,C(\gamma^5)^{\rm T} C^{-1} = \gamma^5]\\
    & \rm since,\,\{\gamma^\mu,\gamma^5\} = 0 \notag  \\
    & = \gamma^5\gamma^\mu C\,b^{\rm T}_\mu\,C^{-1} \,, \\
    \Rightarrow b_\mu       & = C\,b^{\rm T}_\mu\,C^{-1}\,. \label{eq:bL}
    \end{align}
  \end{subequations}
  Following the same procedure for \ref{eq:c} and \ref{eq:d}, we obtain,
  \begin{subequations}
    \begin{align}
    & c_{\mu\nu} = C\,c^{\rm T}_{\mu\nu}\,C^{-1}\,, \label{eq:cL} \\
    & d_{\mu\nu} = -C\,d^{\rm T}_{\mu\nu}\,C^{-1}\,. \label{eq:dL}
    \end{align}
\end{subequations}
The experiments to probe LIV parameters can be categorized on the phenomena of i) Coherent, ii) Interferometric, or iii) Extreme effects. For the mass-induced neutrino oscillations, the LIV parameters have interference effects. Thus, the individual CPT-violating parameters $a_\mu$ and $b_\mu$, as well as the CPT-conserving parameters $c_{\mu\nu}$ and $d_{\mu\nu}$, are not the observable quantities. These interfering parameters can be redefined as the observable quantities decomposed along with the polarization of active left-handed neutrinos and right-handed antineutrinos. Therefore, we define 
\begin{align}
a_L &= \left(a + b\right), \\
a_R &= \left(a - b\right), \\
c_L &= \left(c + d\right), \\
c_R &= \left(c - d\right),
\end{align}
where, indices are supressed for simplicity. The relation between $a_L$ and $a_R$ can be evaluated using \ref{eq:aL} and \ref{eq:bL}, respectively. The expectation value of $\left\langle a_L \right\rangle$ is $\bra{\chi}a_L\ket{\chi}$. Hence, we obtain 
\begin{subequations}
  \begin{align}
  \left\langle a_L \right\rangle & = \bra{\chi}(C(a_L)^{\rm T} C^{-1})\ket{\chi}     \\
  & = \bra{\chi}(C\,a^{\rm T}\,C^{-1} + C\,b^{\rm T}\,C^{-1})\ket{\chi}  \\
  & = \bra{\chi}(-a + b)\ket{\chi}  \\
  & = -\bra{\chi^c}\mathcal{C}a_R\mathcal{C}\ket{\chi^c} ,   \\
  & {\rm since},\,\mathcal{C}^2 = I \notag  \\
  \Rightarrow \left\langle a_L \right\rangle & =-\left\langle a_R \right\rangle.  
  \end{align}
\end{subequations}
  Similarly using \ref{eq:cL} and \ref{eq:dL},
  \begin{subequations}
  \begin{align}
  \left\langle c_L \right\rangle & = \bra{\chi}(C(c_L)^{\rm T} C^{-1})\ket{\chi}     \\
  & = \bra{\chi}(C\,c^{\rm T}\,C^{-1} + C\,d^{\rm T}\,C^{-1})\ket{\chi}  \\
  & = \bra{\chi}(c - d)\ket{\chi}  \\
  & = \bra{\chi^c}\mathcal{C}c_R\mathcal{C}\ket{\chi^c},    \\
  & {\rm since},\,\mathcal{C}^2 = I \notag  \\
  \Rightarrow \left\langle c_L \right\rangle & = \left\langle c_R \right\rangle. 
  \end{align}
\end{subequations}

In simple terms, we can conclude that when we transform from neutrino to antineutrino, the CPT-violating parameters flip their signs, whereas CPT-conserving parameters do not. This is an important property of LIV parameters for the case of neutrino oscillations. 

Another crucial property of the CPT-conserving LIV parameters can be explored by choosing a rotational invariant coordinate with isotropic condition. Now, using the properties of Eq.~\ref{eq:antisymm_c} and Eq.~\ref{eq:antisymm_d}, we obtain $Tr\left(c^{\mu\nu}_L g_{\mu\nu}\right) = 0$, where $g_{\mu\nu}$ is the metric tensor, that implies $c^{00}_L = \sum_i^{1,3} c^{ii}_L$. Applying the condition for isotropic symmetry, we got $c^{11}_L = c^{22}_L = c^{33}_L = \frac{1}{3}c^{00}_L$. Recalling the last term of Eq.~\ref{eq:liv2},
\begin{align}
\frac{1}{E}c^{\mu\nu}_{L} p_\mu p_\nu & = c^{00}_L E + \frac{1}{E}\big(c^{11}_L p^2_{11} + c^{22}_L p^2_{22} + c^{33}_L p^2_{33}\big) \nonumber \\
& = c^{00}_L E + \frac{1}{3E}c^{00}_L \big(p^2_{11}+p^2_{22}+p^2_{33}\big) \nonumber \\
& \approx c^{00}_L E + \frac{1}{3}c^{00}_L E \quad ({\rm using\, ultra\, relativistic\, approximation})\nonumber\\
& = \frac{4}{3}c^{00}_L E\,.
\label{eq:c00-1}
\end{align}

\section{Effect of $a_{e\mu}$ and $a_{e\tau}$  on appearance channel $P(\nu_e \rightarrow \nu_\mu)$}
\label{app:Peu_emu_etau}

The CPT-violating LIV parameters $a_{e\mu}$ and $a_{e\tau}$ affects the  appearance channel $(\nu_e \rightarrow \nu_\mu)$ significantly and have contribution at leading orders. The authors in Ref.~\cite{Kopp:2007ne} have presented the approximate analytic expression for $P(\nu_\mu \rightarrow \nu_e)$ in the presence of neutral-current NSI parameters $\varepsilon_{e\mu}$ and $\varepsilon_{e\tau}$ which occur during neutrino propagation.  We use the Eq.~[33] in Ref.~\cite{Kopp:2007ne} to obtain approximate expression for $P(\nu_e \rightarrow \nu_\mu)$ in the presence of the LIV parameters $a_{e\mu}$ and $a_{e\tau}$ where we replace $\varepsilon_{\alpha\beta}$ with $a_{\alpha\beta}$ with the help of Eq.~\ref{eq:NSI_LIV},
\begin{align}
  \varepsilon_{\alpha\beta} = \frac{2a_{\alpha\beta}E_\nu}{a_{\rm CC}}, 
\end{align}
where, $a_{\rm CC} = 2 \sqrt{2} G_F N_e E_\nu$. We assume $\delta_{\rm CP} = 0$ and focus on real $a_{\alpha\beta}$ by taking the LIV phase $\phi_{\alpha\beta} = 0$, and $\pi$ which ensures that both positive and negative values of $a_{\alpha\beta}$ are considered and we have,
\begin{align}
  a_{e\mu} &\equiv |a_{e\mu}|\cos(\phi_{e\mu})\\
  a_{e\tau} &\equiv |a_{e\tau}|\cos(\phi_{e\tau}).
\end{align}
The resulting expression for $P(\nu_e \rightarrow \nu_\mu)$ in the presence of the LIV parameters $a_{e\mu}$ and $a_{e\tau}$ is given as,
{ \footnotesize
  \begin{align}
  P(\nu_e \rightarrow \nu_\mu) &\simeq
  4 \tilde{s}_{13}^{2} s_{23}^{2} \sin^{2} \frac{(\ldm - a_{\rm CC})L_\nu}{4E_\nu}
  \nonumber\\
  &\hspace{0.5cm}
  + \Big( \frac{\sdm}{\ldm} \Big)^2 c_{23}^2 s_{2 \times 12}^2
  \Big( \frac{\ldm}{a_{\rm CC}} \Big)^2 \sin^{2} \frac{a_{\rm CC} L_\nu}{4E_\nu}
  \nonumber\\
  &\hspace{0.5cm}
  - \frac{\sdm}{\ldm} \tilde{s}_{13} s_{2 \times 12} s_{2 \times 23}
  \frac{\ldm}{a_{\rm CC}}
  \left[   \sin^{2} \frac{a_{\rm CC} L_\nu}{4E_\nu}
  - \sin^{2} \frac{\ldm L_\nu}{4E_\nu}
  + \sin^{2} \frac{(\ldm - a_{\rm CC})L_\nu}{4E_\nu}
  \right]                                           \nonumber\\
  &\hspace{0.5 cm}
  - 8 \omega \frac{a_{e\mu} E_\nu}{a_{\rm CC}} \tilde{s}_{13} s_{23} c_{23}^{2}
  \left[   \sin^{2} \frac{a_{\rm CC} L_\nu}{4E_\nu}
  - \sin^{2} \frac{\ldm L_\nu}{4E_\nu}
  + \sin^{2} \frac{(\ldm - a_{\rm CC})L_\nu}{4E_\nu}
  \right]                                           \nonumber\\
  &\hspace{0.5 cm}
  + 16\omega a_{e\mu} E_\nu \tilde{s}_{13} s_{23}^{3} \frac{1}{\ldm - a_{\rm CC}}
  \sin^{2} \frac{(\ldm - a_{\rm CC})L_\nu}{4E_\nu} \nonumber \\
  &\hspace{0.5 cm}
  + 8 \omega\frac{a_{e\tau} E_\nu}{a_{\rm CC}} \tilde{s}_{13} s_{23}^{2} c_{23}
  \left[   \sin^{2} \frac{a_{\rm CC} L_\nu}{4E_\nu}
  - \sin^{2} \frac{\ldm L_\nu}{4E_\nu}
  + \sin^{2} \frac{(\ldm - a_{\rm CC})L_\nu}{4E_\nu}
  \right]                                           \nonumber\\
  &\hspace{0.5 cm}
  + 16 \omega a_{e\tau} E_\nu \tilde{s}_{13} s_{23}^{2} c_{23} \frac{1}{\ldm - a_{\rm CC}}
  \sin^{2} \frac{(\ldm - a_{\rm CC})L_\nu}{4E_\nu}  \nonumber\\
  &\hspace{0.5 cm}
  + 8 \omega a_{e\mu} E_\nu \frac{\sdm}{\ldm} s_{2 \times 12} c_{23}^3 \frac{\ldm}{a_{\rm CC}^2}
  \sin^{2} \frac{a_{\rm CC} L_\nu}{4E_\nu}                 \nonumber\\
  &\hspace{0.5 cm}
  - 4 \omega \frac{a_{e\mu} E_\nu}{a_{\rm CC}} \frac{\sdm}{\ldm} s_{2 \times 12} s_{23}^2 c_{23} \frac{\ldm}{\ldm - a_{\rm CC}}\nonumber \\
  &\hspace{0.5 cm}\times\left[   \sin^{2} \frac{a_{\rm CC} L_\nu}{4E_\nu}
  - \sin^{2} \frac{\ldm L_\nu}{4E_\nu}
  + \sin^{2} \frac{(\ldm - a_{\rm CC}) L_\nu}{4E_\nu}
  \right]                                           \nonumber\\
  &\hspace{0.5 cm}
  - 8 \omega a_{e\tau}E_\nu \frac{\sdm}{\ldm} s_{2 \times 12} s_{23} c_{23}^2 \frac{\ldm}{a_{\rm CC}^2}
  \sin^{2} \frac{a_{\rm CC} L_\nu}{4E_\nu}                 \nonumber\\
  &\hspace{0.5 cm}
  - 4 \omega \frac{a_{e\mu} E_\nu}{a_{\rm CC}} \frac{\sdm}{\ldm} s_{2 \times 12} s_{23} c_{23}^2 \frac{\ldm}{\ldm -a_{\rm CC}} \nonumber\\
  &\hspace{0.5 cm}\times\left[   \sin^{2} \frac{a_{\rm CC} L_\nu}{4E_\nu}
  - \sin^{2} \frac{\ldm L_\nu}{4E_\nu}
  + \sin^{2} \frac{(\ldm - a_{\rm CC}) L_\nu}{4E_\nu}
  \right]                                           \nonumber\\
  &\hspace{0.5 cm}
  + \mathcal{O}\Big( \Big[ \frac{\sdm}{\ldm} \Big]^3 \Big)
  + \mathcal{O}\Big( \Big[ \frac{\sdm}{\ldm} \Big]^2 s_{13} \Big)
  + \mathcal{O}\Big( \frac{\sdm}{\ldm} s_{13}^2 \Big)
  + \mathcal{O} ( s_{13}^3 )                        \nonumber\\
  &\hspace{0.5 cm}
  + \mathcal{O}\Big( a \Big[ \frac{\sdm}{\ldm} \Big]^2 \Big)
  + \mathcal{O}\Big( a s_{13} \frac{\sdm}{\ldm} \Big)
  + \mathcal{O} ( a s_{13}^2 )
  + \mathcal{O} ( a^2 ),
  \label{eq:Pmue-mat-LIV}
  \end{align}
}
where, $s_{ij} =  \sin\theta_{ij}$,
$c_{ij} = \cos\theta_{ij}$, $s_{2\times ij} = \sin 2\theta_{ij}$,
$c_{2\times ij} = \cos 2\theta_{ij}$. For CPT-violating parameter $a_{\alpha\beta}$, the parameter $\omega = +1$ for the case of neutrino whereas for antineutrino $\omega = -1$.  The effective mixing angle $\theta_{13}$ in matter is defined as, 
  \begin{align}
  \tilde{s}_{13} \equiv \frac{\ldm}{\ldm - a_{\rm CC}} s_{13} + \mathcal{O}(s_{13}^2).
  \end{align}

In Equation~\ref{eq:Pmue-mat-LIV}, the first three terms are driven by standard interactions where the first term has the dominant effect. The remaining terms contain the effect of CPT violating-LIV parameters $a_{e\mu}$ and $a_{e\tau}$.  Now let us try to understand the effect of $a_{e\mu}$ on appearance channel $P(\nu_e \rightarrow \nu_\mu)$ from the approximate expression in Eq.~\ref{eq:Pmue-mat-LIV}. We observe that the effect of $a_{e\mu}$ is dominantly contributed by the fifth term, 
\begin{equation}
+ 16\omega a_{e\mu} E_\nu \tilde{s}_{13} s_{23}^{3} \frac{1}{\ldm - a_{\rm CC}}
\sin^{2} \frac{(\ldm - a_{\rm CC})L_\nu}{4E_\nu},
\end{equation}  
where, $(\ldm - a_{\rm CC})$ factor in the denominator causes the matter-driven resonance effect for the case of neutrino. Since this term has positive sign for the case of neutrino ($\omega = +1$), the positive (negative) value of $a_{e\mu}$ increases (decreases) $P(\nu_e \rightarrow \nu_\mu)$. We can see that the effect of $a_{e\mu}$ is larger in the case of neutrino than for antineutrino. This happens because $a_{\rm CC}$ becomes negative for antineutrino, and matter-driven resonance condition is not fulfilled, due to which the above-mentioned term does not contribute significantly. A similar effect is also observed for the case of $a_{e\tau}$ which can be explained using the seventh term in Eq.~\ref{eq:Pmue-mat-LIV}.

\section{Effective regions in $(E_\mu^\text{rec}, \cos\theta_\mu^\text{rec})$ plane to constrain $a_{e\mu}$ and $a_{e\tau}$}
\label{app:effective_regions_emu_etau}

\begin{figure}[t]
  \centering
  \includegraphics[width=0.45\textwidth]{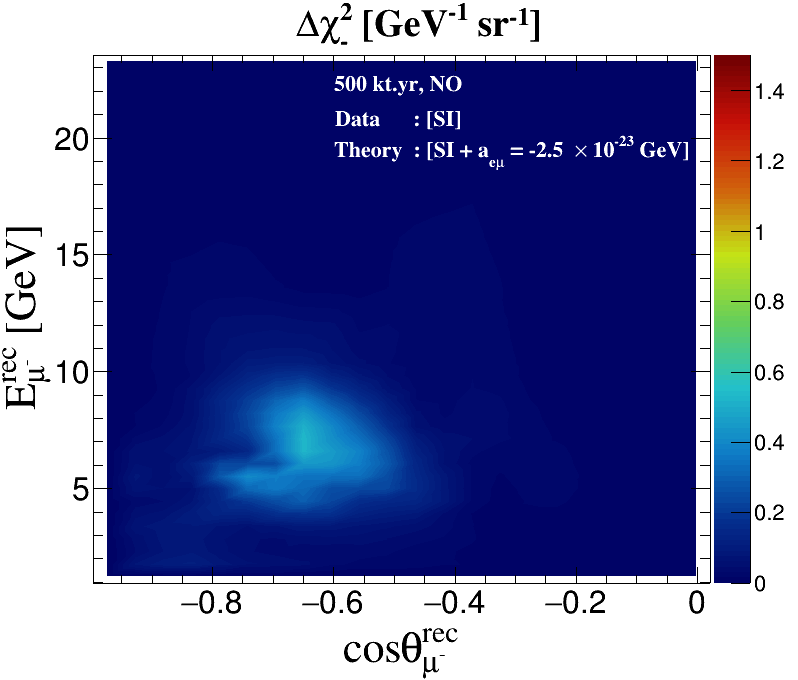}
  \includegraphics[width=0.45\textwidth]{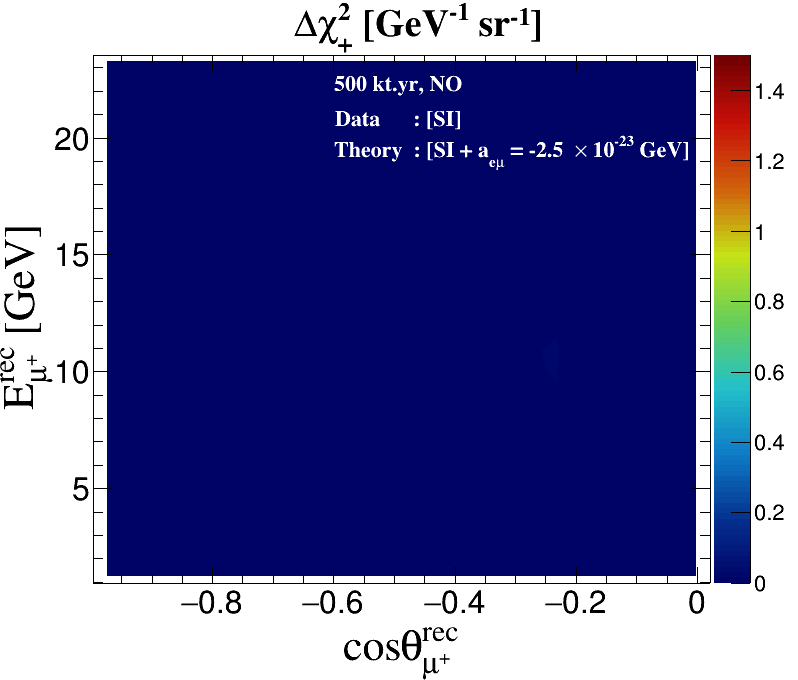}
  \includegraphics[width=0.45\textwidth]{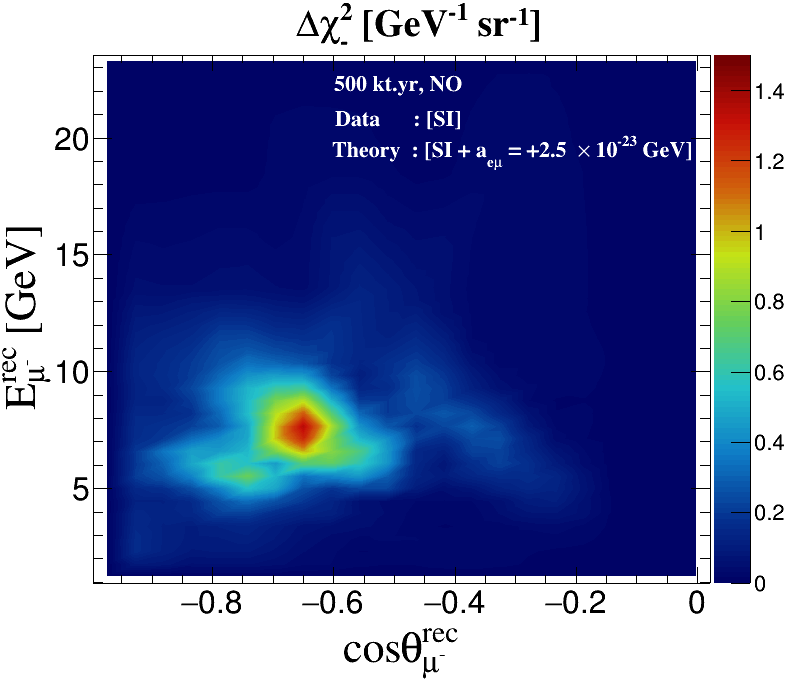}
  \includegraphics[width=0.45\textwidth]{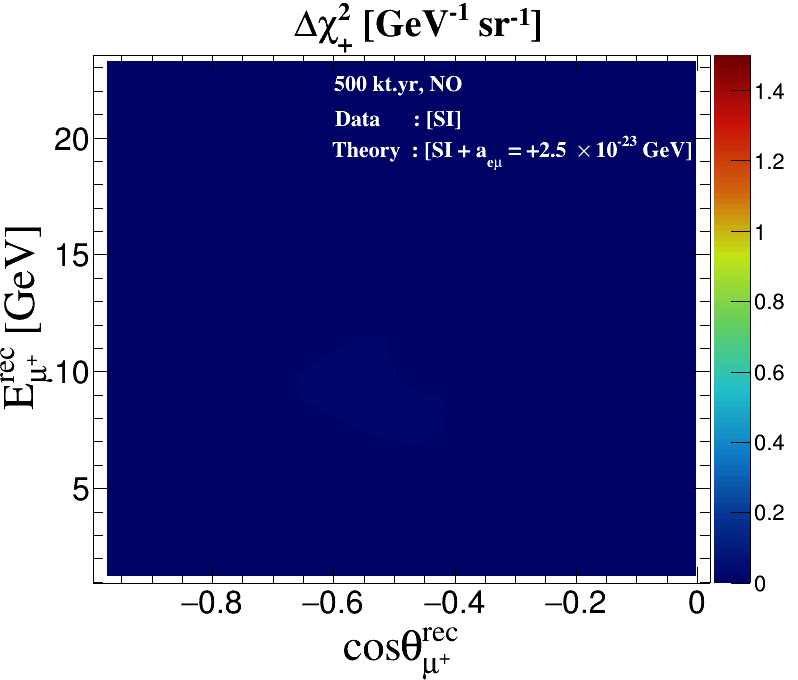}
  \mycaption{The distribution of fixed-parameter $\Delta\chi^2_-$ ($\Delta\chi^2_+$) in the plane of ($E_\mu^\text{rec}$, $\cos\theta_\mu^\text{rec}$) without pull penalty term  using 500 kt$\cdot$yr exposure of the ICAL detector as shown in the left (right) panels. Note that $\Delta\chi^2_-$ and $\Delta\chi^2_+$ are presented in the units of $\text{GeV}^{-1} \text{sr}^{-1}$ where we have divided them by 2$\pi ~ \times$ bin area. In data, $a_{e\mu} = 0$ with NO (true) using the benchmark value of oscillation parameters given in Table~\ref{tab:osc-param-value}. In theory, $a_{e\mu} = -2.5 \times 10^{-23} ~ \text{GeV}$ and $2.5 \times 10^{-23} ~ \text{GeV}$ in the top and bottom panels, respectively. Here, in the fixed-parameter scenario, we do not marginalize over any oscillation parameter in theory, but we do marginalize over systematic uncertainties.
  }
  \label{fig:chisq_contour_emu}
\end{figure}

\begin{figure}[t]
  \centering
  \includegraphics[width=0.45\textwidth]{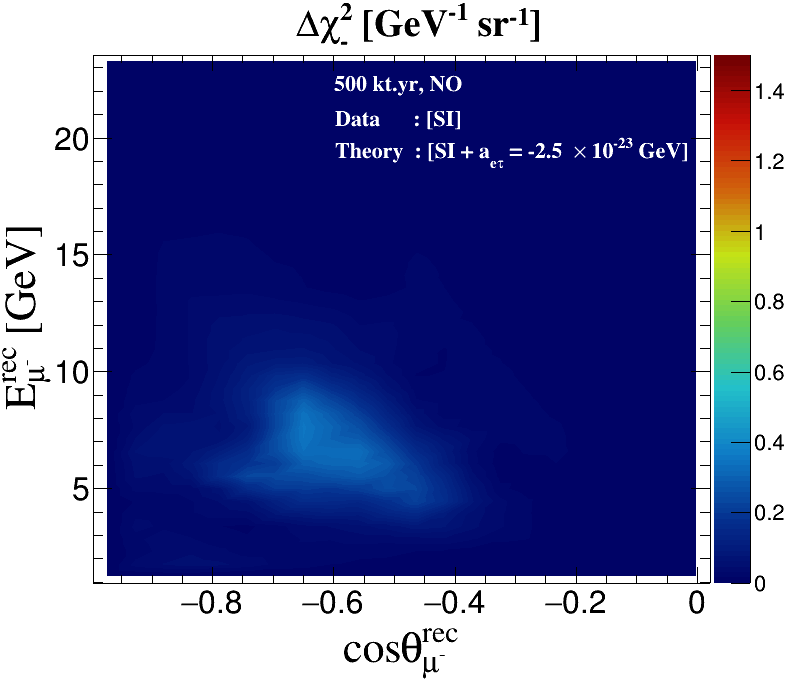}
  \includegraphics[width=0.45\textwidth]{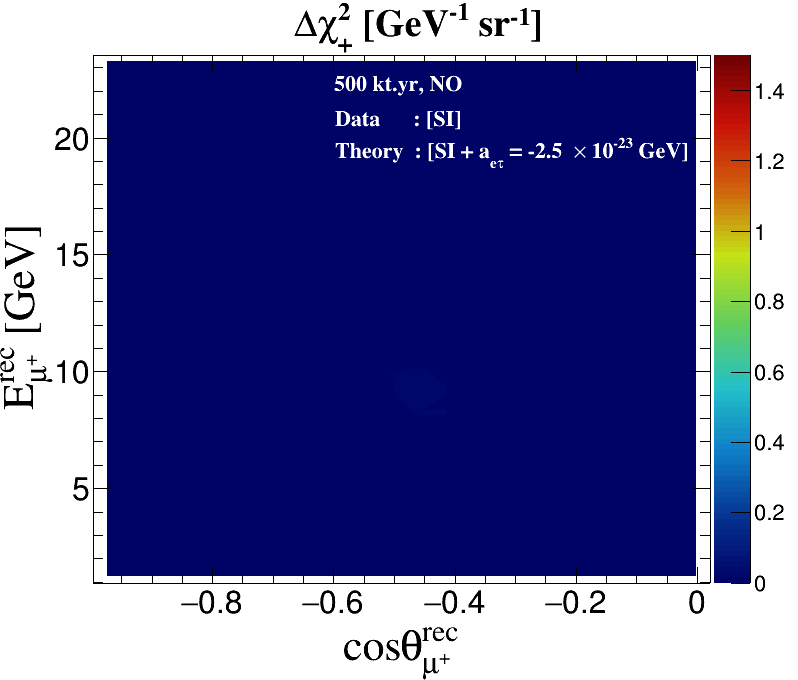}
  \includegraphics[width=0.45\textwidth]{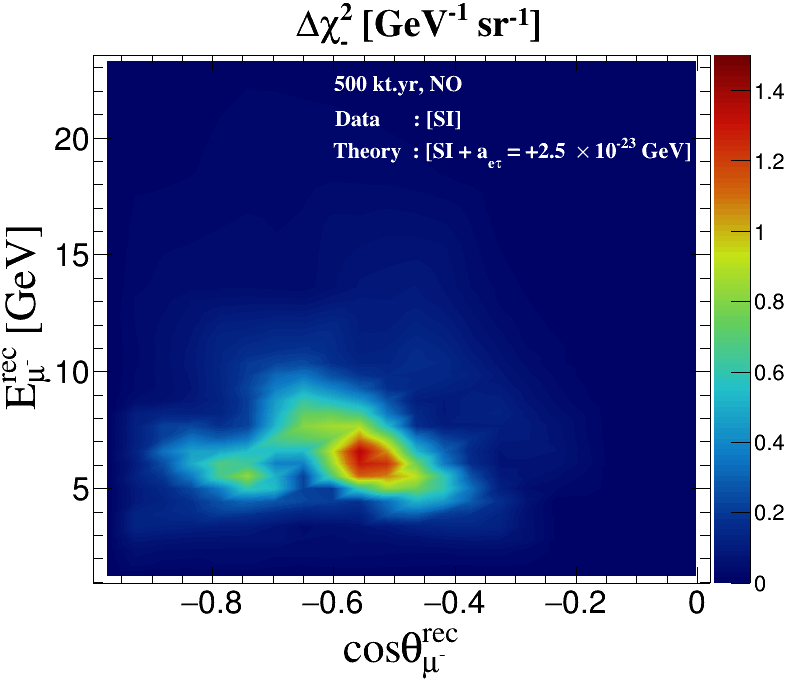}
  \includegraphics[width=0.45\textwidth]{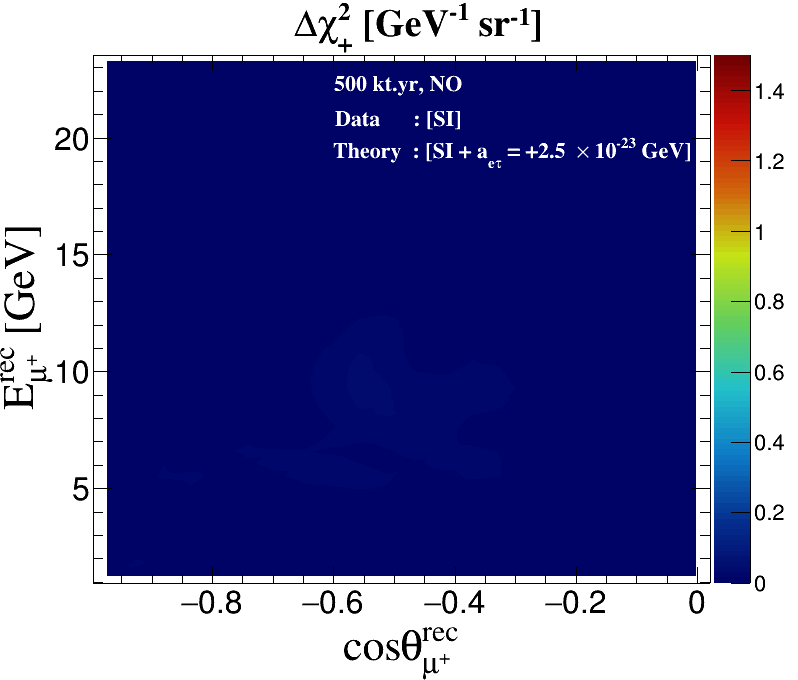}
  \mycaption{The distribution of fixed-parameter $\Delta\chi^2_-$ ($\Delta\chi^2_+$) in the plane of ($E_\mu^\text{rec}$, $\cos\theta_\mu^\text{rec}$) without pull penalty term  using 500 kt$\cdot$yr exposure of the ICAL detector as shown in the left (right) panels. Note that $\Delta\chi^2_-$ and $\Delta\chi^2_+$ are presented in the units of $\text{GeV}^{-1} \text{sr}^{-1}$ where we have divided them by 2$\pi ~ \times$ bin area. In data, $a_{e\tau} = 0$ with NO (true) using the benchmark value of oscillation parameters given in Table~\ref{tab:osc-param-value}. In theory, $a_{e\tau} = -2.5 \times 10^{-23} ~ \text{GeV}$ and $2.5 \times 10^{-23} ~ \text{GeV}$ in the top and bottom panels, respectively. Here, in the fixed-parameter scenario, we do not marginalize over any oscillation parameter in theory, but we do marginalize over systematic uncertainties.
  }
  \label{fig:chisq_contour_etau}
\end{figure}

In this section, we identify the regions in $(E_\mu^\text{rec}, \cos\theta_\mu^\text{rec})$ plane which contribute significantly towards the sensitivity of ICAL for the CPT-violating LIV parameters $a_{e\mu}$ and $a_{e\tau}$ with 500 kt$\cdot$yr exposure. Note that here, we consider one LIV parameter at-a-time. In Fig.~\ref{fig:chisq_contour_emu}, we show the distribution of sensitivity of ICAL for $a_{e\mu}$ in terms of fixed-parameter $\Delta \chi^2_-$ ($\Delta \chi^2_+$) without pull penalty term contributed from reconstructed $\mu^-$ ($\mu^+$) events in $(E_\mu^\text{rec}, \cos\theta_\mu^\text{rec})$ plane following the same procedure as described in section~\ref{sec:effective_regions} for $a_{\mu\tau}$. While estimating sensitivity of ICAL for $a_{e\mu}$ in the top (bottom) panels in Fig.~\ref{fig:chisq_contour_emu}, we assume non-zero value of $a_{e\mu} = -2.5 \times 10^{-23} ~ \text{GeV}$ ($+2.5 \times 10^{-23} ~ \text{GeV}$) in theory while taking $a_{e\mu} = 0$ in data for the case of SI. We keep the oscillation parameters fixed in theory and data at their benchmark values as mentioned in Table~\ref{tab:osc-param-value}. We consider 500 kt$\cdot$yr exposure at ICAL with NO as true mass ordering. We use the binning scheme mentioned in Table~\ref{tab:binning_scheme} where we add contribution from all ${E'}_{\rm had}^{\rm rec}$ bins for each $(E_\mu^\text{rec}, \cos\theta_\mu^\text{rec})$ bins. The left and right panels show the distribution of $\Delta \chi^2_-$ and $\Delta \chi^2_+$, respectively.

We can observe in Fig.~\ref{fig:chisq_contour_emu} that the contribution towards the sensitivity of ICAL for $a_{e\mu}$ mainly stems from $\mu^-$ as shown in left panels, whereas the contribution from $\mu^+$ is negligible as shown in the right panels. This happens because the term with a significant contribution of $a_{e\mu}$ has a matter-driven resonance effect for the case of neutrino as described in section~\ref{sec:oscillograms} while explaining the oscillograms in Fig.~\ref{fig:oscillograms_emu}. We also observe in the left panels in Fig.~\ref{fig:chisq_contour_emu} that $\Delta \chi^2_-$ is significantly larger for the positive value (top left panel) of $a_{e\mu}$ compared to that for negative value (bottom left panel). This observation is also consistent with the asymmetric effect of $a_{e\mu}$ on $\nu_\mu$ survival oscillograms shown in Fig.~\ref{fig:oscillograms_emu}, and the reason behind this effect is explained in detail in section~\ref{sec:oscillograms}. We would like to highlight the observation that the sensitivity of ICAL for $a_{e\mu}$ is significantly contributed by the reconstruction muon energy and direction bins corresponding to the region of matter effect, which is not surprising because the impact of $a_{e\mu}$ is being driven by the resonance effect with the matter as explained in section~\ref{sec:oscillograms}.

Similar to the case of $a_{e\mu}$, we present the distribution of sensitivity of ICAL for $a_{e\tau}$ in terms of fixed-parameter $\Delta \chi^2_-$ ($\Delta \chi^2_+$) with contribution from reconstructed $\mu^-$ ($\mu^+$) events in $(E_\mu^\text{rec}, \cos\theta_\mu^\text{rec})$ plane in Fig.~\ref{fig:chisq_contour_etau}. To estimate the sensitivity of ICAL for $a_{e\tau}$ in the top (bottom) panels in Fig.~\ref{fig:chisq_contour_etau}, we take non-zero value of $a_{e\tau} = -2.5 \times 10^{-23} ~ \text{GeV}$ ($+2.5 \times 10^{-23} ~ \text{GeV}$) in theory while keeping $a_{e\tau} = 0$ in data for the case of SI. Analogous to the case of $a_{e\mu}$, we observe that the sensitivity for $a_{e\tau}$ is more in the case of $\mu^-$ (left panels) than that for $\mu^+$ (right panels). The asymmetric effect of $a_{e\tau}$ leads to higher sensitivity for positive $a_{e\tau}$ as shown in the bottom right panel of Fig.~\ref{fig:chisq_contour_etau}. These features can be explained in the same fashion as we did for $a_{e\mu}$ in the previous paragraph. The reasons behind these effects of $a_{e\tau}$ on $\nu_\mu$ survival oscillograms are described in section~\ref{sec:oscillograms}.

\end{appendix}

\bibliographystyle{JHEP}
\bibliography{LIV-References.bib}

\providecommand{\href}[2]{#2}\begingroup\raggedright\begin{thebibliography}{100}

\bibitem{Gambini:1998it}
R.~Gambini and J.~Pullin, {\it {Nonstandard optics from quantum space-time}},
  {\em Phys. Rev. D} {\bf 59} (1999) 124021,
  [\href{http://arxiv.org/abs/gr-qc/9809038}{{\tt gr-qc/9809038}}].

\bibitem{Alfaro:2002xz}
J.~Alfaro, H.~A. Morales-Tecotl, and L.~F. Urrutia, {\it {Quantum gravity and
  spin 1/2 particles effective dynamics}},  {\em Phys. Rev. D} {\bf 66} (2002)
  124006, [\href{http://arxiv.org/abs/hep-th/0208192}{{\tt hep-th/0208192}}].

\bibitem{Sudarsky:2002ue}
D.~Sudarsky, L.~Urrutia, and H.~Vucetich, {\it {New observational bounds to
  quantum gravity signals}},  {\em Phys. Rev. Lett.} {\bf 89} (2002) 231301,
  [\href{http://arxiv.org/abs/gr-qc/0204027}{{\tt gr-qc/0204027}}].

\bibitem{Amelino-Camelia:2002aqz}
G.~Amelino-Camelia, {\it {Quantum gravity phenomenology: Status and
  prospects}},  {\em Mod. Phys. Lett. A} {\bf 17} (2002) 899--922,
  [\href{http://arxiv.org/abs/gr-qc/0204051}{{\tt gr-qc/0204051}}].

\bibitem{Ng:2003jk}
Y.~J. Ng, {\it {Selected topics in Planck scale physics}},  {\em Mod. Phys.
  Lett. A} {\bf 18} (2003) 1073--1098,
  [\href{http://arxiv.org/abs/gr-qc/0305019}{{\tt gr-qc/0305019}}].

\bibitem{Polyakov:1987ez}
A.~M. Polyakov, {\em {Gauge Fields and Strings}}, vol.~3 of {\em {Contemporary
  concepts in physics}}.
\newblock {Harwood Academic Publishers}, 1987.

\bibitem{Kostelecky:1988zi}
V.~A. Kostelecky and S.~Samuel, {\it {Spontaneous Breaking of Lorentz Symmetry
  in String Theory}},  {\em Phys. Rev. D} {\bf 39} (1989) 683.

\bibitem{Kostelecky:1991ak}
V.~A. Kostelecky and R.~Potting, {\it {CPT and strings}},  {\em Nucl. Phys. B}
  {\bf 359} (1991) 545--570.

\bibitem{Kostelecky:1995qk}
V.~A. Kostelecky and R.~Potting, {\it {Expectation values, Lorentz invariance,
  and CPT in the open bosonic string}},  {\em Phys. Lett. B} {\bf 381} (1996)
  89--96, [\href{http://arxiv.org/abs/hep-th/9605088}{{\tt hep-th/9605088}}].

\bibitem{Kostelecky:2000hz}
V.~A. Kostelecky and R.~Potting, {\it {Analytical construction of a
  nonperturbative vacuum for the open bosonic string}},  {\em Phys. Rev. D}
  {\bf 63} (2001) 046007, [\href{http://arxiv.org/abs/hep-th/0008252}{{\tt
  hep-th/0008252}}].

\bibitem{Kostelecky:1999mu}
V.~A. Kostelecky, M.~Perry, and R.~Potting, {\it {Off-shell structure of the
  string sigma model}},  {\em Phys. Rev. Lett.} {\bf 84} (2000) 4541--4544,
  [\href{http://arxiv.org/abs/hep-th/9912243}{{\tt hep-th/9912243}}].

\bibitem{Carroll:2001ws}
S.~M. Carroll, J.~A. Harvey, V.~A. Kostelecky, C.~D. Lane, and T.~Okamoto, {\it
  {Noncommutative field theory and Lorentz violation}},  {\em Phys. Rev. Lett.}
  {\bf 87} (2001) 141601, [\href{http://arxiv.org/abs/hep-th/0105082}{{\tt
  hep-th/0105082}}].

\bibitem{Mocioiu:2001fx}
I.~Mocioiu, M.~Pospelov, and R.~Roiban, {\it {Breaking CPT by mixed
  noncommutativity}},  {\em Phys. Rev. D} {\bf 65} (2002) 107702,
  [\href{http://arxiv.org/abs/hep-ph/0108136}{{\tt hep-ph/0108136}}].

\bibitem{Greenberg:2002uu}
O.~W. Greenberg, {\it {CPT violation implies violation of Lorentz invariance}},
   {\em Phys. Rev. Lett.} {\bf 89} (2002) 231602,
  [\href{http://arxiv.org/abs/hep-ph/0201258}{{\tt hep-ph/0201258}}].

\bibitem{Myers:2003fd}
R.~C. Myers and M.~Pospelov, {\it {Ultraviolet modifications of dispersion
  relations in effective field theory}},  {\em Phys. Rev. Lett.} {\bf 90}
  (2003) 211601, [\href{http://arxiv.org/abs/hep-ph/0301124}{{\tt
  hep-ph/0301124}}].

\bibitem{Mavromatos:2003qc}
N.~E. Mavromatos, {\it {Theoretical and phenomenological aspects of CPT
  violation}},  {\em Nucl. Instrum. Meth. B} {\bf 214} (2004) 1--6,
  [\href{http://arxiv.org/abs/hep-ph/0305215}{{\tt hep-ph/0305215}}].

\bibitem{Colladay:1998fq}
D.~Colladay and V.~A. Kostelecky, {\it {Lorentz violating extension of the
  standard model}},  {\em Phys. Rev. D} {\bf 58} (1998) 116002,
  [\href{http://arxiv.org/abs/hep-ph/9809521}{{\tt hep-ph/9809521}}].

\bibitem{Kostelecky:2003fs}
V.~A. Kostelecky, {\it {Gravity, Lorentz violation, and the standard model}},
  {\em Phys. Rev. D} {\bf 69} (2004) 105009,
  [\href{http://arxiv.org/abs/hep-th/0312310}{{\tt hep-th/0312310}}].

\bibitem{Bluhm:2005uj}
R.~Bluhm, {\it {Overview of the SME: Implications and phenomenology of Lorentz
  violation}},  {\em Lect. Notes Phys.} {\bf 702} (2006) 191--226,
  [\href{http://arxiv.org/abs/hep-ph/0506054}{{\tt hep-ph/0506054}}].

\bibitem{Colladay:1996iz}
D.~Colladay and V.~A. Kostelecky, {\it {CPT violation and the standard model}},
   {\em Phys. Rev. D} {\bf 55} (1997) 6760--6774,
  [\href{http://arxiv.org/abs/hep-ph/9703464}{{\tt hep-ph/9703464}}].

\bibitem{Pontecorvo:1967fh}
B.~Pontecorvo, {\it {Neutrino Experiments and the Problem of Conservation of
  Leptonic Charge}},  {\em Zh. Eksp. Teor. Fiz.} {\bf 53} (1967) 1717--1725.

\bibitem{Super-Kamiokande:2014exs}
{\bf Super-Kamiokande} Collaboration, K.~Abe et~al., {\it {Test of Lorentz
  invariance with atmospheric neutrinos}},  {\em Phys. Rev. D} {\bf 91} (2015),
  no.~5 052003, [\href{http://arxiv.org/abs/1410.4267}{{\tt arXiv:1410.4267}}].

\bibitem{Diaz:2015dxa}
J.~S. Diaz, {\it {Correspondence between nonstandard interactions and CPT
  violation in neutrino oscillations}},
  \href{http://arxiv.org/abs/1506.01936}{{\tt arXiv:1506.01936}}.

\bibitem{Arguelles:2016rkg}
C.~A. Arg\"uelles, G.~H. Collin, J.~M. Conrad, T.~Katori, and A.~Kheirandish,
  {\it {Search for Lorentz Violation in km$^3$-Scale Neutrino Telescopes}},  in
  {\em {7th Meeting on CPT and Lorentz Symmetry}}, pp.~153--156, 2017.
\newblock \href{http://arxiv.org/abs/1608.02946}{{\tt arXiv:1608.02946}}.

\bibitem{IceCube:2017qyp}
{\bf IceCube} Collaboration, M.~G. Aartsen et~al., {\it {Neutrino
  Interferometry for High-Precision Tests of Lorentz Symmetry with IceCube}},
  {\em Nature Phys.} {\bf 14} (2018), no.~9 961--966,
  [\href{http://arxiv.org/abs/1709.03434}{{\tt arXiv:1709.03434}}].

\bibitem{Katori:2019xpc}
{\bf IceCube} Collaboration, T.~Katori, C.~A. Arg\"uelles, K.~Farrag, and
  S.~Mandalia, {\it {Test of Lorentz Violation with Astrophysical Neutrino
  Flavor at IceCube}},  in {\em {8th Meeting on CPT and Lorentz Symmetry}},
  pp.~166--169, 2020.
\newblock \href{http://arxiv.org/abs/1906.09240}{{\tt arXiv:1906.09240}}.

\bibitem{KumarAgarwalla:2019gdj}
S.~K. Agarwalla and M.~Masud, {\it {Can Lorentz invariance violation affect the
  sensitivity of deep underground neutrino experiment?}},  {\em Eur. Phys. J.
  C} {\bf 80} (2020), no.~8 716, [\href{http://arxiv.org/abs/1912.13306}{{\tt
  arXiv:1912.13306}}].

\bibitem{Fukuda:1998mi}
{\bf Super-Kamiokande} Collaboration, Y.~Fukuda et~al., {\it {Evidence for
  oscillation of atmospheric neutrinos}},  {\em Phys. Rev. Lett.} {\bf 81}
  (1998) 1562, [\href{http://arxiv.org/abs/hep-ex/9807003}{{\tt
  hep-ex/9807003}}].

\bibitem{Wolfenstein:1977ue}
L.~Wolfenstein, {\it {Neutrino Oscillations in Matter}},  {\em Phys. Rev. D}
  {\bf 17} (1978) 2369--2374.

\bibitem{Mikheev:1986gs}
S.~P. Mikheev and A.~Y. Smirnov, {\it {Resonance enhancement of oscillations in
  matter and solar neutrino spectroscopy}},  {\em Sov. J. Nucl. Phys.} {\bf 42}
  (1985) 913. [Yad.Fiz.42:1441-1448,1985].

\bibitem{Mikheev:1986wj}
S.~P. Mikheev and A.~Y. Smirnov, {\it {Resonant amplification of neutrino
  oscillations in matter and solar neutrino spectroscopy}},  {\em Nuovo Cim. C}
  {\bf 9} (1986) 17--26.

\bibitem{Petcov:1998su}
S.~Petcov, {\it {Diffractive - like (or parametric resonance - like?)
  enhancement of the earth (day - night) effect for solar neutrinos crossing
  the earth core}},  {\em Phys. Lett. B} {\bf 434} (1998) 321,
  [\href{http://arxiv.org/abs/hep-ph/9805262}{{\tt hep-ph/9805262}}].

\bibitem{Chizhov:1998ug}
M.~Chizhov, M.~Maris, and S.~T. Petcov, {\it {On the oscillation length
  resonance in the transitions of solar and atmospheric neutrinos crossing the
  earth core}},  \href{http://arxiv.org/abs/hep-ph/9810501}{{\tt
  hep-ph/9810501}}.

\bibitem{Petcov:1998sg}
S.~T. Petcov, {\it {New enhancement mechanism of the transitions in the earth
  of the solar and atmospheric neutrinos crossing the earth core}},  {\em Nucl.
  Phys. B Proc. Suppl.} {\bf 77} (1999) 93--97,
  [\href{http://arxiv.org/abs/hep-ph/9809587}{{\tt hep-ph/9809587}}].

\bibitem{Chizhov:1999az}
M.~Chizhov and S.~Petcov, {\it {New conditions for a total neutrino conversion
  in a medium}},  {\em Phys.Rev.Lett.} {\bf 83} (1999) 1096--1099,
  [\href{http://arxiv.org/abs/hep-ph/9903399}{{\tt hep-ph/9903399}}].

\bibitem{Chizhov:1999he}
M.~V. Chizhov and S.~T. Petcov, {\it {Enhancing mechanisms of neutrino
  transitions in a medium of nonperiodic constant density layers and in the
  earth}},  {\em Phys. Rev. D} {\bf 63} (2001) 073003,
  [\href{http://arxiv.org/abs/hep-ph/9903424}{{\tt hep-ph/9903424}}].

\bibitem{Akhmedov:1998ui}
E.~K. Akhmedov, {\it {Parametric resonance of neutrino oscillations and passage
  of solar and atmospheric neutrinos through the earth}},  {\em Nucl. Phys.}
  {\bf B538} (1999) 25, [\href{http://arxiv.org/abs/hep-ph/9805272}{{\tt
  hep-ph/9805272}}].

\bibitem{Akhmedov:1998xq}
E.~K. Akhmedov, A.~Dighe, P.~Lipari, and A.~Smirnov, {\it {Atmospheric
  neutrinos at Super-Kamiokande and parametric resonance in neutrino
  oscillations}},  {\em Nucl. Phys.} {\bf B542} (1999) 3,
  [\href{http://arxiv.org/abs/hep-ph/9808270}{{\tt hep-ph/9808270}}].

\bibitem{Guzzo:1991hi}
M.~M. Guzzo, A.~Masiero, and S.~T. Petcov, {\it {On the MSW effect with
  massless neutrinos and no mixing in the vacuum}},  {\em Phys. Lett. B} {\bf
  260} (1991) 154--160.

\bibitem{Super-Kamiokande:2011dam}
{\bf Super-Kamiokande} Collaboration, G.~Mitsuka et~al., {\it {Study of
  Non-Standard Neutrino Interactions with Atmospheric Neutrino Data in
  Super-Kamiokande I and II}},  {\em Phys. Rev. D} {\bf 84} (2011) 113008,
  [\href{http://arxiv.org/abs/1109.1889}{{\tt arXiv:1109.1889}}].

\bibitem{Ohlsson:2012kf}
T.~Ohlsson, {\it {Status of non-standard neutrino interactions}},  {\em Rept.
  Prog. Phys.} {\bf 76} (2013) 044201,
  [\href{http://arxiv.org/abs/1209.2710}{{\tt arXiv:1209.2710}}].

\bibitem{Esmaili:2013fva}
A.~Esmaili and A.~Y. Smirnov, {\it {Probing Non-Standard Interaction of
  Neutrinos with IceCube and DeepCore}},  {\em JHEP} {\bf 06} (2013) 026,
  [\href{http://arxiv.org/abs/1304.1042}{{\tt arXiv:1304.1042}}].

\bibitem{Miranda:2015dra}
O.~G. Miranda and H.~Nunokawa, {\it {Non standard neutrino interactions:
  current status and future prospects}},  {\em New J. Phys.} {\bf 17} (2015),
  no.~9 095002, [\href{http://arxiv.org/abs/1505.06254}{{\tt
  arXiv:1505.06254}}].

\bibitem{Salvado:2016uqu}
J.~Salvado, O.~Mena, S.~Palomares-Ruiz, and N.~Rius, {\it {Non-standard
  interactions with high-energy atmospheric neutrinos at IceCube}},  {\em JHEP}
  {\bf 01} (2017) 141, [\href{http://arxiv.org/abs/1609.03450}{{\tt
  arXiv:1609.03450}}].

\bibitem{IceCube:2017zcu}
{\bf IceCube} Collaboration, M.~G. Aartsen et~al., {\it {Search for Nonstandard
  Neutrino Interactions with IceCube DeepCore}},  {\em Phys. Rev. D} {\bf 97}
  (2018), no.~7 072009, [\href{http://arxiv.org/abs/1709.07079}{{\tt
  arXiv:1709.07079}}].

\bibitem{Farzan:2017xzy}
Y.~Farzan and M.~Tortola, {\it {Neutrino oscillations and Non-Standard
  Interactions}},  {\em Front.in Phys.} {\bf 6} (2018) 10,
  [\href{http://arxiv.org/abs/1710.09360}{{\tt arXiv:1710.09360}}].

\bibitem{Esteban:2018ppq}
I.~Esteban, M.~C. Gonzalez-Garcia, M.~Maltoni, I.~Martinez-Soler, and
  J.~Salvado, {\it {Updated constraints on non-standard interactions from
  global analysis of oscillation data}},  {\em JHEP} {\bf 08} (2018) 180,
  [\href{http://arxiv.org/abs/1805.04530}{{\tt arXiv:1805.04530}}]. [Addendum:
  JHEP 12, 152 (2020)].

\bibitem{Bhupal:2019qno}
P.~S. Bhupal~Dev et~al., {\it {Neutrino Non-Standard Interactions: A Status
  Report}},  {\em SciPost Phys. Proc.} (2019) 1,
  [\href{http://arxiv.org/abs/1907.00991}{{\tt arXiv:1907.00991}}].

\bibitem{Khatun:2019tad}
A.~Khatun, S.~S. Chatterjee, T.~Thakore, and S.~K. Agarwalla, {\it {Enhancing
  sensitivity to non-standard neutrino interactions at INO combining muon and
  hadron information}},  {\em Eur. Phys. J. C} {\bf 80} (2020), no.~6 533,
  [\href{http://arxiv.org/abs/1907.02027}{{\tt arXiv:1907.02027}}].

\bibitem{Arguelles:2019xgp}
C.~A. Arg\"uelles et~al., {\it {New opportunities at the next-generation
  neutrino experiments I: BSM neutrino physics and dark matter}},  {\em Rept.
  Prog. Phys.} {\bf 83} (2020), no.~12 124201,
  [\href{http://arxiv.org/abs/1907.08311}{{\tt arXiv:1907.08311}}].

\bibitem{KhanChowdhury:2020xev}
{\bf KM3NeT} Collaboration, N.~Khan~Chowdhury, T.~Thakore, J.~B. Coelho,
  J.~Zornoza, and S.~Navas, {\it {Sensitivity to Non-Standard Interactions
  (NSI) with KM3NeT-ORCA}},  {\em PoS} {\bf ICRC2019} (2020) 931.

\bibitem{Blot:2020ekg}
{\bf IceCube} Collaboration, S.~Blot, {\it {Probing Beyond Standard Model
  Physics via Oscillations with IceCube DeepCore}},  {\em J. Phys. Conf. Ser.}
  {\bf 1468} (2020), no.~1 012168.

\bibitem{Kumar:2021lrn}
A.~Kumar, A.~Khatun, S.~K. Agarwalla, and A.~Dighe, {\it {A New Approach to
  Probe Non-Standard Interactions in Atmospheric Neutrino Experiments}},  {\em
  JHEP} {\bf 04} (2021) 159, [\href{http://arxiv.org/abs/2101.02607}{{\tt
  arXiv:2101.02607}}].

\bibitem{KumarAgarwalla:2021twp}
S.~K. Agarwalla, S.~Das, M.~Masud, and P.~Swain, {\it {Evolution of Neutrino
  Mass-Mixing Parameters in Matter with Non-Standard Interactions}},
  \href{http://arxiv.org/abs/2103.13431}{{\tt arXiv:2103.13431}}.

\bibitem{IceCube:2021abg}
{\bf IceCube} Collaboration, R.~Abbasi et~al., {\it {All-flavor constraints on
  nonstandard neutrino interactions and generalized matter potential with three
  years of IceCube DeepCore data}},
  \href{http://arxiv.org/abs/2106.07755}{{\tt arXiv:2106.07755}}.

\bibitem{HernandezRey:2021qac}
{\bf ANTARES, KM3NeT} Collaboration, J.~J. Hern\'andez~Rey, N.~R.
  Khan~Chowdhury, J.~Manczak, S.~Navas, and J.~D. Zornoza, {\it {Search for
  neutrino non-standard interactions with ANTARES and KM3NeT-ORCA}},  in {\em
  {9th Very Large Volume Neutrino Telescopes Workshop 2021}}, 7, 2021.
\newblock \href{http://arxiv.org/abs/2107.14296}{{\tt arXiv:2107.14296}}.

\bibitem{Razzaque:2011ab}
S.~Razzaque and A.~{\relax Yu}. Smirnov, {\it {Searching for sterile neutrinos
  in ice}},  {\em JHEP} {\bf 07} (2011) 084,
  [\href{http://arxiv.org/abs/1104.1390}{{\tt arXiv:1104.1390}}].

\bibitem{Abazajian:2012ys}
K.~N. Abazajian et~al., {\it {Light Sterile Neutrinos: A White Paper}},
  \href{http://arxiv.org/abs/1204.5379}{{\tt arXiv:1204.5379}}.

\bibitem{Super-Kamiokande:2014ndf}
{\bf Super-Kamiokande} Collaboration, K.~Abe et~al., {\it {Limits on sterile
  neutrino mixing using atmospheric neutrinos in Super-Kamiokande}},  {\em
  Phys. Rev. D} {\bf 91} (2015) 052019,
  [\href{http://arxiv.org/abs/1410.2008}{{\tt arXiv:1410.2008}}].

\bibitem{IceCube:2016rnb}
{\bf IceCube} Collaboration, M.~G. Aartsen et~al., {\it {Searches for Sterile
  Neutrinos with the IceCube Detector}},  {\em Phys. Rev. Lett.} {\bf 117}
  (2016), no.~7 071801, [\href{http://arxiv.org/abs/1605.01990}{{\tt
  arXiv:1605.01990}}].

\bibitem{IceCube:2017ivd}
{\bf IceCube} Collaboration, M.~G. Aartsen et~al., {\it {Search for sterile
  neutrino mixing using three years of IceCube DeepCore data}},  {\em Phys.
  Rev. D} {\bf 95} (2017), no.~11 112002,
  [\href{http://arxiv.org/abs/1702.05160}{{\tt arXiv:1702.05160}}].

\bibitem{Blennow:2018hto}
M.~Blennow, E.~Fernandez-Martinez, J.~Gehrlein, J.~Hernandez-Garcia, and
  J.~Salvado, {\it {IceCube bounds on sterile neutrinos above 10 eV}},  {\em
  Eur. Phys. J. C} {\bf 78} (2018), no.~10 807,
  [\href{http://arxiv.org/abs/1803.02362}{{\tt arXiv:1803.02362}}].

\bibitem{Denton:2018dqq}
P.~B. Denton, Y.~Farzan, and I.~M. Shoemaker, {\it {Activating the fourth
  neutrino of the 3+1 scheme}},  {\em Phys. Rev. D} {\bf 99} (2019), no.~3
  035003, [\href{http://arxiv.org/abs/1811.01310}{{\tt arXiv:1811.01310}}].

\bibitem{Miranda:2018buo}
L.~S. Miranda and S.~Razzaque, {\it {Revisiting constraints on 3 + 1
  active-sterile neutrino mixing using IceCube data}},  {\em JHEP} {\bf 03}
  (2019) 203, [\href{http://arxiv.org/abs/1812.00831}{{\tt arXiv:1812.00831}}].

\bibitem{ANTARES:2018rtf}
{\bf ANTARES} Collaboration, A.~Albert et~al., {\it {Measuring the atmospheric
  neutrino oscillation parameters and constraining the 3+1 neutrino model with
  ten years of ANTARES data}},  {\em JHEP} {\bf 06} (2019) 113,
  [\href{http://arxiv.org/abs/1812.08650}{{\tt arXiv:1812.08650}}].

\bibitem{Moulai:2019gpi}
M.~H. Moulai, C.~A. Arg\"uelles, G.~H. Collin, J.~M. Conrad, A.~Diaz, and M.~H.
  Shaevitz, {\it {Combining Sterile Neutrino Fits to Short Baseline Data with
  IceCube Data}},  {\em Phys. Rev. D} {\bf 101} (2020), no.~5 055020,
  [\href{http://arxiv.org/abs/1910.13456}{{\tt arXiv:1910.13456}}].

\bibitem{KhanChowdhury:2020qqu}
{\bf KM3NeT} Collaboration, N.~R. Khan~Chowdhury, {\it {Neutrino Oscillations
  and Non-standard Interactions with KM3NeT-ORCA}},  in {\em {Prospects in
  Neutrino Physics}}, 4, 2020.
\newblock \href{http://arxiv.org/abs/2004.05004}{{\tt arXiv:2004.05004}}.

\bibitem{IceCube:2020phf}
{\bf IceCube} Collaboration, M.~G. Aartsen et~al., {\it {eV-Scale Sterile
  Neutrino Search Using Eight Years of Atmospheric Muon Neutrino Data from the
  IceCube Neutrino Observatory}},  {\em Phys. Rev. Lett.} {\bf 125} (2020),
  no.~14 141801, [\href{http://arxiv.org/abs/2005.12942}{{\tt
  arXiv:2005.12942}}].

\bibitem{IceCube:2020tka}
{\bf IceCube} Collaboration, M.~G. Aartsen et~al., {\it {Searching for eV-scale
  sterile neutrinos with eight years of atmospheric neutrinos at the IceCube
  Neutrino Telescope}},  {\em Phys. Rev. D} {\bf 102} (2020), no.~5 052009,
  [\href{http://arxiv.org/abs/2005.12943}{{\tt arXiv:2005.12943}}].

\bibitem{Razzaque:2021cft}
S.~Razzaque and L.~S. Miranda, {\it {Revisiting constraints on sterile neutrino
  mixing parameters using IceCube atmospheric neutrino data}},  {\em PoS} {\bf
  ICRC2019} (2021) 987.

\bibitem{Schneider:2021wzs}
A.~Schneider, B.~Skrzypek, C.~A. Arg\"uelles, and J.~M. Conrad, {\it {Closing
  the Neutrino ''BSM Gap'': Physics Potential of Atmospheric Through-Going
  Muons at DUNE}},  \href{http://arxiv.org/abs/2106.01508}{{\tt
  arXiv:2106.01508}}.

\bibitem{KM3NeT:2021uez}
{\bf KM3NeT} Collaboration, S.~Aiello et~al., {\it {Sensitivity to light
  sterile neutrino mixing parameters with KM3NeT/ORCA}},
  \href{http://arxiv.org/abs/2107.00344}{{\tt arXiv:2107.00344}}.

\bibitem{Chikashige:1980ui}
Y.~Chikashige, R.~N. Mohapatra, and R.~D. Peccei, {\it {Are There Real
  Goldstone Bosons Associated with Broken Lepton Number?}},  {\em Phys. Lett.
  B} {\bf 98} (1981) 265--268.

\bibitem{Gelmini:1980re}
G.~B. Gelmini and M.~Roncadelli, {\it {Left-Handed Neutrino Mass Scale and
  Spontaneously Broken Lepton Number}},  {\em Phys. Lett. B} {\bf 99} (1981)
  411--415.

\bibitem{Gelmini:1983ea}
G.~B. Gelmini and J.~W.~F. Valle, {\it {Fast Invisible Neutrino Decays}},  {\em
  Phys. Lett. B} {\bf 142} (1984) 181--187.

\bibitem{Super-Kamiokande:2011dgc}
{\bf Super-Kamiokande} Collaboration, K.~Abe et~al., {\it {Search for
  Differences in Oscillation Parameters for Atmospheric Neutrinos and
  Antineutrinos at Super-Kamiokande}},  {\em Phys. Rev. Lett.} {\bf 107} (2011)
  241801, [\href{http://arxiv.org/abs/1109.1621}{{\tt arXiv:1109.1621}}].

\bibitem{Coelho:2017cwp}
{\bf KM3NeT} Collaboration, J.~a. A.~B. Coelho, {\it {Probing new physics with
  atmospheric neutrinos at KM3NeT-ORCA}},  {\em J. Phys. Conf. Ser.} {\bf 888}
  (2017), no.~1 012115, [\href{http://arxiv.org/abs/1702.04508}{{\tt
  arXiv:1702.04508}}].

\bibitem{Kumar:2017sdq}
{\bf ICAL} Collaboration, S.~Ahmed et~al., {\it {Physics Potential of the ICAL
  detector at the India-based Neutrino Observatory (INO)}},  {\em Pramana} {\bf
  88} (2017), no.~5 79, [\href{http://arxiv.org/abs/1505.07380}{{\tt
  arXiv:1505.07380}}].

\bibitem{Behera:2014zca}
S.~P. Behera, M.~S. Bhatia, V.~M. Datar, and A.~K. Mohanty, {\it {Simulation
  Studies for Electromagnetic Design of INO ICAL Magnet and its Response to
  Muons}},  {\em IEEE Trans. Magnetics} {\bf 51} (2015) 4624,
  [\href{http://arxiv.org/abs/1406.3965}{{\tt arXiv:1406.3965}}].

\bibitem{Chatterjee:2014vta}
A.~Chatterjee, K.~Meghna, K.~Rawat, T.~Thakore, V.~Bhatnagar, et~al., {\it {A
  Simulations Study of the Muon Response of the Iron Calorimeter Detector at
  the India-based Neutrino Observatory}},  {\em JINST} {\bf 9} (2014) P07001,
  [\href{http://arxiv.org/abs/1405.7243}{{\tt arXiv:1405.7243}}].

\bibitem{GOSWAMI2009198}
S.~Goswami, {\it Physics program of india based neutrino observatory},  {\em
  Nuclear Physics B - Proceedings Supplements} {\bf 188} (2009) 198--200.
  Proceedings of the Neutrino Oscillation Workshop.

\bibitem{Ghosh:2012px}
A.~Ghosh, T.~Thakore, and S.~Choubey, {\it {Determining the Neutrino Mass
  Hierarchy with INO, T2K, NOvA and Reactor Experiments}},  {\em JHEP} {\bf
  1304} (2013) 009, [\href{http://arxiv.org/abs/1212.1305}{{\tt
  arXiv:1212.1305}}].

\bibitem{Thakore:2013xqa}
T.~Thakore, A.~Ghosh, S.~Choubey, and A.~Dighe, {\it {The Reach of INO for
  Atmospheric Neutrino Oscillation Parameters}},  {\em JHEP} {\bf 1305} (2013)
  058, [\href{http://arxiv.org/abs/1303.2534}{{\tt arXiv:1303.2534}}].

\bibitem{Ghosh:2013mga}
A.~Ghosh and S.~Choubey, {\it {Measuring the Mass Hierarchy with Muon and
  Hadron Events in Atmospheric Neutrino Experiments}},  {\em JHEP} {\bf 1310}
  (2013) 174, [\href{http://arxiv.org/abs/1306.1423}{{\tt arXiv:1306.1423}}].

\bibitem{Devi:2014yaa}
M.~M. Devi, T.~Thakore, S.~K. Agarwalla, and A.~Dighe, {\it {Enhancing
  sensitivity to neutrino parameters at INO combining muon and hadron
  information}},  {\em JHEP} {\bf 10} (2014) 189,
  [\href{http://arxiv.org/abs/1406.3689}{{\tt arXiv:1406.3689}}].

\bibitem{Mohan:2016gxm}
L.~S. Mohan and D.~Indumathi, {\it {Pinning down neutrino oscillation
  parameters in the 2--3 sector with a magnetised atmospheric neutrino
  detector: a new study}},  {\em Eur. Phys. J.} {\bf C77} (2017), no.~1 54,
  [\href{http://arxiv.org/abs/1605.04185}{{\tt arXiv:1605.04185}}].

\bibitem{Rebin:2018fdl}
K.~R. Rebin, J.~Libby, D.~Indumathi, and L.~S. Mohan, {\it {Study of neutrino
  oscillation parameters at the INO-ICAL detector using event-by-event
  reconstruction}},  {\em Eur. Phys. J. C} {\bf 79} (2019), no.~4 295,
  [\href{http://arxiv.org/abs/1804.02138}{{\tt arXiv:1804.02138}}].

\bibitem{Datta:2019uwv}
J.~Datta, M.~Nizam, A.~Ajmi, and S.~U. Sankar, {\it {Matter vs vacuum
  oscillations in atmospheric neutrinos}},  {\em Nucl. Phys. B} {\bf 961}
  (2020) 115251, [\href{http://arxiv.org/abs/1907.08966}{{\tt
  arXiv:1907.08966}}].

\bibitem{Kumar:2020wgz}
A.~Kumar, A.~Khatun, S.~K. Agarwalla, and A.~Dighe, {\it {From oscillation dip
  to oscillation valley in atmospheric neutrino experiments}},  {\em Eur. Phys.
  J. C} {\bf 81} (2021), no.~2 190,
  [\href{http://arxiv.org/abs/2006.14529}{{\tt arXiv:2006.14529}}].

\bibitem{Kumar:2021faw}
A.~Kumar and S.~K. Agarwalla, {\it {Validating the Earth\textquoteright{}s core
  using atmospheric neutrinos with ICAL at INO}},  {\em JHEP} {\bf 08} (2021)
  139, [\href{http://arxiv.org/abs/2104.11740}{{\tt arXiv:2104.11740}}].

\bibitem{Thakore:2018lgn}
T.~Thakore, M.~M. Devi, S.~K. Agarwalla, and A.~Dighe, {\it {Active-sterile
  neutrino oscillations at INO-ICAL over a wide mass-squared range}},  {\em
  JHEP} {\bf 08} (2018) 022, [\href{http://arxiv.org/abs/1804.09613}{{\tt
  arXiv:1804.09613}}].

\bibitem{Dash:2014fba}
N.~Dash, V.~M. Datar, and G.~Majumder, {\it {Sensitivity of the INO-ICAL
  detector to magnetic monopoles}},  {\em Astropart. Phys.} {\bf 70} (2015)
  33--38, [\href{http://arxiv.org/abs/1406.3938}{{\tt arXiv:1406.3938}}].

\bibitem{Behera:2016kwr}
S.~P. Behera, A.~Ghosh, S.~Choubey, V.~M. Datar, D.~K. Mishra, and A.~K.
  Mohanty, {\it {Search for the sterile neutrino mixing with the ICAL detector
  at INO}},  {\em Eur. Phys. J.} {\bf C77} (2017), no.~5 307,
  [\href{http://arxiv.org/abs/1605.08607}{{\tt arXiv:1605.08607}}].

\bibitem{Khatun:2017adx}
A.~Khatun, R.~Laha, and S.~K. Agarwalla, {\it {Indirect searches of Galactic
  diffuse dark matter in INO-MagICAL detector}},  {\em JHEP} {\bf 06} (2017)
  057, [\href{http://arxiv.org/abs/1703.10221}{{\tt arXiv:1703.10221}}].

\bibitem{Choubey:2017eyg}
S.~Choubey, S.~Goswami, C.~Gupta, S.~M. Lakshmi, and T.~Thakore, {\it
  {Sensitivity to neutrino decay with atmospheric neutrinos at the INO-ICAL
  detector}},  {\em Phys. Rev. D} {\bf 97} (2018), no.~3 033005,
  [\href{http://arxiv.org/abs/1709.10376}{{\tt arXiv:1709.10376}}].

\bibitem{Khatun:2018lzs}
A.~Khatun, T.~Thakore, and S.~K. Agarwalla, {\it {Can INO be Sensitive to
  Flavor-Dependent Long-Range Forces?}},  {\em JHEP} {\bf 04} (2018) 023,
  [\href{http://arxiv.org/abs/1801.00949}{{\tt arXiv:1801.00949}}].

\bibitem{Choubey:2017vpr}
S.~Choubey, A.~Ghosh, and D.~Tiwari, {\it {Prospects of Indirect Searches for
  Dark Matter at INO}},  {\em JCAP} {\bf 05} (2018) 006,
  [\href{http://arxiv.org/abs/1711.02546}{{\tt arXiv:1711.02546}}].

\bibitem{Tiwari:2018gxz}
D.~Tiwari, S.~Choubey, and A.~Ghosh, {\it {Prospects of indirect searches for
  dark matter annihilations in the earth with ICAL@INO}},  {\em JHEP} {\bf 05}
  (2019) 039, [\href{http://arxiv.org/abs/1806.05058}{{\tt arXiv:1806.05058}}].

\bibitem{Chatterjee:2014oda}
A.~Chatterjee, R.~Gandhi, and J.~Singh, {\it {Probing Lorentz and CPT Violation
  in a Magnetized Iron Detector using Atmospheric Neutrinos}},  {\em JHEP} {\bf
  1406} (2014) 045, [\href{http://arxiv.org/abs/1402.6265}{{\tt
  arXiv:1402.6265}}].

\bibitem{Coleman:1998ti}
S.~R. Coleman and S.~L. Glashow, {\it {High-energy tests of Lorentz
  invariance}},  {\em Phys. Rev. D} {\bf 59} (1999) 116008,
  [\href{http://arxiv.org/abs/hep-ph/9812418}{{\tt hep-ph/9812418}}].

\bibitem{Kaur:2017dpd}
D.~Kaur, Z.~A. Dar, S.~Kumar, and M.~Naimuddin, {\it {Search for the
  differences in atmospheric neutrino and antineutrino oscillation parameters
  at the INO-ICAL experiment}},  {\em Phys. Rev. D} {\bf 95} (2017), no.~9
  093005, [\href{http://arxiv.org/abs/1703.06710}{{\tt arXiv:1703.06710}}].

\bibitem{Dar:2019mnk}
Z.~A. Dar, D.~Kaur, S.~Kumar, and M.~Naimuddin, {\it {Independent measurement
  of muon neutrino and antineutrino oscillations at the INO\textendash{}ICAL
  experiment}},  {\em J. Phys. G} {\bf 46} (2019), no.~6 065001,
  [\href{http://arxiv.org/abs/2004.01127}{{\tt arXiv:2004.01127}}].

\bibitem{Pontecorvo:1957qd}
B.~Pontecorvo, {\it {Inverse beta processes and nonconservation of lepton
  charge}},  {\em Zh. Eksp. Teor. Fiz.} {\bf 34} (1957) 247.

\bibitem{Maki:1962mu}
Z.~Maki, M.~Nakagawa, and S.~Sakata, {\it {Remarks on the unified model of
  elementary particles}},  {\em Prog. Theor. Phys.} {\bf 28} (1962) 870--880.

\bibitem{Mikheyev:1985zog}
S.~P. Mikheyev and A.~Y. Smirnov, {\it {Resonance Amplification of Oscillations
  in Matter and Spectroscopy of Solar Neutrinos}},  {\em Sov. J. Nucl. Phys.}
  {\bf 42} (1985) 913--917.

\bibitem{Kostelecky:2008ts}
V.~A. Kostelecky and N.~Russell, {\it {Data Tables for Lorentz and CPT
  Violation}},  \href{http://arxiv.org/abs/0801.0287}{{\tt arXiv:0801.0287}}.

\bibitem{Kopp:2007ne}
J.~Kopp, M.~Lindner, T.~Ota, and J.~Sato, {\it {Non-standard neutrino
  interactions in reactor and superbeam experiments}},  {\em Phys. Rev. D} {\bf
  77} (2008) 013007, [\href{http://arxiv.org/abs/0708.0152}{{\tt
  arXiv:0708.0152}}].

\bibitem{DUNE:2015lol}
{\bf DUNE} Collaboration, R.~Acciarri et~al., {\it {Long-Baseline Neutrino
  Facility (LBNF) and Deep Underground Neutrino Experiment (DUNE)}: {Conceptual
  Design Report, Volume 2: The Physics Program for DUNE at LBNF}},
  \href{http://arxiv.org/abs/1512.06148}{{\tt arXiv:1512.06148}}.

\bibitem{Dziewonski:1981xy}
A.~Dziewonski and D.~Anderson, {\it {Preliminary Reference Earth Model}},  {\em
  Phys.Earth Planet.Interiors} {\bf 25} (1981) 297--356.

\bibitem{GonzalezGarcia:2004wg}
M.~C. Gonzalez-Garcia and M.~Maltoni, {\it {Atmospheric neutrino oscillations
  and new physics}},  {\em Phys. Rev.} {\bf D70} (2004) 033010,
  [\href{http://arxiv.org/abs/hep-ph/0404085}{{\tt hep-ph/0404085}}].

\bibitem{Mocioiu:2014gua}
I.~Mocioiu and W.~Wright, {\it {Non-standard neutrino interactions in the
  mu--tau sector}},  {\em Nucl.Phys.} {\bf B893} (2015) 376--390,
  [\href{http://arxiv.org/abs/1410.6193}{{\tt arXiv:1410.6193}}].

\bibitem{Kostelecky:2003cr}
V.~A. Kostelecky and M.~Mewes, {\it {Lorentz and CPT violation in neutrinos}},
  {\em Phys. Rev. D} {\bf 69} (2004) 016005,
  [\href{http://arxiv.org/abs/hep-ph/0309025}{{\tt hep-ph/0309025}}].

\bibitem{Kostelecky:2004hg}
V.~A. Kostelecky and M.~Mewes, {\it {Lorentz violation and short-baseline
  neutrino experiments}},  {\em Phys. Rev. D} {\bf 70} (2004) 076002,
  [\href{http://arxiv.org/abs/hep-ph/0406255}{{\tt hep-ph/0406255}}].

\bibitem{Kostelecky:2003xn}
V.~A. Kostelecky and M.~Mewes, {\it {Lorentz and CPT violation in the neutrino
  sector}},  {\em Phys. Rev. D} {\bf 70} (2004) 031902,
  [\href{http://arxiv.org/abs/hep-ph/0308300}{{\tt hep-ph/0308300}}].

\bibitem{Katori:2006mz}
T.~Katori, V.~A. Kostelecky, and R.~Tayloe, {\it {Global three-parameter model
  for neutrino oscillations using Lorentz violation}},  {\em Phys. Rev. D} {\bf
  74} (2006) 105009, [\href{http://arxiv.org/abs/hep-ph/0606154}{{\tt
  hep-ph/0606154}}].

\bibitem{Barger:2007dc}
V.~Barger, D.~Marfatia, and K.~Whisnant, {\it {Challenging Lorentz noninvariant
  neutrino oscillations without neutrino masses}},  {\em Phys. Lett. B} {\bf
  653} (2007) 267--277, [\href{http://arxiv.org/abs/0706.1085}{{\tt
  arXiv:0706.1085}}].

\bibitem{Diaz:2009qk}
J.~S. Diaz, V.~A. Kostelecky, and M.~Mewes, {\it {Perturbative Lorentz and CPT
  violation for neutrino and antineutrino oscillations}},  {\em Phys. Rev. D}
  {\bf 80} (2009) 076007, [\href{http://arxiv.org/abs/0908.1401}{{\tt
  arXiv:0908.1401}}].

\bibitem{Diaz:2011tx}
J.~S. Diaz, {\it {Overview of Lorentz Violation in Neutrinos}},  in {\em
  {Meeting of the APS Division of Particles and Fields}}, 9, 2011.
\newblock \href{http://arxiv.org/abs/1109.4620}{{\tt arXiv:1109.4620}}.

\bibitem{Barger:2011qj}
V.~Barger, J.~Liao, D.~Marfatia, and K.~Whisnant, {\it {Lorentz noninvariant
  oscillations of massless neutrinos are excluded}},  {\em Phys. Rev. D} {\bf
  84} (2011) 056014, [\href{http://arxiv.org/abs/1106.6023}{{\tt
  arXiv:1106.6023}}].

\bibitem{Diaz:2010ft}
J.~S. Diaz and V.~A. Kostelecky, {\it {Three-parameter Lorentz-violating
  texture for neutrino mixing}},  {\em Phys. Lett. B} {\bf 700} (2011) 25--28,
  [\href{http://arxiv.org/abs/1012.5985}{{\tt arXiv:1012.5985}}].

\bibitem{Kostelecky:2011gq}
A.~Kostelecky and M.~Mewes, {\it {Neutrinos with Lorentz-violating operators of
  arbitrary dimension}},  {\em Phys. Rev. D} {\bf 85} (2012) 096005,
  [\href{http://arxiv.org/abs/1112.6395}{{\tt arXiv:1112.6395}}].

\bibitem{Diaz:2011ia}
J.~S. Diaz and A.~Kostelecky, {\it {Lorentz- and CPT-violating models for
  neutrino oscillations}},  {\em Phys. Rev. D} {\bf 85} (2012) 016013,
  [\href{http://arxiv.org/abs/1108.1799}{{\tt arXiv:1108.1799}}].

\bibitem{Auerbach:2005tq}
{\bf LSND} Collaboration, L.~B. Auerbach et~al., {\it {Tests of Lorentz
  violation in anti-nu(mu) ---\ensuremath{>} anti-nu(e) oscillations}},  {\em
  Phys. Rev. D} {\bf 72} (2005) 076004,
  [\href{http://arxiv.org/abs/hep-ex/0506067}{{\tt hep-ex/0506067}}].

\bibitem{Adamson:2008aa}
{\bf MINOS} Collaboration, P.~Adamson et~al., {\it {Testing Lorentz Invariance
  and CPT Conservation with NuMI Neutrinos in the MINOS Near Detector}},  {\em
  Phys. Rev. Lett.} {\bf 101} (2008) 151601,
  [\href{http://arxiv.org/abs/0806.4945}{{\tt arXiv:0806.4945}}].

\bibitem{Adamson:2010rn}
{\bf MINOS} Collaboration, P.~Adamson et~al., {\it {A Search for Lorentz
  Invariance and CPT Violation with the MINOS Far Detector}},  {\em Phys. Rev.
  Lett.} {\bf 105} (2010) 151601, [\href{http://arxiv.org/abs/1007.2791}{{\tt
  arXiv:1007.2791}}].

\bibitem{Abbasi:2010kx}
{\bf IceCube} Collaboration, R.~Abbasi et~al., {\it {Search for a
  Lorentz-violating sidereal signal with atmospheric neutrinos in IceCube}},
  {\em Phys. Rev. D} {\bf 82} (2010) 112003,
  [\href{http://arxiv.org/abs/1010.4096}{{\tt arXiv:1010.4096}}].

\bibitem{AguilarArevalo:2011yi}
{\bf MiniBooNE} Collaboration, A.~A. Aguilar-Arevalo et~al., {\it {Test of
  Lorentz and CPT violation with Short Baseline Neutrino Oscillation
  Excesses}},  {\em Phys. Lett. B} {\bf 718} (2013) 1303--1308,
  [\href{http://arxiv.org/abs/1109.3480}{{\tt arXiv:1109.3480}}].

\bibitem{Adamson:2012hp}
{\bf MINOS} Collaboration, P.~Adamson et~al., {\it {Search for Lorentz
  invariance and CPT violation with muon antineutrinos in the MINOS Near
  Detector}},  {\em Phys. Rev. D} {\bf 85} (2012) 031101,
  [\href{http://arxiv.org/abs/1201.2631}{{\tt arXiv:1201.2631}}].

\bibitem{Abe:2012gw}
{\bf Double Chooz} Collaboration, Y.~Abe et~al., {\it {First Test of Lorentz
  Violation with a Reactor-based Antineutrino Experiment}},  {\em Phys. Rev. D}
  {\bf 86} (2012) 112009, [\href{http://arxiv.org/abs/1209.5810}{{\tt
  arXiv:1209.5810}}].

\bibitem{Rebel:2013vc}
B.~Rebel and S.~Mufson, {\it {The Search for Neutrino-Antineutrino Mixing
  Resulting from Lorentz Invariance Violation using neutrino interactions in
  MINOS}},  {\em Astropart. Phys.} {\bf 48} (2013) 78--81,
  [\href{http://arxiv.org/abs/1301.4684}{{\tt arXiv:1301.4684}}].

\bibitem{Diaz:2013iba}
J.~S. D\'\i{}az, T.~Katori, J.~Spitz, and J.~M. Conrad, {\it {Search for
  neutrino-antineutrino oscillations with a reactor experiment}},  {\em Phys.
  Lett. B} {\bf 727} (2013) 412--416,
  [\href{http://arxiv.org/abs/1307.5789}{{\tt arXiv:1307.5789}}].

\bibitem{Diaz:2016fqd}
J.~S. Diaz and T.~Schwetz, {\it {Limits on CPT violation from solar
  neutrinos}},  {\em Phys. Rev. D} {\bf 93} (2016), no.~9 093004,
  [\href{http://arxiv.org/abs/1603.04468}{{\tt arXiv:1603.04468}}].

\bibitem{Abe:2017eot}
{\bf T2K} Collaboration, K.~Abe et~al., {\it {Search for Lorentz and CPT
  violation using sidereal time dependence of neutrino flavor transitions over
  a short baseline}},  {\em Phys. Rev. D} {\bf 95} (2017), no.~11 111101,
  [\href{http://arxiv.org/abs/1703.01361}{{\tt arXiv:1703.01361}}].

\bibitem{Adey:2018qsd}
{\bf Daya Bay} Collaboration, D.~Adey et~al., {\it {Search for a time-varying
  electron antineutrino signal at Daya Bay}},  {\em Phys. Rev. D} {\bf 98}
  (2018), no.~9 092013, [\href{http://arxiv.org/abs/1809.04660}{{\tt
  arXiv:1809.04660}}].

\bibitem{Antonelli:2020nhn}
V.~Antonelli, L.~Miramonti, and M.~D.~C. Torri, {\it {Phenomenological Effects
  of CPT and Lorentz Invariance Violation in Particle and Astroparticle
  Physics}},  {\em Symmetry} {\bf 12} (2020), no.~11 1821,
  [\href{http://arxiv.org/abs/2110.09185}{{\tt arXiv:2110.09185}}].

\bibitem{Marrone:2021}
A.~Marrone, {\it {Phenomenology of Three Neutrino Oscillations}},  2021.
\newblock Talk given at the XIX International Workshop on Neutrino Telescopes,
  18th to 26th February, 2021, Padova, Italy, {\tt
  https://agenda.infn.it/event/24250/overview}.

\bibitem{NuFIT}
NuFIT 5.0 (2020), http://www.nu-fit.org/.

\bibitem{Esteban:2020cvm}
I.~Esteban, M.~C. Gonzalez-Garcia, M.~Maltoni, T.~Schwetz, and A.~Zhou, {\it
  {The fate of hints: updated global analysis of three-flavor neutrino
  oscillations}},  {\em JHEP} {\bf 09} (2020) 178,
  [\href{http://arxiv.org/abs/2007.14792}{{\tt arXiv:2007.14792}}].

\bibitem{deSalas:2020pgw}
P.~F. de~Salas, D.~V. Forero, S.~Gariazzo, P.~Mart\'\i{}nez-Mirav\'e, O.~Mena,
  C.~A. Ternes, M.~T\'ortola, and J.~W.~F. Valle, {\it {2020 global
  reassessment of the neutrino oscillation picture}},  {\em JHEP} {\bf 02}
  (2021) 071, [\href{http://arxiv.org/abs/2006.11237}{{\tt arXiv:2006.11237}}].

\bibitem{deGouvea:2005hk}
A.~de~Gouvea, J.~Jenkins, and B.~Kayser, {\it {Neutrino mass hierarchy, vacuum
  oscillations, and vanishing U(e3)}},  {\em Phys.Rev.} {\bf D71} (2005)
  113009, [\href{http://arxiv.org/abs/hep-ph/0503079}{{\tt hep-ph/0503079}}].

\bibitem{Nunokawa:2005nx}
H.~Nunokawa, S.~J. Parke, and R.~Zukanovich~Funchal, {\it {Another possible way
  to determine the neutrino mass hierarchy}},  {\em Phys.Rev.} {\bf D72} (2005)
  013009, [\href{http://arxiv.org/abs/hep-ph/0503283}{{\tt hep-ph/0503283}}].

\bibitem{Gandhi:2004md}
R.~Gandhi, P.~Ghoshal, S.~Goswami, P.~Mehta, and S.~U. Sankar, {\it {Large
  matter effects in nu(mu) to nu(tau) oscillations}},  {\em Phys.Rev.Lett.}
  {\bf 94} (2005) 051801, [\href{http://arxiv.org/abs/hep-ph/0408361}{{\tt
  hep-ph/0408361}}].

\bibitem{SANTONICO1981377}
R.~Santonico and R.~Cardarelli, {\it Development of resistive plate counters},
  {\em Nuclear Instruments and Methods in Physics Research} {\bf 187} (1981),
  no.~2 377 -- 380.

\bibitem{Bheesette:2009yrp}
S.~Bheesette, {\em {Design and Characterisation Studies of Resistive Plate
  Chambers}}.
\newblock PhD thesis, Indian Inst. Tech., Mumbai, 2009.

\bibitem{Bhuyan:2012zzc}
M.~Bhuyan et~al., {\it {Development of 2m x 2m size glass RPCs for INO}},  {\em
  Nucl. Instrum. Meth. A} {\bf 661} (2012) S64--S67.

\bibitem{Dash:2014ifa}
N.~Dash, V.~M. Datar, and G.~Majumder, {\it {A Study on the time resolution of
  Glass RPC}},  \href{http://arxiv.org/abs/1410.5532}{{\tt arXiv:1410.5532}}.

\bibitem{Bhatt:2016rek}
A.~D. Bhatt, V.~M. Datar, G.~Majumder, N.~K. Mondal, Pathaleswar, and
  B.~Satyanarayana, {\it {Improvement of time measurement with the INO-ICAL
  resistive plate chambers}},  {\em JINST} {\bf 11} (2016), no.~11 C11001.

\bibitem{Gaur:2017uaf}
A.~Gaur, A.~Kumar, and M.~Naimuddin, {\it {Study of timing response and charge
  spectra of glass based Resistive Plate Chamber detectors for INO-ICAL
  experiment}},  {\em JINST} {\bf 12} (2017), no.~03 C03081.

\bibitem{Casper:2002sd}
D.~Casper, {\it {The Nuance neutrino physics simulation, and the future}},
  {\em Nucl. Phys. Proc. Suppl.} {\bf 112} (2002) 161,
  [\href{http://arxiv.org/abs/hep-ph/0208030}{{\tt hep-ph/0208030}}].

\bibitem{Athar:2012it}
M.~Sajjad~Athar, M.~Honda, T.~Kajita, K.~Kasahara, and S.~Midorikawa, {\it
  {Atmospheric neutrino flux at INO, South Pole and Pyh\'asalmi}},  {\em Phys.
  Lett.} {\bf B718} (2013) 1375, [\href{http://arxiv.org/abs/1210.5154}{{\tt
  arXiv:1210.5154}}].

\bibitem{Honda:2015fha}
M.~Honda, M.~Sajjad~Athar, T.~Kajita, K.~Kasahara, and S.~Midorikawa, {\it
  {Atmospheric neutrino flux calculation using the NRLMSISE-00 atmospheric
  model}},  {\em Phys. Rev. D} {\bf 92} (2015), no.~2 023004,
  [\href{http://arxiv.org/abs/1502.03916}{{\tt arXiv:1502.03916}}].

\bibitem{Dash:2015blu}
N.~Dash, {\em {Feasibility studies for the detection of exotic particles using
  ICAL at INO}}.
\newblock PhD thesis, HBNI, Mumbai, 2015.

\bibitem{Devi:2013wxa}
M.~M. Devi, A.~Ghosh, D.~Kaur, L.~S. Mohan, S.~Choubey, et~al., {\it {Hadron
  energy response of the Iron Calorimeter detector at the India-based Neutrino
  Observatory}},  {\em JINST} {\bf 8} (2013) P11003,
  [\href{http://arxiv.org/abs/1304.5115}{{\tt arXiv:1304.5115}}].

\bibitem{Bhattacharya:2015bsp}
K.~Bhattacharya, {\em {Event Reconstruction for ICAL Detector and Neutrino Mass
  Hierarchy Sensitivity Analysis at India-based Neutrino Observatory (INO)}}.
\newblock PhD thesis, TIFR, Mumbai, 2015.

\bibitem{Pal:2014tre}
S.~Pal, {\em {Development of the INO-ICAL detector and its physics potential}}.
\newblock PhD thesis, HBNI, Mumbai, 2014.

\bibitem{Blennow:2013oma}
M.~Blennow, P.~Coloma, P.~Huber, and T.~Schwetz, {\it {Quantifying the
  sensitivity of oscillation experiments to the neutrino mass ordering}},  {\em
  JHEP} {\bf 1403} (2014) 028, [\href{http://arxiv.org/abs/1311.1822}{{\tt
  arXiv:1311.1822}}].

\bibitem{Baker:1983tu}
S.~Baker and R.~D. Cousins, {\it {Clarification of the Use of Chi Square and
  Likelihood Functions in Fits to Histograms}},  {\em Nucl. Instrum. Meth.}
  {\bf 221} (1984) 437--442.

\bibitem{Kameda:2002fx}
J.~Kameda, {\em {Detailed studies of neutrino oscillations with atmospheric
  neutrinos of wide energy range from 100 MeV to 1000 GeV in
  Super-Kamiokande}}.
\newblock PhD thesis, Tokyo U., 2002.

\bibitem{Huber:2002mx}
P.~Huber, M.~Lindner, and W.~Winter, {\it {Superbeams versus neutrino
  factories}},  {\em Nucl. Phys.} {\bf B645} (2002) 3--48,
  [\href{http://arxiv.org/abs/hep-ph/0204352}{{\tt hep-ph/0204352}}].

\bibitem{Fogli:2002pt}
G.~L. Fogli, E.~Lisi, A.~Marrone, D.~Montanino, and A.~Palazzo, {\it {Getting
  the most from the statistical analysis of solar neutrino oscillations}},
  {\em Phys. Rev. D} {\bf 66} (2002) 053010,
  [\href{http://arxiv.org/abs/hep-ph/0206162}{{\tt hep-ph/0206162}}].

\bibitem{Kostelecky:2000mm}
V.~A. Kostelecky and R.~Lehnert, {\it {Stability, causality, and Lorentz and
  CPT violation}},  {\em Phys. Rev. D} {\bf 63} (2001) 065008,
  [\href{http://arxiv.org/abs/hep-th/0012060}{{\tt hep-th/0012060}}].

\bibitem{Kostelecky:2001jc}
V.~A. Kostelecky, C.~D. Lane, and A.~G.~M. Pickering, {\it {One loop
  renormalization of Lorentz violating electrodynamics}},  {\em Phys. Rev. D}
  {\bf 65} (2002) 056006, [\href{http://arxiv.org/abs/hep-th/0111123}{{\tt
  hep-th/0111123}}].

\end{thebibliography}\endgroup

\end{document}